\documentclass[aps,prb,showpacs,twocolumn]{revtex4-1}
\usepackage{amsmath,amssymb,epsfig,graphicx,longtable,lineno}
\begin{document}
\title{Comparison between exact and semilocal exchange potentials:
An all-electron study for solids}
\author{Fabien Tran}
\author{Peter Blaha}
\affiliation{Institute of Materials Chemistry, Vienna University of Technology,
Getreidemarkt 9/165-TC, A-1060 Vienna, Austria}
\author{Markus Betzinger}
\author{Stefan Bl\"{u}gel}
\affiliation{Peter-Gr\"{u}nberg Institut and Institute for Advanced Simulation,
Forschungszentrum J\"{u}lich and JARA, D-52425 J\"{u}lich, Germany}

\begin{abstract}

The exact-exchange (EXX) potential, which is obtained by solving the
optimized-effective potential (OEP) equation, is compared to various approximate
semilocal exchange potentials for a set of selected solids
(C, Si, BN, MgO, Cu$_{2}$O, and NiO).
This is done in the framework of the linearized augmented plane-wave method,
which allows for a very accurate all-electron solution of electronic structure
problems in solids. In order to assess the ability of the semilocal potentials
to approximate the EXX-OEP, we considered the EXX total energy,
electronic structure, electric-field gradient, and magnetic moment.
An attempt to parameterize a semilocal exchange potential is also reported.

\end{abstract}

\maketitle

\section{\label{introduction}Introduction}

Given an expression for the total energy of an atom, molecule, or solid,
\begin{equation}
E_{\text{tot}} = T_{\text{s}} + E_{\text{en}} + E_{\text{H}} +
E_{\text{xc}} + E_{\text{nn}},
\label{Etot}
\end{equation}
where the terms on the right-hand side represent the noninteracting kinetic,
electron-nucleus, Hartree, exchange-correlation, and nucleus-nucleus energies,
respectively, the search for the Slater determinant which minimizes
$E_{\text{tot}}$ leads to one-electron Schr\"{o}dinger equations
\begin{equation}
\left(-\frac{1}{2}\nabla^{2}+v_{\text{en}}(\mathbf{r})+
v_{\text{H}}(\mathbf{r})+\hat{v}_{\text{xc}}(\mathbf{r})\right)\psi_{i}(\mathbf{r}) =
\varepsilon_{i}\psi_{i}(\mathbf{r})
\label{KS}
\end{equation}
for the orbitals $\psi_{i}$ and their energies $\varepsilon_{i}$.
[For ease of notation, all formulas are given in spin-unpolarized form
and for non-zero gap systems. $N$ will denote the number of (doubly) occupied
orbitals.] In the Kohn-Sham (KS)\cite{KohnPR65} version of density functional
theory (DFT),\cite{HohenbergPR64} the exchange-correlation potential
$\hat{v}_{\text{xc}}$ is calculated as the
functional derivative of $E_{\text{xc}}$ with respect to the electron density
$\rho$ ($\hat{v}_{\text{xc}}=\delta E_{\text{xc}}/\delta\rho$) and, as a
consequence, $\hat{v}_{\text{xc}}$ is a multiplicative potential
($\hat{v}_{\text{xc}}\psi_{i}=v_{\text{xc}}\psi_{i}$), i.e., it is the
same for all orbitals.
Instead, in the generalized KS (gKS) framework, formally introduced in
Ref.~\onlinecite{SeidlPRB96}, the derivative of $E_{\text{xc}}$ is taken with
respect to $\psi_{i}$
($\hat{v}_{\text{xc}}\psi_{i}=\delta E_{\text{xc}}/\delta\psi_{i}^{*}$)
as in the Hartree-Fock (HF) method. In other words, in the KS method the
minimization of the total energy [Eq.~(\ref{Etot})] is done with the constraint
that the orbitals forming the (single) Slater determinant are solutions to a
Schr\"{o}dinger equation with a
multiplicative potential, whereas in the gKS scheme this constraint is dropped.
For exchange-correlation functionals $E_{\text{xc}}$ which depend
explicitly only on $\rho$ (and eventually its derivatives) like in the local
density approximation (LDA) \cite{KohnPR65} or generalized gradient approximation
(GGA),\cite{BeckePRA88,PerdewPRL96} the gKS method
leads to the same multiplicative potential $v_{\text{xc}}$ as the KS method.
However, for an orbital-dependent functional $E_{\text{xc}}$, i.e.,
a functional which depends on the orbitals
$\psi_{i}$ not only via $\rho$, such as meta-GGA
(see Ref.~\onlinecite{SunPRB11b} and references therein),
self-interaction corrected,\cite{PerdewPRB81} or hybrid\cite{BeckeJCP93}
functionals, the gKS method leads to a non-multiplicative
(i.e., orbital-dependent) potential $\hat{v}_{\text{xc}}=v_{\text{xc},i}$ as in
the HF method. For such functionals,
the calculation of $\delta E_{\text{xc}}/\delta\rho$, as required in the KS
formalism, is highly non-trivial, but possible by solving the
optimized effective potential (OEP) equation.\cite{SharpPR53}

The focus of the present work will be on the multiplicative exchange potential
$v_{\text{x}}$. More specifically, approximate semilocal exchange potentials
will be compared to the exact exchange (EXX) potential obtained by means of the
OEP method (called EXX-OEP thereafter), which has been implemented very recently
\cite{BetzingerPRB11,BetzingerPRB12,BetzingerPRB13} within the linearized
augmented plane-wave\cite{AndersenPRB75,Singh,Blugel} (LAPW) method for solids.

The advantage of semilocal potentials, which depend on the local quantities
$\rho$, $\nabla\rho$, $\nabla^{2}\rho$, or the kinetic-energy density
$t=\sum_{i=1}^{N}\nabla\psi_{i}^{*}\cdot\nabla\psi_{i}$ is that they are rather
simple to implement and lead to calculations which are much faster than
EXX-OEP calculations or HF/hybrid calculations with a non-multiplicative
potential. The LDA exchange potential is for example a simple function of
$\rho$, whereas in the case of GGA functionals, the corresponding potential
becomes a function of $\rho$ and its first two derivatives.

The problem with the LDA and vast majority of GGA exchange functionals
$E_{\text{x}}$ is that their functional derivative $v_{\text{x}}$ barely
resembles the EXX-OEP potential.\cite{EngelPRA93,EngelPRB93}
Therefore several studies
have focused on the search for better semilocal approximations for
$v_{\text{x}}$ rather than $E_{\text{x}}$. Among these studies there are the
early works of Engel and Vosko,\cite{EngelPRB93} Baerends and co-workers,
\cite{vanLeeuwenPRA94,GritsenkoPRA95,GritsenkoIJQC96,SchipperJCP00}
and the more recent works of Becke and Johnson,\cite{BeckeJCP06}
Staroverov and co-workers,
\cite{StaroverovJCP08,GaidukPRA11,GaidukJCP12,GaidukCJC15}
Armiento \textit{et al}., \cite{ArmientoPRB08,ArmientoPRL13} and
others.\cite{UmezawaPRA06,KuismaPRB10}

It is worth mentioning that all these approximations for
$v_{\text{x}}$ can be categorized in one of these two groups,
namely, those which are functional derivative of a functional
$E_{\text{x}}$ and those which are not
(such potentials were termed \textit{stray} in Ref.~\onlinecite{GaidukJCTC09}).
Examples of exchange potentials which were modelled with
the constrained to be a functional derivative are the ones
from Engel and Vosko\cite{EngelPRB93} (EV93) and Armiento and K\"{u}mmel
\cite{ArmientoPRL13} (AK13), while the potentials from
van Leeuwen and Baerends\cite{vanLeeuwenPRA94} (LB94) and Becke and
Johnson\cite{BeckeJCP06} (BJ) are stray potentials.
Not constraining a potential to be a functional derivative means much more
freedom for its analytical form, however
it has been shown that stray potentials have undesirable features
both at the fundamental and practical level.
\cite{GaidukJCP09,GaidukJCTC09,KarolewskiPRA13}
Furthermore, attempts to turn a stray potential into a functional derivative
without loosing too much of its original features have been rather unsuccessful
up to now (see Refs.~\onlinecite{GaidukPRA11,GaidukJCP12,KarolewskiPRA13}).

As already mentioned above, various semilocal exchange potentials $v_{\text{x}}$
will be studied and compared to the EXX-OEP which will serve as reference.
We will focus in particular on the BJ potential,
which has been shown to reproduce quite well the EXX-OEP potential in
atoms\cite{BeckeJCP06,StaroverovJCP08} and has been applied to molecules
\cite{GaidukJCP08,OliveiraJCTC10} and solids,\cite{TranJPCM07} as well as
modified to improve the results in various cases.
\cite{ArmientoPRB08,StaroverovJCP08,KarolewskiJCTC09,TranPRL09,
RasanenJCP10,PittalisPRB10,CerqueiraJCTC}
Furthermore, in an attempt to be as close as possible to the EXX-OEP potential,
a more general form of the BJ potential will be proposed.

The paper is organized as follows. Section~\ref{theory} gives a summary of the
EXX-OEP method and introduces the tested semilocal exchange potentials, while the
computational details are given in Sec.~\ref{computationaldetails}. Then, the
results are presented and discussed in Sec.~\ref{results}.
Finally, Sec.~\ref{summaryoutlook} gives the
summary and an outlook for possible improvements.

\section{\label{theory}Theory}

\subsection{\label{Optimizedeffectivepotential}Optimized effective potential}

As mentioned in the Introduction, the calculation of the multiplicative
exchange-correlation potential
$v_{\text{xc}}=\delta E_{\text{xc}}/\delta\rho$ for any
orbital-dependent functional $E_{\text{xc}}$ can be achieved
by solving the integro-differential OEP equation\cite{SharpPR53} for
$v_{\text{xc}}$ (see Ref.~\onlinecite{KuemmelRMP08} for a review),
which is given in general by
\begin{equation}
\int\chi(\mathbf{r},\mathbf{r}')v_{\text{xc}}(\mathbf{r}')d^{3}r' =
\Lambda_{\text{xc}}(\mathbf{r}),
\label{OEP}
\end{equation}
with
\begin{eqnarray}
\Lambda_{\text{xc}}(\mathbf{r}) & = & \sum_{i}\left[\int 
\left(\frac{\delta E_{\text{xc}}}{\delta \psi_i(\mathbf{r}')}
\frac{\delta\psi_i(\mathbf{r}')}{\delta v_{\text{eff}}(\mathbf{r})} + \text{c.c.}
\right)d^3r'\right. \nonumber \\
& & \left.+ \frac{\delta E_{\text{xc}}}{\delta \varepsilon_i} \frac{\delta
\varepsilon_i}{\delta v_{\text{eff}}(\mathbf{r})}\right],
\label{lambda}
\end{eqnarray}
where $v_{\text{eff}}=v_{\text{en}}+v_{\text{H}}+v_{\text{xc}}$ is the
KS effective potential and
\begin{eqnarray}
\chi(\mathbf{r},\mathbf{r}') & = &
\frac{\delta\rho(\mathbf{r})}{\delta v_{\text{eff}}(\mathbf{r}')} \nonumber \\
& = & 2\sum_{i=1}^{N}\sum_{j=N+1}^{\infty}
\frac{\psi_{i}^{*}(\mathbf{r})\psi_{j}(\mathbf{r})
\psi_{j}^{*}(\mathbf{r}')\psi_{i}(\mathbf{r}')}
{\varepsilon_{i}-\varepsilon_{j}} + \text{c.c.}
\label{chi}
\end{eqnarray}
is the KS (non-interacting) density response function.

So far, the OEP method has been applied mostly to the EXX energy
\begin{equation}
E_{\text{x}}^{\text{EXX}} = -\sum_{i=1}^{N}\sum_{j=1}^{N}\int\int
\frac{
\psi_{i}^{*}(\mathbf{r})\psi_{j}(\mathbf{r})
\psi_{j}^{*}(\mathbf{r}')\psi_{i}(\mathbf{r}')}
{\left\vert\mathbf{r}-\mathbf{r}'\right\vert}d^{3}rd^{3}r',
\label{ExEXX}
\end{equation}
which has the same analytic form as the HF exchange energy, but is evaluated
with the KS orbitals instead of the HF orbitals. In the case of EXX,
the sum over $i$ in Eq.~(\ref{lambda}) runs over the occupied orbitals only
[Eq.~(\ref{ExEXX}) does not depend on unoccupied orbitals] and
$\delta E_{\text{xc}}/\delta\varepsilon_{i}=0$.

Talman and Shadwick\cite{TalmanPRA76} were the first who reported
EXX-OEP calculations (on spherical atoms). Initially, EXX-OEP was proposed
as an approximation to HF in order to get rid of the non-multiplicative
HF potential. Later, it was recognized that the EXX-OEP method
represents also the exact exchange method within the KS DFT framework
(see, e.g., Ref.~\onlinecite{EngelPRA93} and
references therein). Since then
EXX-OEP has attracted more and more attention. For solids
the first EXX-OEP implementation was reported
by Kotani.\cite{KotaniPRB94} Subsequent reports of OEP
calculations on solids (the focus of the present work) can
be found in Refs.
\onlinecite{KotaniPRL95,KotaniPRB96,KotaniJPCM98,KotaniJMMM98,
StaedelePRL97,StaedelePRB99,AulburPRB00,FleszarPRB01,
MagyarPRB04,RinkeNJP05,QteishPRB05,QteishPRB06,RinkeAPL06,
GruningJCP06,GruningPRB06,EngelPRL09,BetzingerPRB11,BetzingerPRB12,
BetzingerPRB13,KlimesJCP14}.

The implementation of the OEP equation is rather complicated and its solution
prone to instabilities in particular if localized basis functions are
used.\cite{StaroverovJCP06,GoerlingJCP08,GidopoulosPRA12} Recently, the
implementation of the EXX-OEP method within the LAPW method has been
reported.\cite{BetzingerPRB11,BetzingerPRB12,BetzingerPRB13,FriedrichPRA13}
It employs an auxiliary basis, the mixed product basis, for representing the
OEP. As discussed in detail in Refs.~\onlinecite{BetzingerPRB11,FriedrichPRA13},
in order to obtain a stable and physical EXX-OEP potential, the orbital
(LAPW) and auxiliary (mixed product) basis sets have to satisfy a
basis-set balance condition. This condition is
fulfilled when the orbital basis set is converged with respect to the
auxilary basis set and usually demands large orbital basis sets. The usage of
(uneconomically) large LAPW basis sets can be avoided if the response of the
LAPW basis functions is explicitly taken into account in the calculation of
the KS orbital response $\delta\psi_{i}/\delta v_{\text{eff}}$ in Eq.~(\ref{lambda})
and the calculation of the KS density response [Eq.~(\ref{chi})].
\cite{BetzingerPRB12,BetzingerPRB13}
In this way, a much faster convergence
is achieved with respect to basis set size (and number of unoccupied states)
and the basis balance condition is fulfilled with smaller orbital basis
sets.

\subsection{\label{semilocal}Semilocal potentials}

\begin{figure}
\includegraphics[width=\columnwidth]{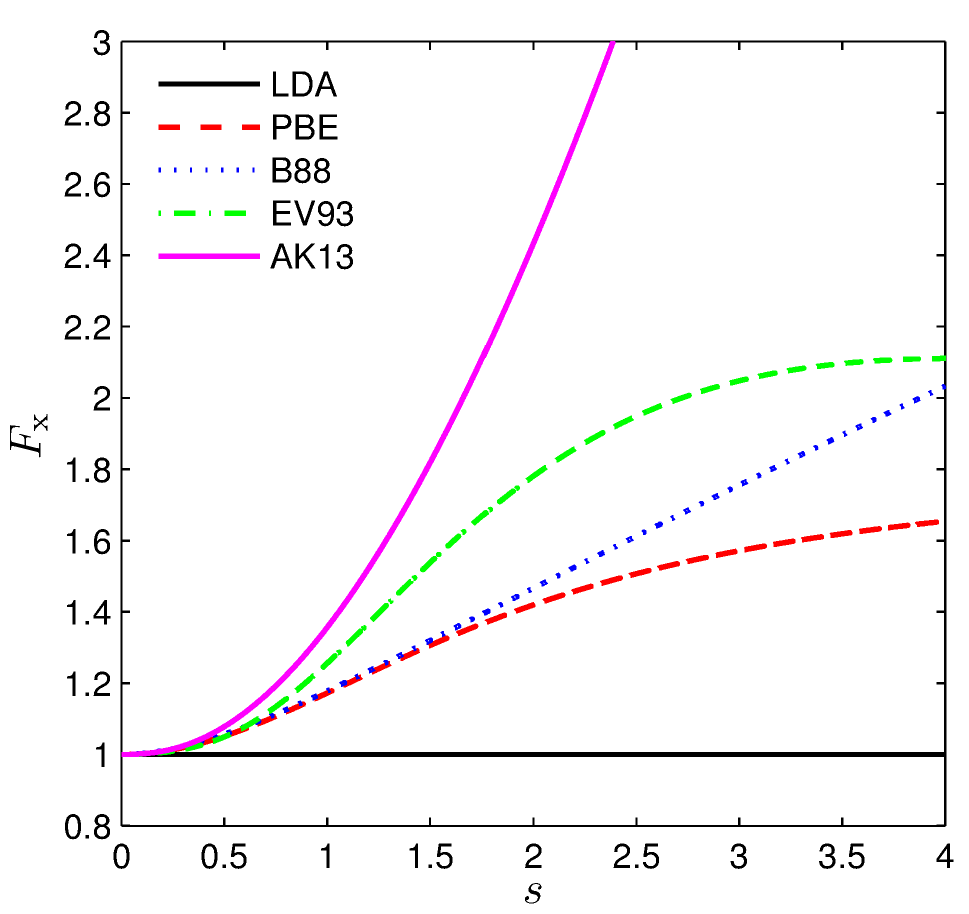}
\caption{\label{fig1}(Color online) The enhancement factors $F_{\text{x}}(s)$
[see Eq.~(\ref{ExGGA})] of the different exchange functionals considered in this work.}
\end{figure}

Among the considered semilocal exchange potentials, LDA \cite{KohnPR65} as well
as the GGAs B88, \cite{BeckePRA88} PBE,\cite{PerdewPRL96} EV93,\cite{EngelPRB93}
and AK13\cite{ArmientoPRL13} are functional derivatives of
exchange-energy functionals that have the generic form
\begin{equation}
E_{\text{x}} = -\frac{3}{4}\left(\frac{3}{\pi}\right)^{1/3}
\int\rho^{4/3}(\mathbf{r})F_{\text{x}}\left(s(\mathbf{r})\right)d^{3}r,
\label{ExGGA}
\end{equation}
where $F_{\text{x}}\left(s\right)$ is the so-called exchange enhancement factor
which depends on the reduced density gradient
$s=\left\vert\nabla\rho\right\vert/\left(2\left(3\pi^{2}\right)^{1/3}\rho^{4/3}\right)$.
LDA is the exact form for the homogeneous electron gas and corresponds to
$F_{\text{x}}(s)=1$. As shown in Fig.~\ref{fig1}, the enhancement factors of
the GGA functionals are larger than one, thus correcting the tendency of LDA
to underestimate the magnitude of the exchange energy. Compared to the
\textit{standard} PBE functional, EV93 and AK13 are much stronger.
Note that at $s=0$ all factors $F_{\text{x}}(s)$ reduce to one in order to satisfy
the homogeneous electron gas limit given by LDA. B88 and PBE, which are among the
most popular GGA functionals for calculating the
properties of molecules and solids, respectively, were constructed without
considering the quality of the potential $v_{\text{x}}$.
For EV93 and AK13, however, the emphasis was put on $v_{\text{x}}$.
The parameters in EV93 were determined by a fit to EXX-OEP potentials in
atoms,\cite{EngelPRB93} while Armiento and K\"{u}mmel were able
to find an analytical form for AK13 such that $v_{\text{x}}$ changes
discontinuously at integer particle numbers.\cite{ArmientoPRL13}
Both EV93 and AK13 were shown to improve over the standard LDA and PBE
functionals for the band gaps in solids.
\cite{DufekPRB94,TranJPCM07,ArmientoPRL13,VlcekPRB15}

In addition to these potentials, we consider in this work the BJ potential\cite{BeckeJCP06}
which is of stray type (see Refs.~\onlinecite{KarolewskiJCTC09,GaidukJCP09})
and has the form
\begin{equation}
v_{\text{x}}^{\text{BJ}}(\mathbf{r}) =
v_{\text{x}}^{\text{S/BR}}(\mathbf{r}) +
\frac{1}{\pi}\sqrt{\frac{5}{6}}\sqrt{\frac{t(\mathbf{r})}{\rho(\mathbf{r})}},
\label{vxBJ}
\end{equation}
where $v_{\text{x}}^{\text{S/BR}}$ is either the Slater (S) potential
\cite{SlaterPR51}
\begin{equation}
v_{\text{x}}^\text{S}(\mathbf{r}) =
-\frac{2}{\rho(\mathbf{r})}
\sum_{i=1}^{N}\sum_{j=1}^{N}
\psi_{i}^{*}(\mathbf{r})\psi_{j}(\mathbf{r})
\int\frac{\psi_{j}^{*}(\mathbf{r}')\psi_{i}(\mathbf{r}')}
{\left\vert\mathbf{r}-\mathbf{r}'\right\vert}d^{3}r'
\label{vxS}
\end{equation}
or the Becke-Roussel potential\cite{BeckePRA89}
\begin{equation}
v_{\text{x}}^{\text{BR}}(\mathbf{r}) =
-\frac{1}{b(\mathbf{r})}\left(1 - e^{-x(\mathbf{r})} -
\frac{1}{2}x(\mathbf{r})e^{-x(\mathbf{r})}\right).
\label{vxBR}
\end{equation}
The function $x$ in Eq.~(\ref{vxBR}) is calculated by solving (at each point
of space) the nonlinear equation (or using the analytical interpolation formula
for $x$ from Ref.~\onlinecite{ProynovCPL08})
\begin{equation}
\frac{x(\mathbf{r})e^{-2x(\mathbf{r})/3}}{x(\mathbf{r})-2} =
\frac{1}{3}\left(\frac{\pi}{2}\right)^{2/3}\frac{\rho^{5/3}(\mathbf{r})}{Q(\mathbf{r})},
\label{xeq}
\end{equation}
where
\begin{equation}
Q(\mathbf{r}) =
\frac{1}{12}\left(\nabla^{2}\rho(\mathbf{r}) - 4\gamma D(\mathbf{r})\right)
\label{Q}
\end{equation}
with
\begin{equation}
D(\mathbf{r}) = t(\mathbf{r}) -
\frac{1}{8}\frac{\left\vert\nabla\rho(\mathbf{r})\right\vert^{2}}
{\rho(\mathbf{r})}.
\label{D}
\end{equation}
After $x$ is calculated, $b$ in Eq.~(\ref{vxBR}) is given by
\begin{equation}
b(\mathbf{r}) =
\left(\frac{x^{3}(\mathbf{r})e^{-x(\mathbf{r})}}
{4\pi\rho(\mathbf{r})}\right)^{1/3}.
\label{b}
\end{equation}
The parameter $\gamma$ in Eq.~(\ref{Q}) has to be set to 1 or 0.8 in order to
recover the exact exchange potential of the hydrogen atom or the homogeneous
electron gas, respectively.\cite{BeckePRA89} Note that since the BR potential
and the second term of Eq.~(\ref{vxBJ}) depend on the kinetic-energy density
$t$, they are of the semilocal meta-GGA form,\cite{PerdewJCP05} while the
Slater potential [Eq.~(\ref{vxS})] is nonlocal
in the sense that the calculation of $v_{\text{x}}^{\text{S}}$ at
$\mathbf{r}$ requires the value of quantities (the occupied orbitals) at all
points of space $\mathbf{r}'$. For closed-shell atoms it was shown that the BR
potential is very close to the Slater potential.\cite{BeckePRA89,BeckeJCP06}
In the rest of this work, we focus on the BJ-based potentials
using the BR potential.

Several modifications of the BJ potential have been proposed.
\cite{ArmientoPRB08,StaroverovJCP08,TranPRL09,RasanenJCP10}
For instance, in Ref.~\onlinecite{TranPRL09} we proposed the modified
BJ (mBJ) potential 
\begin{equation}
v_{\text{x}}^{\text{mBJ}}(\mathbf{r}) =
cv_{\text{x}}^{\text{BR}}(\mathbf{r}) +
\left(3c-2\right)\frac{1}{\pi}\sqrt{\frac{5}{6}}
\sqrt{\frac{t(\mathbf{r})}{\rho(\mathbf{r})}},
\label{vxmBJ}
\end{equation}
where $c$ is a parameter that was introduced to improve the agreement with
experiment for the band gaps in solids and that was parameterized using the
average of $\left\vert\nabla\rho\right\vert/\rho$ in the unit cell.

As pointed out by R\"{a}s\"{a}nen \textit{et al}. in
Ref.~\onlinecite{RasanenJCP10}, the BJ potential is not gauge-invariant,
does not show the correct asymptotic behavior at
$\mathbf{r}\rightarrow\infty$ in finite systems, and the correction to the
Slater (or BR) term is not zero for one-electron systems as it should be.
In order to cure these deficiencies, they proposed an universal correction (UC)
to the BJ potential. For systems with zero current density $\mathbf{J}$
(like those considered in this work), the UC consists of replacing $t$ by $D$
[Eq.~(\ref{D})] in the second term of Eq.~(\ref{vxBJ}).

In an attempt to propose in the present work a semilocal exchange potential
which can reproduce accurately the EXX-OEP, we will consider a more general
form of the BJ and mBJ potentials, called generalized BJ (gBJ) thereafter:
\begin{equation}
v_{\text{x}}^{\text{gBJ}}(\mathbf{r}) =
cv_{\text{x}}^{\text{BR}}(\mathbf{r}) +
\left(3c-2\right)
\frac{\frac{1}{2}\left(\frac{3}{\pi}\right)^{1/3}}
{\left(\frac{3}{10}\left(3\pi^{2}\right)^{2/3}\right)^{p}}
\frac{t^{p}(\mathbf{r})}{\rho^{\left(5p-1\right)/3}(\mathbf{r})},
\label{vxgBJ}
\end{equation}
which, in addition to the two parameters $\gamma$ [in
$v_{\text{x}}^{\text{BR}}$, Eq.~(\ref{Q})] and $c$ as in mBJ, contains a third
one ($p$) whose value is $0.5$ in BJ and mBJ. The form of the second term of
Eq.~(\ref{vxgBJ}) was chosen such that the LDA exchange potential is recovered
for constant electron densities (and if $\gamma=0.8$, see above). As a
modification of Eq.~(\ref{vxgBJ}), we will also consider its variant where $t$
in the second term is replaced by $D$ (UC, Ref.~\onlinecite{RasanenJCP10}),
leading to the gBJUC potential. The parameters $\gamma$, $c$, and $p$ of the
gBJ and gBJUC potentials were varied around the standard BJ values
within the intervals
$\left[0.4,1.4\right]$, $\left[1.0,1.4\right]$, and $\left[0.35,0.65\right]$
and in steps of $0.2$, $0.1$, and $0.05$, respectively. In the following, BJ
will denote the unmodified potential given by Eq.~(\ref{vxBJ}) using BR with
$\gamma=0.8$ and similarly for BJUC.
The values of the parameters in the gBJ and gBJUC potentials will be specified
in this order: $(\gamma,c,p)$.

\begin{figure}
\begin{picture}(8,17)(0,0)
\put(0,11.2){\epsfxsize=8cm \epsfbox{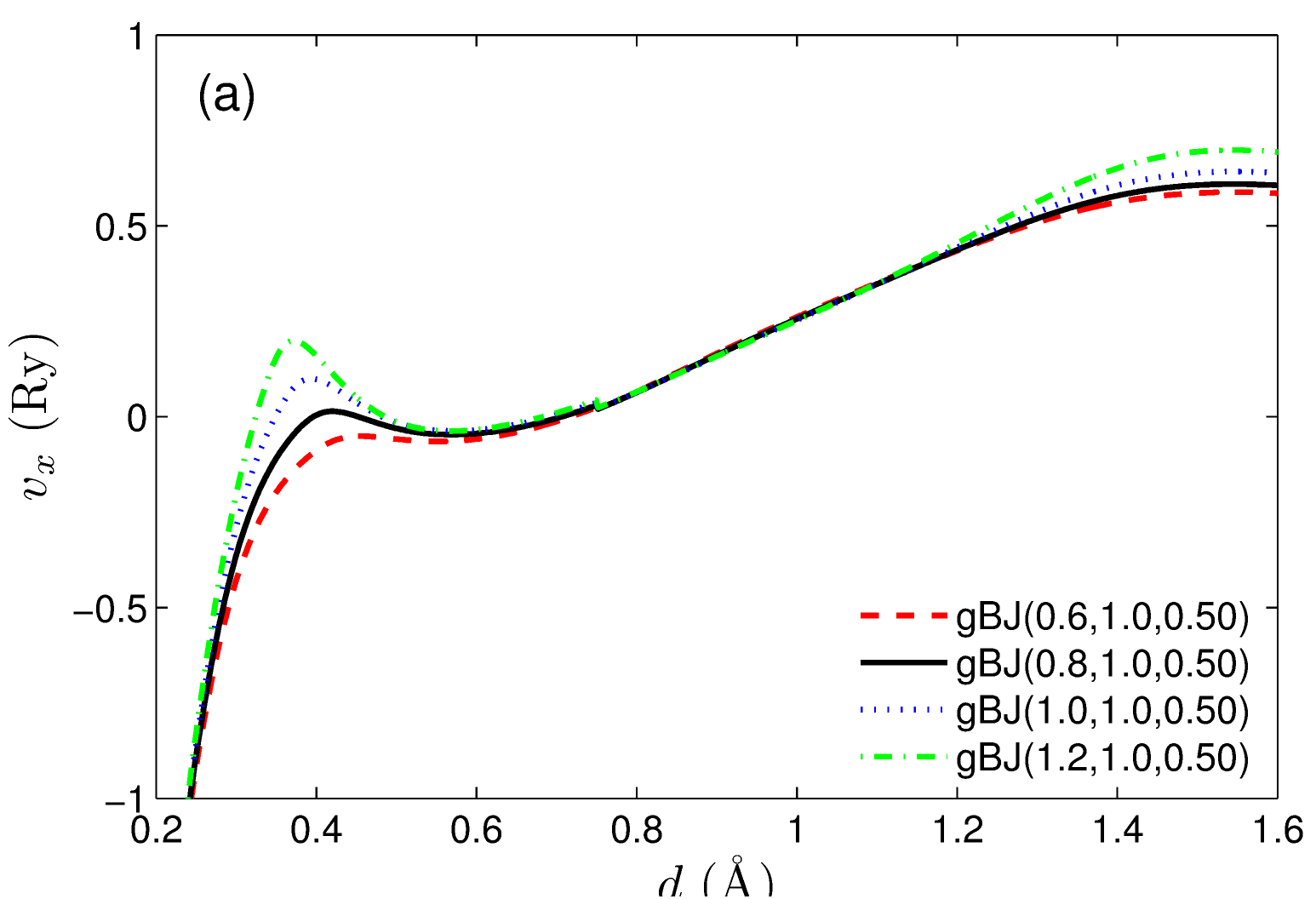}}
\put(0,5.6){\epsfxsize=8cm \epsfbox{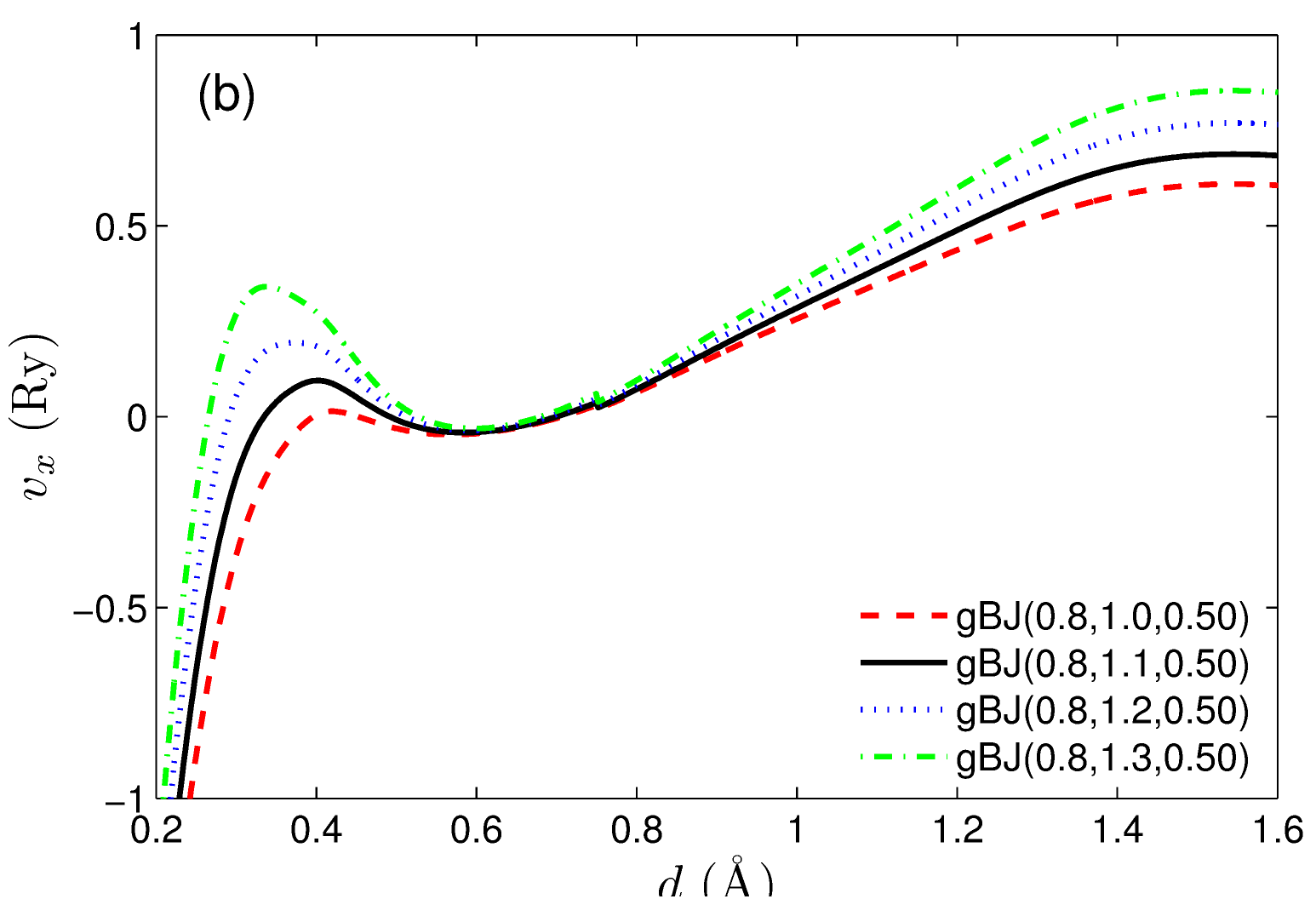}}
\put(0,0){\epsfxsize=8cm \epsfbox{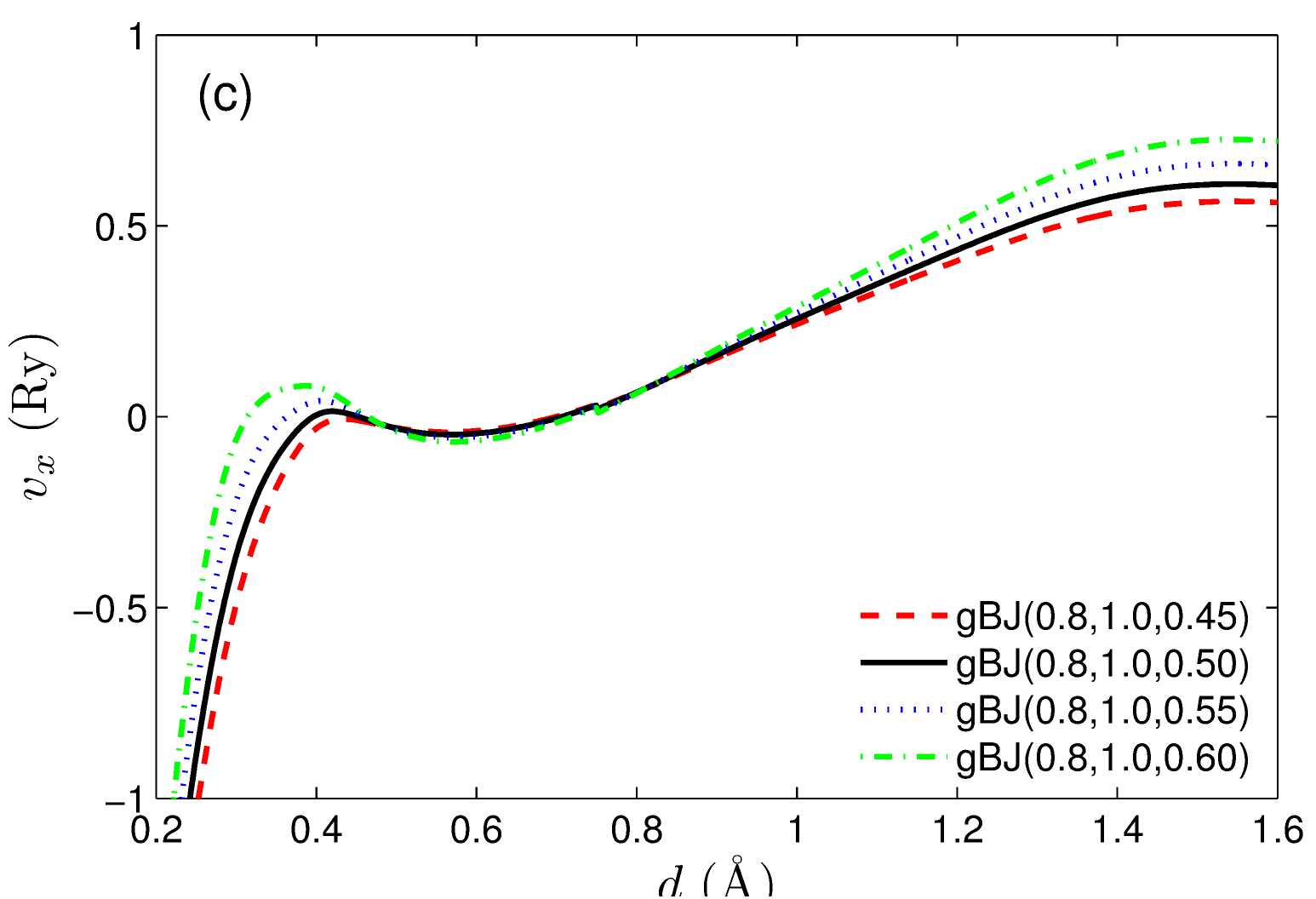}}
\end{picture}
\caption{\label{fig2}(Color online) gBJ$(\gamma,c,p)$ exchange potentials in C
plotted starting at a distance of 0.2 \AA~from the atom at site $(1/4,1/4,1/4)$
in the [111] direction. The center of the unit cell is at $d=2.32$ \AA.
The potentials were shifted such that
$\int_{\text{cell}}v_{\text{x}}(\mathbf{r})d^{3}r=0$.}
\end{figure}

\begin{figure}
\begin{picture}(8,11.3)(0,0)
\put(0,5.6){\epsfxsize=8cm \epsfbox{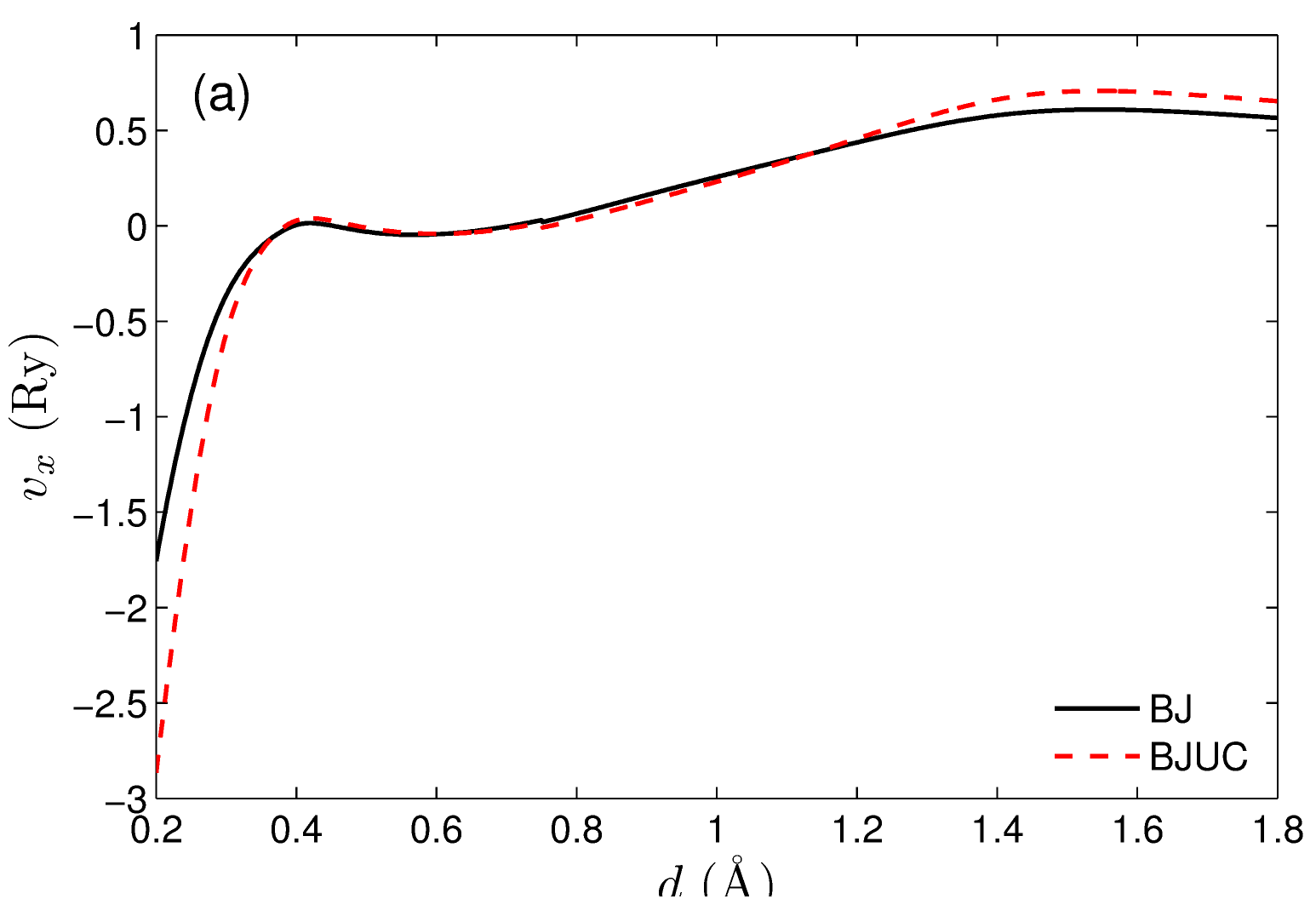}}
\put(0,0){\epsfxsize=8cm \epsfbox{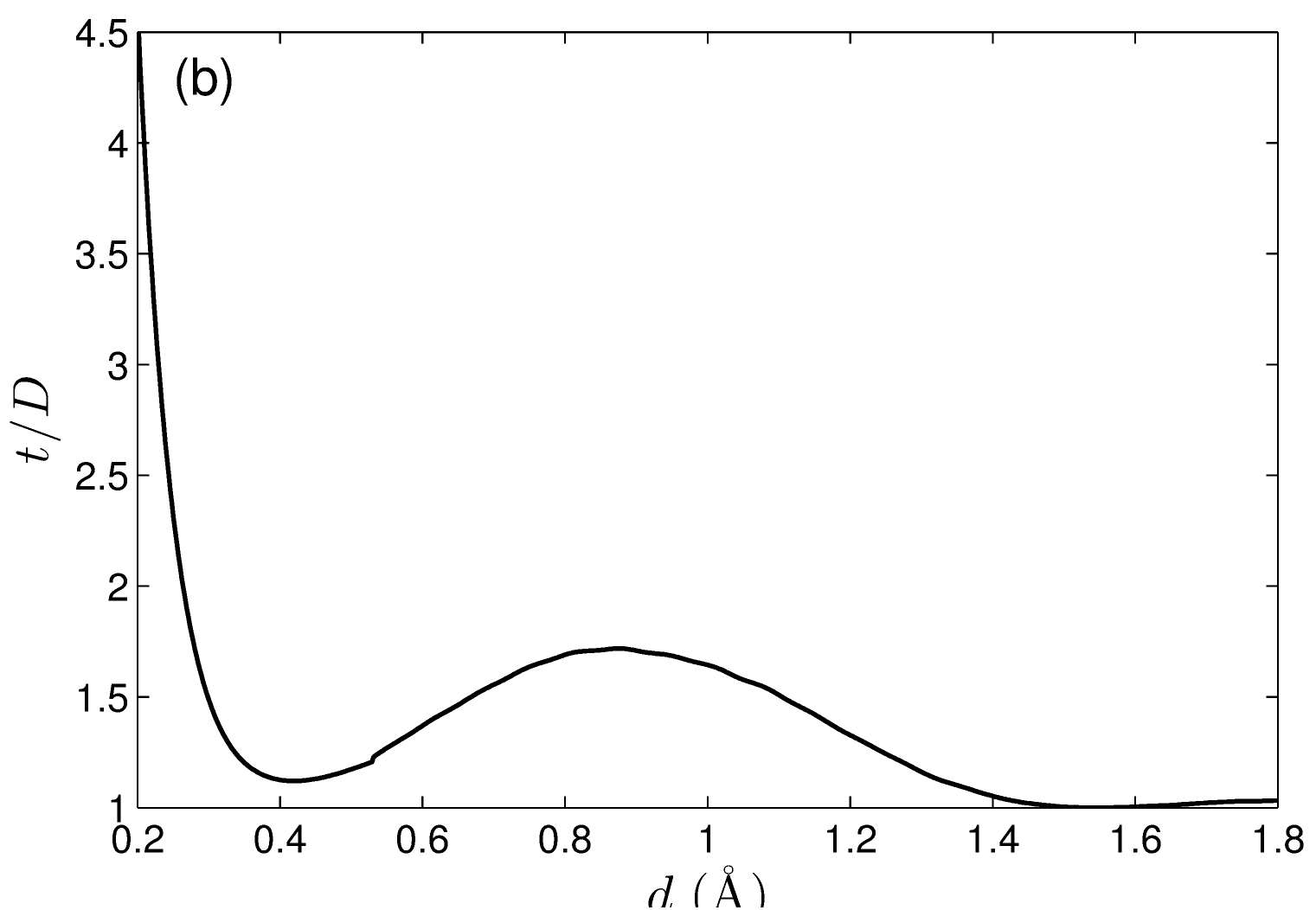}}
\end{picture}
\caption{\label{fig3}(Color online)
(a) BJ and BJUC exchange potentials in C. (b) Ratio $t/D$ for C.
The path and potential shift are as in Fig.~\ref{fig1}.}
\end{figure}

Regarding the influence of the parameters $\gamma$, $c$, and $p$ on the shape
of the gBJ potential, we have generally observed that an increase of one or
another of the parameters leads to more pronounced variations,
and this effect is rather similar for the three parameters
(see Fig.~\ref{fig2} for an illustrative example in the diamond phase of C).
Nevertheless, by looking more closely at Fig.~\ref{fig2}, we can notice some
differences in the way the parameters $\gamma$, $c$, and $p$ modify the
potential. For instance, when $\gamma$ is increased [Fig.~\ref{fig2}(a)],
the intershell peak at $d\sim0.4$ \AA~gets more spiky, while the value of $c$
affects the potential in a broader region of space [Fig.~\ref{fig2}(b)].
This is the reason why we have found it useful to use
three parameters instead of only one or two in our attempt for reproducing
at best the EXX-OEP results with the gBJ(UC) potential.

The effect of the UC is shown in Fig.~\ref{fig3}(a) by comparing the
BJ and BJUC potentials in C. Rather large differences between the two
potentials are visible in the core region ($d<0.4$ \AA~in this example) where
the BJUC potential is much more attractive than BJ. As a consequence,
the core states are bound stronger when the UC is used.
We made the same observation for all other investigated solids
(see also Fig.~3 of Ref.~\onlinecite{RasanenJCP10} for the Ne atom).
In Fig.~\ref{fig3}(b), the kinetic-energy density $t$ and $D$ [Eq.~(\ref{D})]
are compared by showing the ratio $t/D$, where we can see that it is indeed
in the core region that $t/D$ differs the most from 1.
Actually, the term $\left\vert\nabla\rho\right\vert^{2}/\left(8\rho\right)$ in
$D$ is the von Weizs\"{a}cker\cite{vonWeizsackerZP35} kinetic-energy density
which is equal to the exact kinetic-energy density
$t$ in regions of space dominated by a single orbital.
Therefore, in such regions $D=0$ and the BJUC potential reduces to the BR
(or Slater) potential which is by far too negative compared to
EXX-OEP.\cite{BeckeJCP06} Note that the (smaller) differences between
$t$ and $D$ in the region 0.4$-$1.4 \AA~also affect the potentials, however, due
to the alignment of the different potentials
$\left(\int_{\text{cell}}v_{\text{x}}(\mathbf{r})d^{3}r=0\right)$, the
differences between BJ and BJUC are transferred to $d>1.2$ \AA.

\section{\label{computationaldetails}Computational details}

The calculations with the semilocal potentials and EXX-OEP were done with the
WIEN2k\cite{WIEN2k} and FLEUR\cite{FLEUR} codes, respectively. Since the two
codes use the same basis set (LAPW\cite{AndersenPRB75,Singh,Blugel}), it was
also possible to calculate the EXX-OEP orbitals with WIEN2k by fixing the
potential $v_{\text{x}}$ to the EXX-OEP read from a file (containing the radial
functions and Fourier coefficients of the spherical harmonics and plane-wave
expansions of the potential) generated by FLEUR. Despite some (small) technical
differences between the two codes and different computational parameters used
for the calculations (e.g., basis sets), we observed for all solids, that
running a PBE calculation with WIEN2k as usual or with the FLEUR-generated PBE
potential leads to very close results (e.g., the transition energies differ by
less than 0.02 eV). Therefore, we are convinced that this
procedure of calculating orbitals using a potential generated from another code
is reliable in terms of accuracy. In addition, HF calculations
with the WIEN2k code were also done. Note, however, that in
the current implementation of the HF method,\cite{TranPRB11} the core
electrons experience a semilocal potential, like in other
implementations of the HF method with the LAPW basis
set.\cite{MassiddaPRB93,BetzingerPRB10}
The $\mathbf{k}$-mesh for the integration of the Brillouin zone and size
of the basis sets were chosen to be converged for the purpose of our work.

The set of solids that we will consider consists of the nonmagnetic cubic
(the space group, number of atoms in the primitive unit cell, and
cubic lattice constant are indicated in parenthesis)
C ($Fd\overline{3}m$, two atoms, 3.57 \AA), Si ($Fd\overline{3}m$, two atoms,
5.43 \AA), MgO ($Fm\overline{3}m$, two atoms, 4.23 \AA), BN ($F\overline{4}3m$,
two atoms, 3.62 \AA), and Cu$_{2}$O ($Pn\overline{3}m$, six atoms, 4.27 \AA).
C, Si, and BN are covalent, while MgO and Cu$_{2}$O are ionic.
Also included in our test set is NiO whose type-II antiferromagnetic order
along the [111] direction reduces the symmetry from cubic ($Fm\overline{3}m$,
two atoms, 4.17 \AA) to rhombohedral ($R\overline{3}m$, four atoms).
All these solids are nonmetallic and while most of them are simple $sp$-type
semiconductors or insulators, two of them, namely Cu$_{2}$O and NiO, represent
more stringent tests.

NiO is a rather difficult system to describe theoretically\cite{TerakuraPRB84}
since the Ni-$3d$ electrons are strongly correlated as it is generally the case
in magnetic $3d$-transition-metal oxides. Due to their inherent self-interaction
error,\cite{PerdewPRB81} the semilocal functionals perform particularly bad in
such systems and more advanced methods like DFT+$U$\cite{AnisimovPRB91} are
commonly used. In Refs.~\onlinecite{TranPRB11,KollerPRB11} we showed that a
correct description of the band gap and electric-field gradient (EFG) in Cu$_{2}$O
could only be achieved with the hybrid functionals, while the results obtained
with the LDA, GGA, LDA+$U$, and mBJ methods were qualitatively wrong.

\section{\label{results}Results and discussion}

We quantify the difference between the semilocal exchange potentials and
the reference EXX-OEP by comparing the EXX total energy and the electronic
structure for our test set of solids. The electronic structure of the solids
is assessed in terms of the band transition across the band gap, the position
of the core electrons, and the density of states (DOS). Furthermore, as a
measure of the similarity of the electron density we compare the resulting
EFG in Cu$_{2}$O and the magnetic moment in NiO.
We start with the discussion of the EXX total energy.

\subsection{\label{totalenergy}EXX total energy}

\begin{table*}
\caption{\label{table1}EXX total energy $E_{\text{tot}}^{\text{EXX}}$ (in Ry/cell)
calculated with orbitals
generated from various exchange potentials. The values for the semilocal potentials
are the differences with respect to EXX-OEP, and a positive value
indicates that EXX-OEP leads to a more negative energy as it always should.}
\begin{ruledtabular}
\begin{tabular}{lcccccc}
Potential                             & C        & Si        & BN       & MgO      & Cu$_{2}$O  & NiO       \\
\hline
EXX-OEP                               & $-151.592$ & $-1158.353$ & $-158.623$ & $-550.129$ & $-13527.744$ & $-6377.723$ \\
LDA                                   &     0.042  &      0.079  &     0.047  &     0.079  &       0.576  &      0.949  \\
PBE                                   &     0.026  &      0.040  &     0.027  &     0.037  &       0.351  &      0.619  \\
B88                                   &     0.026  &      0.040  &     0.026  &     0.036  &       0.357  &      0.620  \\
EV93                                  &     0.017  &      0.015  &     0.015  &     0.010  &       0.206  &      0.420  \\
AK13                                  &     0.029  &      0.027  &     0.023  &     0.014  &       0.152  &      0.312  \\
BJ                                    &     0.008  &      0.019  &     0.008  &     0.015  &       0.177  &      0.395  \\
BJUC                                  &     0.074  &      0.096  &     0.067  &     0.072  &       0.256  &      0.642  \\
gBJ$(0.6,1.0,0.60)$\footnotemark[1]   &     0.003  &      0.000  &     0.002  &     0.001  &       0.154  &      0.264  \\
gBJ$(1.4,1.1,0.50)$\footnotemark[2]   &     0.014  &      0.054  &     0.013  &     0.037  &       0.286  &      0.257  \\
gBJ$(0.4,1.3,0.65)$\footnotemark[3]   &     0.159  &      0.240  &     0.148  &     0.199  &       0.726  &      0.335  \\
gBJUC$(1.4,1.2,0.50)$\footnotemark[4] &     0.202  &      0.272  &     0.202  &     0.257  &       0.786  &      0.757  \\
\end{tabular}
\end{ruledtabular}
\footnotetext[1]{Good compromise for the EXX total energy of C, Si, BN, MgO, and Cu$_{2}$O.}
\footnotetext[2]{Good compromise for transition energies in C, Si, BN, and MgO.}
\footnotetext[3]{Good compromise for transition energies and Ni magnetic moment in NiO.}
\footnotetext[4]{Good compromise for transition energies and EFG in Cu$_{2}$O.}
\end{table*}

The EXX total energy $E_{\text{tot}}^{\text{EXX}}$ is calculated with orbitals
either generated from the multiplicative EXX-OEP or the semilocal exchange
potentials. The obtained total energies are shown in Table~\ref{table1}.
The lowest EXX total energy is obtained by using the EXX-OEP orbitals,
which was expected since the EXX-OEP is also the solution of the equation
$\delta E_{\text{tot}}^{\text{EXX}}/\delta v_{\text{eff}}=0$.\cite{TalmanPRA76}
The LDA orbitals lead to energies which are higher by 0.04-0.08 Ry
for C, Si, BN, and MgO, 0.6 Ry for Cu$_{2}$O, and 0.9 Ry for NiO.
All sets of GGA (PBE, B88, EV93, and AK13) orbitals improve upon LDA by
reducing the difference with respect to EXX-OEP by a factor of two up to four.
On average EV93 and AK13 yield total energies which are closer to the EXX-OEP
energy than PBE and B88. The BJ potential
[Eq.~(\ref{vxBJ})] shows a rather similar performance as EV93 and AK13,
while BJUC leads to total energies that are sometimes even worse than LDA.

The results for the gBJ potential [Eq.~(\ref{vxgBJ})] are shown for a few
selected sets of parameters $(\gamma$,$c$,$p$). With the parameters
$(0.6,1.0,0.60)$ the results are close to optimal (within the space of
parameters) for $E_{\text{tot}}^{\text{EXX}}$ and all solids except NiO for
which an increase of $c$ to 1.2 or 1.3 would further reduce the difference with
respect to EXX-OEP by a factor of two. It should be stressed that
the error obtained with gBJ$(0.6,1.0,0.60)$ is only of the order of $0.001$\%,
i.e., below 1-3 mRy for the light solids without transition-metal atoms.
Nevertheless, as shown below, such a good agreement for the total energy does not
necessarily mean a good agreement with EXX-OEP for other quantities
like the transition energies, which require other sets of parameters
$(\gamma$,$c$,$p$) (also shown in Table~\ref{table1}).
It should be also mentioned that
showing the results for the parameters $(0.6,1.0,0.60)$ is only one choice among
a few others which lead to a similar (albeit maybe slightly worse overall)
agreement with EXX-OEP. For instance, by varying only $c$
with respect to the original BJ potential (i.e., considering mBJ)
the results for $E_{\text{tot}}^{\text{EXX}}$ are also very good with $c=1.1$.
The gBJUC orbitals lead systematically to very high EXX total energies whatever
the parameters $(\gamma$,$c$,$p$) are. Actually, this is related to the poor
reproduction of the EXX-OEP potential by gBJUC in the region close to the
nuclei (see below) which substantially affects the total energy.

\subsection{\label{electronicstructure}Electronic structure}

We now turn to the discussion of the electronic structure and focus first on
the comparison of the direct transition energies across the band gap.

\subsubsection{\label{transitionenergies}Transition energies}

\begin{table*}
\caption{\label{table2}Statistics on direct transition energies
$\Delta\varepsilon_{\mathbf{k}}=
\varepsilon^{\mathbf{k}}_{N+1}-\varepsilon^{\mathbf{k}}_{N}$
at three different $\mathbf{k}$-points (specified in the text). The values for
EXX-OEP are the mean over the three $\mathbf{k}$-points of the transition energy
($\sum_{\mathbf{k}}\Delta\varepsilon_{\mathbf{k}}^{\text{EXX-OEP}}$), while for
the semilocal potentials the values are the MAE\protect\footnotemark[1]
and ME\protect\footnotemark[2] with respect to EXX-OEP. All values are in eV.}
\begin{ruledtabular}
\begin{tabular}{lcccccccccccc}
\multicolumn{1}{l}{Potential} &
\multicolumn{2}{c}{C} &
\multicolumn{2}{c}{Si} &
\multicolumn{2}{c}{BN} &
\multicolumn{2}{c}{MgO} &
\multicolumn{2}{c}{Cu$_{2}$O} &
\multicolumn{2}{c}{NiO} \\
\hline
\multicolumn{1}{l}{EXX-OEP} &
\multicolumn{2}{c}{9.80} &
\multicolumn{2}{c}{3.49} &
\multicolumn{2}{c}{10.93} &
\multicolumn{2}{c}{9.50} &
\multicolumn{2}{c}{3.08} &
\multicolumn{2}{c}{5.37} \\
\cline{2-3}\cline{4-5}\cline{6-7}\cline{8-9}\cline{10-11}\cline{12-13}
\multicolumn{1}{l}{} &
\multicolumn{1}{c}{MAE} &
\multicolumn{1}{c}{ME} &
\multicolumn{1}{c}{MAE} &
\multicolumn{1}{c}{ME} &
\multicolumn{1}{c}{MAE} &
\multicolumn{1}{c}{ME} &
\multicolumn{1}{c}{MAE} &
\multicolumn{1}{c}{ME} &
\multicolumn{1}{c}{MAE} &
\multicolumn{1}{c}{ME} &
\multicolumn{1}{c}{MAE} &
\multicolumn{1}{c}{ME} \\
\hline
LDA                                   & 0.62 & $-0.62$ & 0.68 & $-0.68$ & 0.97 & $-0.97$ & 1.95 & $-1.95$ & 1.23 & $-1.23$ & 3.68 & $-3.68$ \\
PBE                                   & 0.32 & $-0.32$ & 0.41 & $-0.41$ & 0.62 & $-0.62$ & 1.30 & $-1.30$ & 1.05 & $-1.05$ & 3.11 & $-3.11$ \\
B88                                   & 0.32 & $-0.32$ & 0.40 & $-0.40$ & 0.60 & $-0.60$ & 1.26 & $-1.26$ & 1.04 & $-1.04$ & 3.08 & $-3.08$ \\
EV93                                  & 0.31 & $-0.18$ & 0.23 & $-0.17$ & 0.41 & $-0.41$ & 0.81 & $-0.81$ & 0.98 & $-0.98$ & 2.78 & $-2.78$ \\
AK13                                  & 0.30 & $-0.03$ & 0.38 &   0.18  & 0.27 & $-0.11$ & 0.19 &   0.19  & 0.82 & $-0.82$ & 2.27 & $-2.27$ \\
BJ                                    & 0.27 & $-0.27$ & 0.38 & $-0.38$ & 0.45 & $-0.45$ & 1.01 & $-1.01$ & 0.95 & $-0.95$ & 2.53 & $-2.53$ \\
BJUC                                  & 0.38 & $-0.38$ & 0.53 & $-0.53$ & 0.45 & $-0.45$ & 0.44 & $-0.44$ & 0.42 & $-0.42$ & 2.64 & $-2.64$ \\
gBJ$(0.6,1.0,0.60)$\footnotemark[3]   & 0.13 & $-0.13$ & 0.17 & $-0.17$ & 0.29 & $-0.29$ & 0.71 & $-0.71$ & 1.01 & $-1.01$ & 2.22 & $-2.22$ \\
gBJ$(1.4,1.1,0.50)$\footnotemark[4]   & 0.14 &   0.14  & 0.06 &   0.02  & 0.07 &   0.07  & 0.19 & $-0.19$ & 0.87 & $-0.87$ & 1.49 & $-1.35$ \\
gBJ$(0.4,1.3,0.65)$\footnotemark[5]   & 0.86 &   0.86  & 1.42 &   1.42  & 1.13 &   1.13  & 1.67 &   1.67  & 1.19 & $-1.19$ & 0.28 &   0.11  \\
gBJUC$(1.4,1.2,0.50)$\footnotemark[6] & 0.10 &   0.10  & 0.01 & $-0.01$ & 0.30 &   0.30  & 0.88 &   0.88  & 0.06 &   0.06  & 1.70 & $-1.54$ \\
\end{tabular}
\end{ruledtabular}
\footnotetext[1]{$\text{MAE}=\sum_{\mathbf{k}}\left\vert
\Delta\varepsilon_{\mathbf{k}}^{\text{semilocal}}-
\Delta\varepsilon_{\mathbf{k}}^{\text{EXX-OEP}}\right\vert$.}
\footnotetext[2]{$\text{ME}=\sum_{\mathbf{k}}\left(
\Delta\varepsilon_{\mathbf{k}}^{\text{semilocal}}-
\Delta\varepsilon_{\mathbf{k}}^{\text{EXX-OEP}}\right)$.}
\footnotetext[3]{Good compromise for the EXX total energy of C, Si, BN, MgO, and Cu$_{2}$O.}
\footnotetext[4]{Good compromise for transition energies in C, Si, BN, and MgO.}
\footnotetext[5]{Good compromise for transition energies and Ni magnetic moment in NiO.}
\footnotetext[6]{Good compromise for transition energies and EFG in Cu$_{2}$O.}
\end{table*}

For each solid the direct transition energies are calculated at three
$\mathbf{k}$-points in the Brillouin zone (expressed in primitive basis for NiO
and conventional basis for the other solids):
$\Gamma=\left(0,0,0\right)$, $X=\left(0,1,0\right)$, and
$L=\left(1/2,1/2,1/2\right)$ for C, Si, BN, and MgO.
$\Gamma=\left(0,0,0\right)$, $X=\left(0,1/2,0\right)$, and
$M=\left(1/2,1/2,0\right)$ for Cu$_{2}$O.
$\Gamma=\left(0,0,0\right)$, $L=\left(0,1/2,0\right)$, and
$F=\left(0,1/2,1/2\right)$ for NiO.
The mean
error (ME) and mean absolute error (MAE) with respect to the EXX-OEP is shown
in Table~\ref{table2} for the different solids and potentials. Applying the LDA
the MAE is in the range of 0.6-3.7 eV where the largest error is for NiO. 
Actually, it is well known\cite{StaedelePRL97} that LDA strongly underestimates
the band gap with respect to experiment and EXX-OEP. The GGA, and in particular
EV93 and AK13, improve over LDA by reducing the MAE by a few 0.1 eV for C, Si,
BN, and Cu$_{2}$O or more than 1 eV for MgO and NiO, but overall the errors
remain rather substantial. BJ and BJUC perform similarly to EV93 or AK13.
For all these potentials except AK13, the negative sign of the ME and its
magnitude which is equal to the MAE in most cases indicate that the deviation
from EXX-OEP corresponds to an systematic underestimation of the transition
energies.

For gBJ, we found that the parameters $(1.4,1.1,0.50)$ lead to a very small MAE
(below 0.2 eV) for C, Si, BN, and MgO. For Cu$_{2}$O and NiO different sets
of parameters are required. For Cu$_{2}$O it was not possible to find a
combination of the parameters (within the considered ranges) that
reduces the MAE below 0.7 eV. However, by considering the gBJ potential
with the UC (gBJUC), we were able to improve substantially the
results for the transitions energies. For instance (see
Table~\ref{table2}), with the parameters $(1.4,1.2,0.50)$, gBJUC leads to a
MAE of 0.06 eV for the transition energies of Cu$_{2}$O.

In the case of NiO, a substantial improvement for the transition energies can
be obtained if the parameter $c$ is increased to at least 1.2. For instance,
with the parameters $(0.4,1.3,0.65)$ the MAE on the transition energies is
below 0.3 eV, which is one order of magnitude smaller than with the other
methods. We mention that for NiO, the values of the parameters $\gamma$ and
$p$ have little influence on the results and only an increase of $c$ can lead
to a clear improvement.

\subsubsection{\label{corestates}Core states}

\begin{table*}
\caption{\label{table3}MARE\protect\footnotemark[1] and
MRE\protect\footnotemark[2] (with respect to EXX-OEP and in \%) for the energy
position of the core states with respect to the valence band maximum
($\Delta\varepsilon_{\text{core},i}=\varepsilon_{\text{core},i}-\varepsilon_{\text{VBM}}$).
The MARE and MRE are over all core states in the solid: C ($1s$), Si ($1s$),
BN (B: $1s$; N: $1s$), MgO (Mg: $1s$; O: $1s$),
Cu$_{2}$O (Cu: $1s$, $2s$, $2p$; O: $1s$), NiO (Ni: $1s$, $2s$, $2p$; O: $1s$).
A negative MRE means that on average the core states are deeper in energy than
EXX-OEP.}
\begin{ruledtabular}
\begin{tabular}{lcccccccccccc}
\multicolumn{1}{l}{} &
\multicolumn{2}{c}{C} &
\multicolumn{2}{c}{Si} &
\multicolumn{2}{c}{BN} &
\multicolumn{2}{c}{MgO} &
\multicolumn{2}{c}{Cu$_{2}$O} &
\multicolumn{2}{c}{NiO} \\
\cline{2-3}\cline{4-5}\cline{6-7}\cline{8-9}\cline{10-11}\cline{12-13}
\multicolumn{1}{l}{Potential} &
\multicolumn{1}{c}{MARE} &
\multicolumn{1}{c}{MRE} &
\multicolumn{1}{c}{MARE} &
\multicolumn{1}{c}{MRE} &
\multicolumn{1}{c}{MARE} &
\multicolumn{1}{c}{MRE} &
\multicolumn{1}{c}{MARE} &
\multicolumn{1}{c}{MRE} &
\multicolumn{1}{c}{MARE} &
\multicolumn{1}{c}{MRE} &
\multicolumn{1}{c}{MARE} &
\multicolumn{1}{c}{MRE} \\
\hline
LDA                                   &  1.67 &    1.67  & 0.83 &   0.83  &  2.46 &    2.46  & 1.28 &   1.28  & 0.70 &   0.70  & 0.72 &   0.70  \\
PBE                                   &  0.28 &    0.28  & 0.31 &   0.31  &  0.95 &    0.95  & 0.49 &   0.49  & 0.37 &   0.37  & 0.52 &   0.45  \\
B88                                   &  0.20 &    0.20  & 0.28 &   0.28  &  0.85 &    0.85  & 0.45 &   0.45  & 0.34 &   0.34  & 0.51 &   0.42  \\
EV93                                  &  0.27 &  $-0.27$ & 0.21 &   0.21  &  0.54 &    0.40  & 0.27 &   0.26  & 0.35 &   0.32  & 0.48 &   0.43  \\
AK13                                  &  1.25 &  $-1.25$ & 0.07 & $-0.07$ &  0.66 &  $-0.66$ & 0.36 & $-0.19$ & 0.39 &   0.12  & 0.50 &   0.28  \\
BJ                                    &  1.13 &  $-1.13$ & 0.11 & $-0.11$ &  0.45 &  $-0.42$ & 0.46 & $-0.24$ & 0.27 & $-0.04$ & 0.53 &   0.08  \\
BJUC                                  &  7.82 &  $-7.82$ & 2.08 & $-2.08$ &  7.44 &  $-7.44$ & 3.56 & $-3.56$ & 1.50 & $-1.50$ & 1.17 & $-1.17$ \\
gBJ$(0.6,1.0,0.60)$\footnotemark[3]   &  0.74 &  $-0.74$ & 0.03 & $-0.03$ &  0.56 &    0.00  & 0.39 & $-0.05$ & 0.20 & $-0.02$ & 0.48 &   0.08  \\
gBJ$(1.4,1.1,0.50)$\footnotemark[4]   &  1.61 &  $-1.61$ & 0.06 & $-0.06$ &  1.03 &  $-1.03$ & 0.56 & $-0.40$ & 0.64 &   0.20  & 0.63 &   0.43  \\
gBJ$(0.4,1.3,0.65)$\footnotemark[5]   &  0.11 &  $-0.11$ & 0.04 &   0.04  &  0.67 &    0.62  & 0.24 &   0.19  & 0.17 &   0.00  & 0.40 &   0.15  \\
gBJUC$(1.4,1.2,0.50)$\footnotemark[6] & 12.96 & $-12.96$ & 3.35 & $-3.35$ & 12.96 & $-12.96$ & 6.06 & $-6.06$ & 2.21 & $-2.21$ & 1.72 & $-1.62$ \\
\end{tabular}
\end{ruledtabular}
\footnotetext[1]{$\text{MARE}=\sum_{i}^{\text{core}}100\left\vert
\Delta\varepsilon_{\text{core},i}^{\text{semilocal}}-
\Delta\varepsilon_{\text{core},i}^{\text{EXX-OEP}}\right\vert
/\left\vert\Delta\varepsilon_{\text{core},i}^{\text{EXX-OEP}}\right\vert$.}
\footnotetext[2]{$\text{MRE}=\sum_{i}^{\text{core}}100\left(
\Delta\varepsilon_{\text{core},i}^{\text{semilocal}}-
\Delta\varepsilon_{\text{core},i}^{\text{EXX-OEP}}\right)/
\left\vert\Delta\varepsilon_{\text{core},i}^{\text{EXX-OEP}}\right\vert$.}
\footnotetext[3]{Good compromise for the EXX total energy of C, Si, BN, MgO, and Cu$_{2}$O.}
\footnotetext[4]{Good compromise for transition energies in C, Si, BN, and MgO.}
\footnotetext[5]{Good compromise for transition energies and Ni magnetic moment in NiO.}
\footnotetext[6]{Good compromise for transition energies and EFG in Cu$_{2}$O.}
\end{table*}

We proceed by discussing the effect of the different potentials on the binding
energies of the core electrons. Table~\ref{table3} shows the averaged energetic
position of the core states with respect to the Fermi energy for the different
solids and potentials. For the definition of the mean absolute relative error (MARE)
and mean relative error (MRE) see caption of Table~\ref{table3}. As observed  for the
EXX total energy and the transition energies, all GGA exchange potentials
improve over LDA  by reducing the MARE below 0.5\% for most solids.
The positive MRE for LDA, PBE, and B88 indicate that these potentials
lead to core states which are typically bound to loosely with respect to EXX-OEP.
For EV93, AK13, and BJ there is no systematic trend. Among the four selected
parameterizations of the gBJ potential it turns out that $(0.4,1.3,0.65)$
(optimized for NiO) leads overall to a rather clear improvement over the LDA
and GGA potentials. The accuracy obtained with the set of parameters
$(0.6,1.0,0.60)$ (optimized for the EXX total energy) is satisfying
except for C. The results obtained with the gBJUC-based potentials are
extremely inaccurate, which is, as already mentioned above, due to the very poor
reproduction of the EXX-OEP close to the nuclei, leading to core states that are
too low in energy.

\subsubsection{\label{dos}Density of states}

\begin{figure}
\includegraphics[width=\columnwidth]{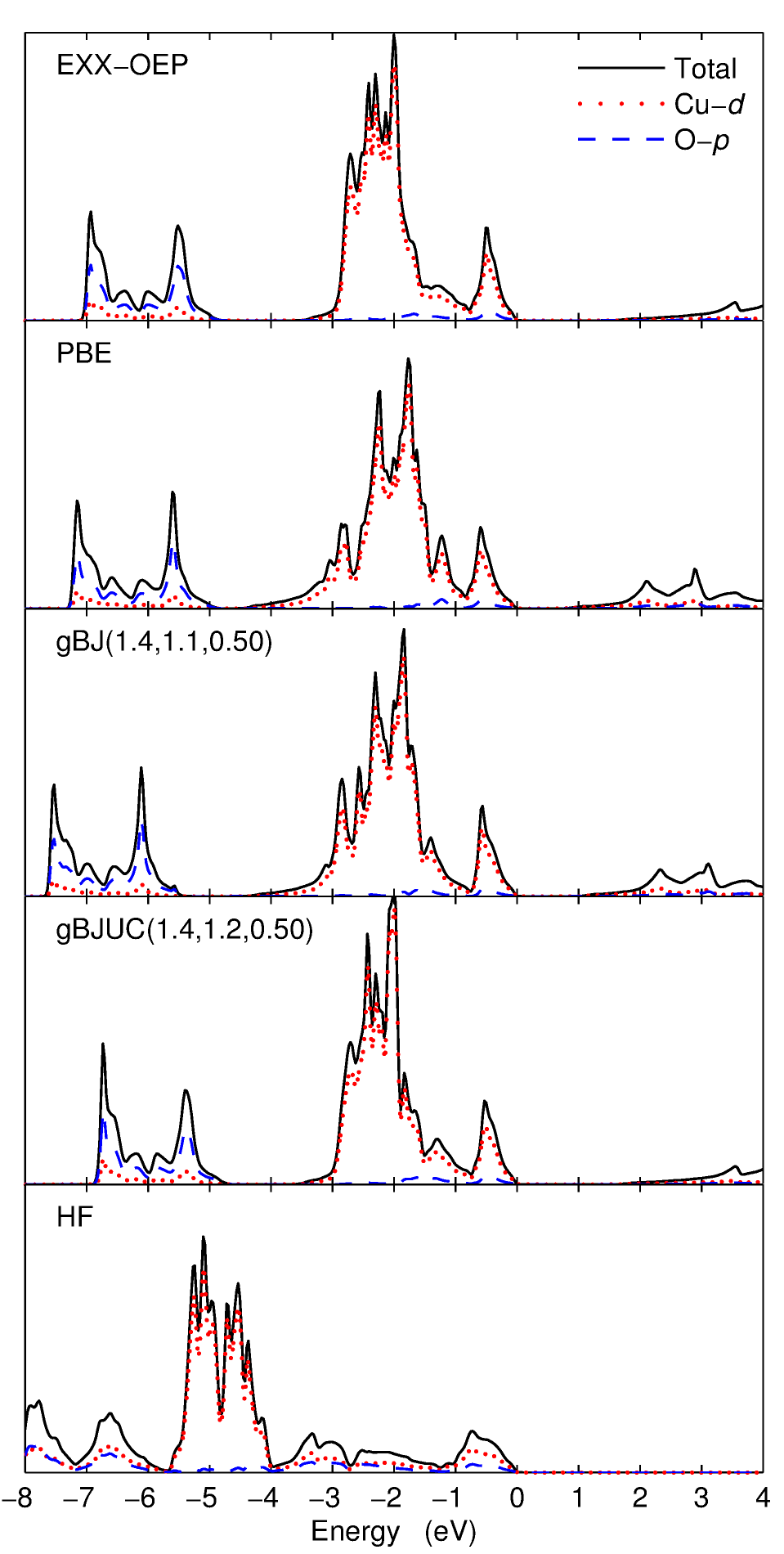}
\caption{\label{fig4}(Color online) DOS of Cu$_{2}$O.
The Fermi energy is set at zero.}
\end{figure}

\begin{figure}
\includegraphics[width=\columnwidth]{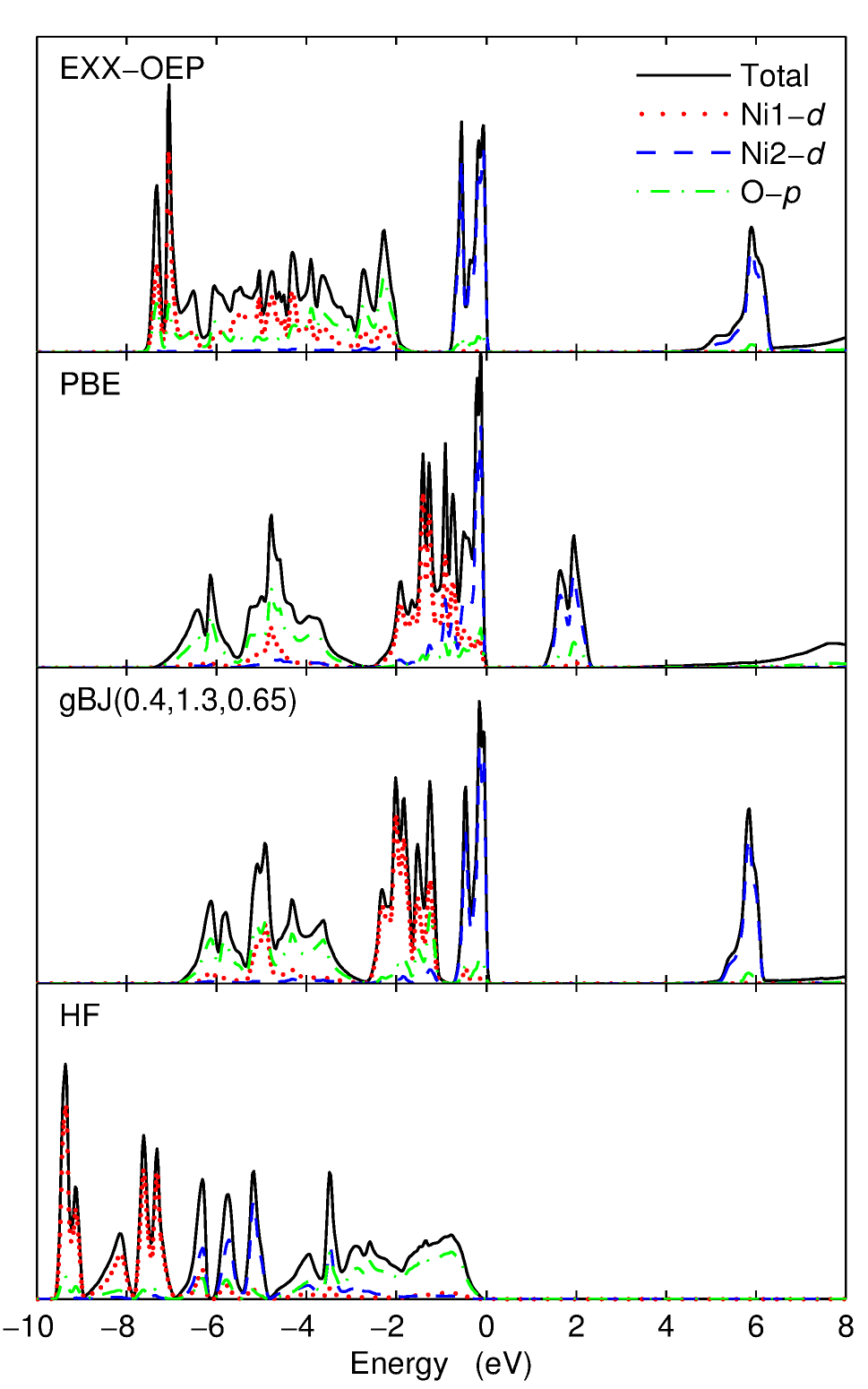}
\caption{\label{fig5}(Color online) Spin-up DOS of NiO.
The Fermi energy is set at zero.}
\end{figure}

The comparison of the electronic structure obtained with the different exchange
potentials focussed so far on the transition energies and the core states.
In order to assess the differences in the electronic structure on a wider
energy spectrum of the valence states,
we compare the density of states of NiO and Cu$_{2}$O around
the Fermi energy. We picked out these two solids from our test set, since the
largest changes in the DOS with respect to the applied potential
can be observed for these two solids. For C, Si, BN, and MgO the basic
structure of the DOS remains very similar independent from the applied
potential (of course, apart from a rigid shift of the conduction bands).

Figures~\ref{fig4} and \ref{fig5} show the DOS of Cu$_{2}$O and NiO,
respectively, for a few selected potentials. In the case of Cu$_{2}$O, we can
clearly see that the gBJUC$(1.4,1.2,0.50)$ potential (very good for the
transition energies and EFG, see below) leads to the best agreement with EXX-OEP,
which is particularly true for the partial Cu-$3d$ DOS in the range from $-3$ to
0 eV below the Fermi energy. The Cu-$3d$ DOS obtained with the other semilocal
potentials, including gBJ without UC, are too broad by about 1 eV.
gBJUC$(1.4,1.2,0.50)$ also leads to correct positions of both the O-$2p$
(extending from $-7$ to $-5$ eV) and the conduction band Cu-$4s$ states.
The HF DOS differs significantly from the other calculations employing a
local exchange potential including EXX-OEP. It is well known that the HF
method systematically leads to band gaps which are by far too large compared
to experiment. In the case of Cu$_{2}$O, the HF band gap amounts to 10.4 eV,
while it is only 2.17 eV in experiment\cite{BaumeisterPR61} and 1.44 eV with
EXX-OEP. In the occupied part of the spectrum, it is obvious that the main
position of the Cu-$3d$ peaks are much lower in energy (below $-4$ eV),
while the bands in the energy range from $-4$ to 0 eV are a mixture of
Cu-$3d$ and O-$2p$ states. For other systems, a comparison between the
HF and EXX-OEP occupied spectrum can be found in, e.g.,
Refs.~\onlinecite{KuemmelRMP08,LuoPRB12,KohutJCP14}.

For NiO, the semilocal methods lead to DOS that differ markedly from the
EXX-OEP DOS. As already discussed in Refs.~\onlinecite{EngelPRL09,BetzingerPRB13}
the spin-up highest valence bands (between $-0.8$ and 0 eV) and lowest conduction
bands in the EXX-OEP DOS are of Ni-$3d$ character coming from the Ni atom with
more spin-down electrons (Ni2 with $t_{2g}^{\uparrow}$ occupied and
$e_{g}^{\uparrow}$ emtpy), while the spin-up states between $-7.5$ and $-2$ eV
are mixtures of Ni-$3d$ from the Ni atom Ni1 ($t_{2g}^{\uparrow}$ and
$e_{g}^{\uparrow}$ fully occupied) and O-$2p$. Therefore, EXX-OEP leads to a clear
energy separation between the spin-up and spin-down Ni-$3d$ states of
the same Ni atom. In the PBE DOS the position of the conduction states is much
too low and there is no Ni-$3d$ peak similar to the one at $\sim-7.5$ eV in the
EXX-OEP DOS. In addition, there is very little energy separation between
the spin-up $3d$ states coming from the two Ni atoms.
The gBJ$(0.4,1.3,0.65)$ potential leads to a good band gap and
a separation of $\sim0.5$ eV between the spin-up $3d$ states from the two Ni
atoms, however, there is still no Ni-$3d$ peak at the lower part of the valence
DOS, which can only be obtained by the LDA+$U$\cite{AnisimovPRB91}
or HF/hybrid\cite{TowlerPRB94,BredowPRB00,MoreiraPRB02,TranPRB06} methods.
We mention that the gBJ potentials with small values of $c$ (1.0 or 1.1) and the
gBJUC potentials do not produce any energy separation between the
$3d$ states of the two Ni atoms. The valence part of
the HF DOS starts at $-10$ eV and about five sharp Ni-$3d$ peaks are equally
distributed in the energy range $-10$ to $-5$ eV, while the DOS from $-3$ to
0 eV is exclusively of O-$2p$ character. The HF band gap is 13.9 eV, which is
in fair agreement with previous HF
results.\cite{TowlerPRB94,BredowPRB00,MoreiraPRB02} In experiment, a gap of
4.0-4.3 eV is observed,\cite{HufnerSSC84,SawatzkyPRL84} while EXX-OEP gives
rise to 3.54 eV.

\subsection{\label{efg}EFG of Cu$_{2}$O and magnetic moment in NiO}

\begin{table}
\caption{\label{table4} Spin magnetic moment $\mu_{S}^{\text{Ni}}$
(in $\mu_{\text{B}}$) inside the Ni atomic sphere of radius 1.016 \AA~in NiO
and EFG of Cu (in $10^{21}$ V/m$^{2}$) in Cu$_{2}$O.}
\begin{ruledtabular}
\begin{tabular}{lcccc}
\multicolumn{1}{l}{} &
\multicolumn{1}{c}{} &
\multicolumn{3}{c}{EFG$_{\text{Cu}}$} \\
\cline{3-5}
\multicolumn{1}{l}{Method} &
\multicolumn{1}{c}{$\mu_{S}^{\text{Ni}}$} &
\multicolumn{1}{c}{Total} &
\multicolumn{1}{c}{$p$-$p$} &
\multicolumn{1}{c}{$d$-$d$} \\
\hline
EXX-OEP                               & 1.91                         & $-17.7$             & $-25.0$ &  7.2 \\
LDA                                   & 1.30                         & $ -4.7$             & $-15.7$ & 10.8 \\
PBE                                   & 1.43                         & $ -5.6$             & $-16.3$ & 10.5 \\
B88                                   & 1.43                         & $ -5.6$             & $-16.3$ & 10.5 \\
EV93                                  & 1.51                         & $ -6.8$             & $-17.5$ & 10.4 \\
AK13                                  & 1.58                         & $ -8.1$             & $-18.5$ & 10.1 \\
BJ                                    & 1.53                         & $ -7.4$             & $-17.7$ & 10.2 \\
BJUC                                  & 1.41                         & $-11.3$             & $-19.4$ &  7.9 \\
gBJ$(0.6,1.0,0.60)$\footnotemark[1]   & 1.61                         & $ -7.0$             & $-17.6$ & 10.5 \\
gBJ$(1.4,1.1,0.50)$\footnotemark[2]   & 1.66                         & $ -8.3$             & $-19.3$ & 10.8 \\
gBJ$(0.4,1.3,0.65)$\footnotemark[3]   & 1.86                         & $ -5.1$             & $-18.2$ & 13.0 \\
gBJUC$(1.4,1.2,0.50)$\footnotemark[4] & 1.59                         & $-15.1$             & $-22.2$ &  6.8 \\
HF                                    & 1.88                         & $-17.0$             & $-25.0$ &  7.9 \\
Expt.                                 & $1.90\pm0.2$\footnotemark[5] & 9.8\footnotemark[6]                  \\
\end{tabular}
\end{ruledtabular}
\footnotetext[1]{Good compromise for the EXX total energy of C, Si, BN, MgO, and Cu$_{2}$O.}
\footnotetext[2]{Good compromise for transition energies in C, Si, BN, and MgO.}
\footnotetext[3]{Good compromise for transition energies and Ni magnetic moment in NiO.}
\footnotetext[4]{Good compromise for transition energies and EFG in Cu$_{2}$O.}
\footnotetext[5]{Does not include the orbital contribution $\mu_{L}^{\text{Ni}}=0.32\pm0.05$
$\mu_{\text{B}}$ (Refs.~\onlinecite{FernandezPRB98,NeubeckJAP99}).}
\footnotetext[6]{Only the magnitude is known. Calculated using the quadrupole
moment $Q\left(^{63}\text{Cu}\right)=0.22$
(Refs.~\onlinecite{KushidaPR56,PyykkoMP01}).}
\end{table}

As shown in Refs.~\onlinecite{BlahaPRB88,SchwarzPRB90} the EFG is mainly
determined by the non-spherical electron density close to the nucleus.
Since the density of the core electrons is (by construction) purely spherical,
it is the electron density of the valence states that determines the EFG.
Moreover, in the case of a $3d$-transition metal like Cu, the valence electron
density in a region of a few tenths of an Angstrom from the nucleus is decisive.
Hence, by comparing the EFG of Cu$_{2}$O for the different potentials we
indirectly measure the difference in the non-spherical part of the valence
electron density. The results for the EFG of Cu are shown in Table~\ref{table4}.
Similarly to the transition energies in Cu$_{2}$O (see
Sec.~\ref{transitionenergies}), only the gBJ potential
with the UC (gBJUC) is able to reproduce the EXX-OEP EFG qualitatively.
For instance, with the parameters $(1.4,1.2,0.50)$,
gBJUC leads to an EFG of $-15.1\times10^{21}$ V/m$^{2}$, while for all other
potentials (except BJUC), the magnitude of the EFG does not exceed
$10\times10^{21}$ V/m$^{2}$. For gBJUC$(1.4,1.2,0.50)$, not only the total EFG,
but also its two main components ($p$-$p$ and $d$-$d$) agree
rather well with the EXX-OEP (and HF) values (see Table~\ref{table4}).
A detailed analysis of the UC will be provided in Sec.~\ref{analysispot}.
A calculation with the non-multiplicative HF
potential leads to an EFG of $-17.0\times10^{21}$ V/m$^{2}$ which is
relatively close to the EXX-OEP value and expected since the first-order change
in the electron density due to the replacement
$v_{\text{x},i}^{\text{HF}}\rightarrow v_{\text{x}}^{\text{EXX-OEP}}$
is zero.\cite{KriegerPRA92b,Grabo00,KuemmelPRB03} The magnitude of the
experimental EFG amounts to $9.8\times10^{21}$ V/m$^{2}$
(Refs.~\onlinecite{KushidaPR56,PyykkoMP01}), which is much smaller than the
EXX-OEP or HF values. Thus, the impact of the electron correlation on the
EFG is significant: the EXX-OEP value has to be reduced by the exact correlation
functional nearly by its half. Furthermore, it was shown in
Refs.~\onlinecite{HuPRB08,ScanlonJCP09,ScanlonJPCL10,TranPRB11,KollerPRB11}
that the LDA, GGA, LDA+$U$, onsite-hybrid,\cite{NovakPSSB06,TranPRB06}
and mBJ methods lead to an EFG in Cu$_{2}$O which is by far too small.
The experimental value could only be approached with full hybrid functionals.

Next we turn the discussion to the spin magnetic moment $\mu_{S}^{\text{Ni}}$ of
Ni in NiO (results in Table~\ref{table4}).
In contrast to the EFG, $\mu_{S}^{\text{Ni}}$ is determined by the
difference of the spherical spin-up and -down electron densities in the atomic
spheres. The best agreement with the EXX-OEP Ni spin magnetic moment is
obtained by the gBJ potential with a $c$ parameter of at least 1.2, which is
in accordance with the observations for the EXX total energy and transition
energies in Secs.~\ref{totalenergy} and \ref{transitionenergies}. In fact, the
parameters $(0.4,1.3,0.65)$ of the gBJ potential lead to
$\mu_{S}^{\text{Ni}}=1.86$ $\mu_{\text{B}}$,
which is very close to the EXX-OEP, HF, and experimental values. All
other investigated potentials substantially underestimate the EXX-OEP
spin-magnetic moment.

\subsection{\label{analysispot}Analysis of the potentials}

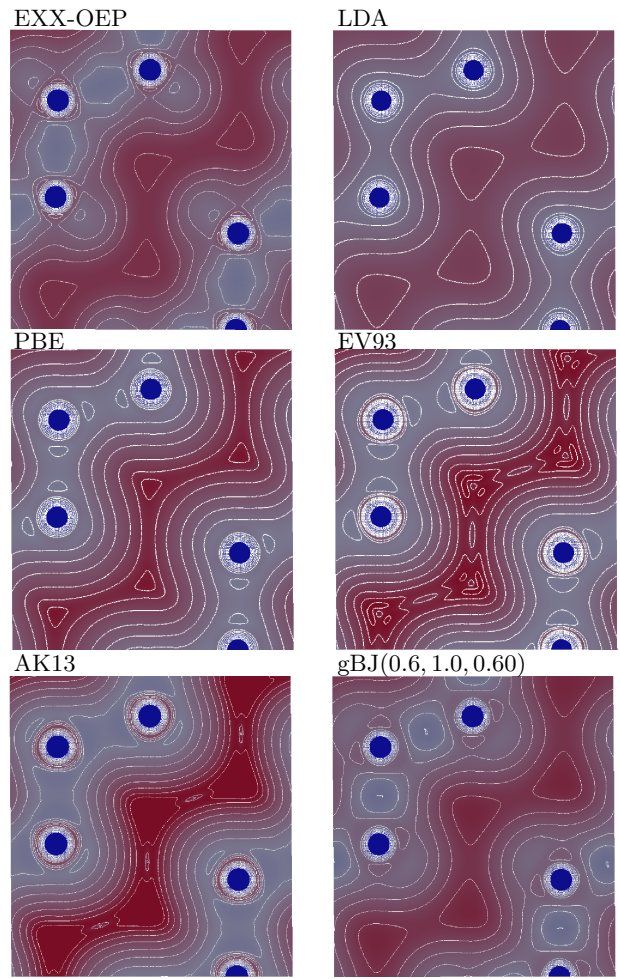
\begin{figure}
\begin{picture}(8.6,13)(0,0)
\put(0,8.6){\epsfxsize=4.0cm \epsfbox{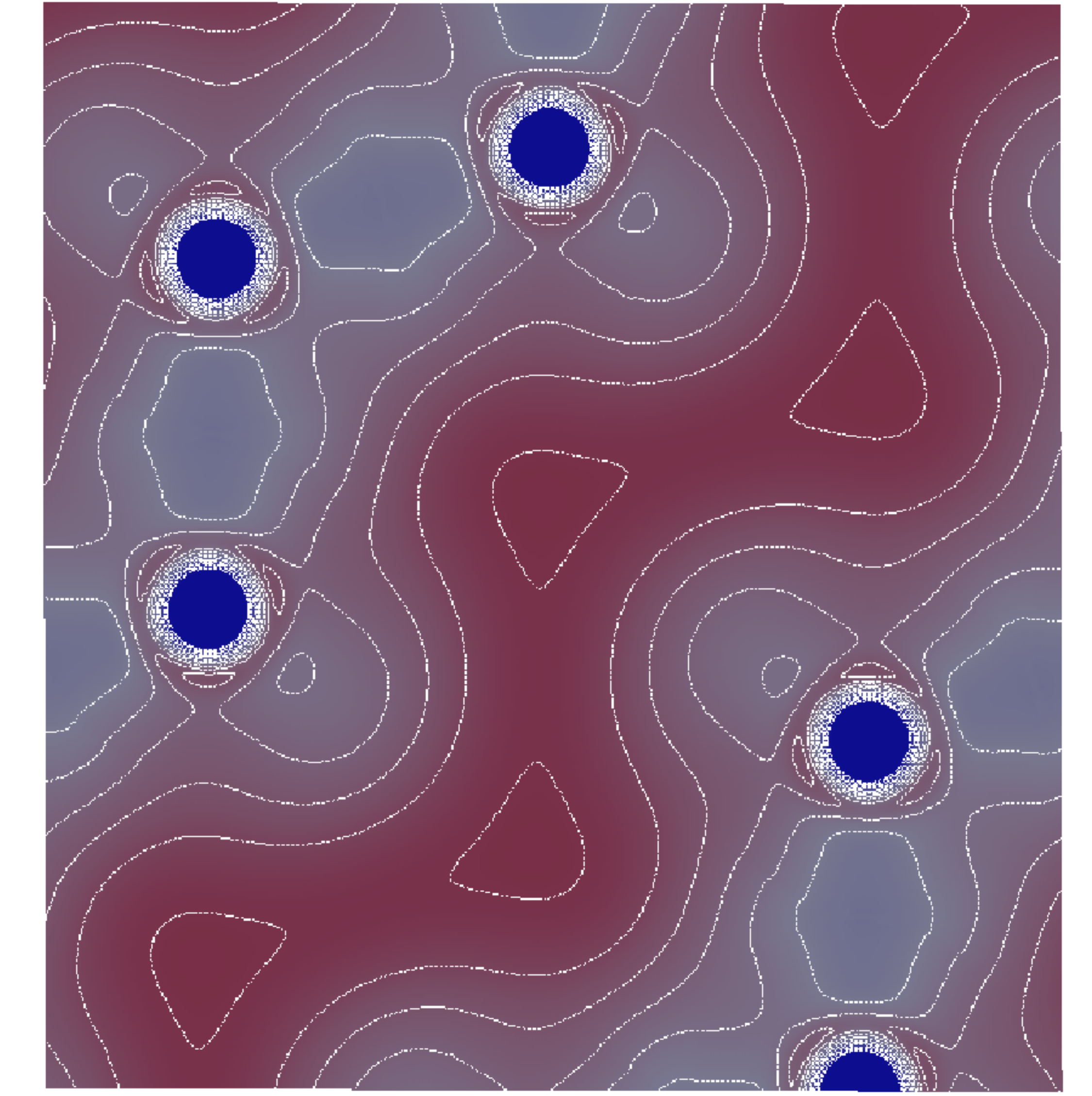}}
\put(4.3,8.6){\epsfxsize=4.0cm \epsfbox{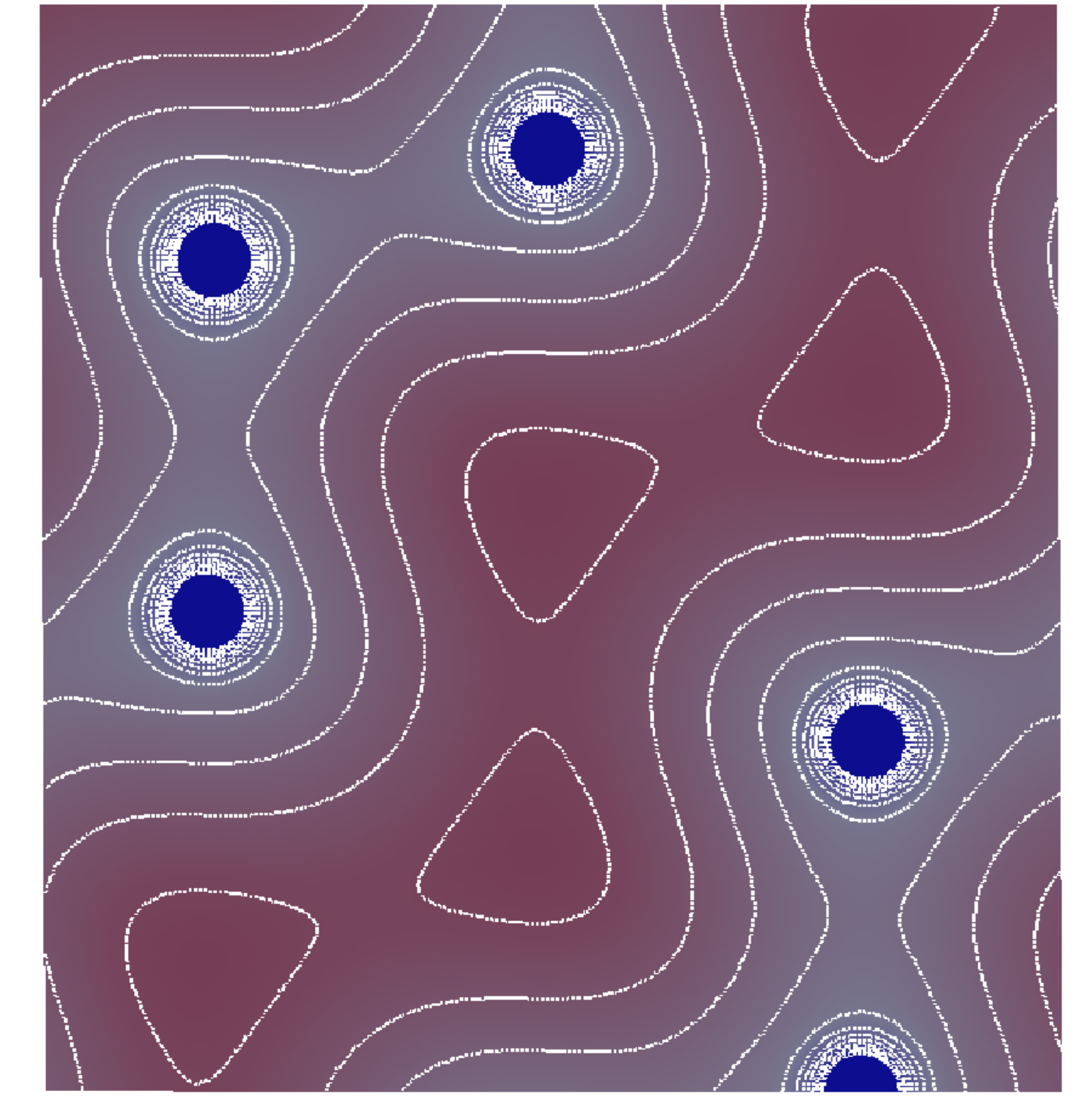}}
\put(0,4.35){\epsfxsize=4.0cm \epsfbox{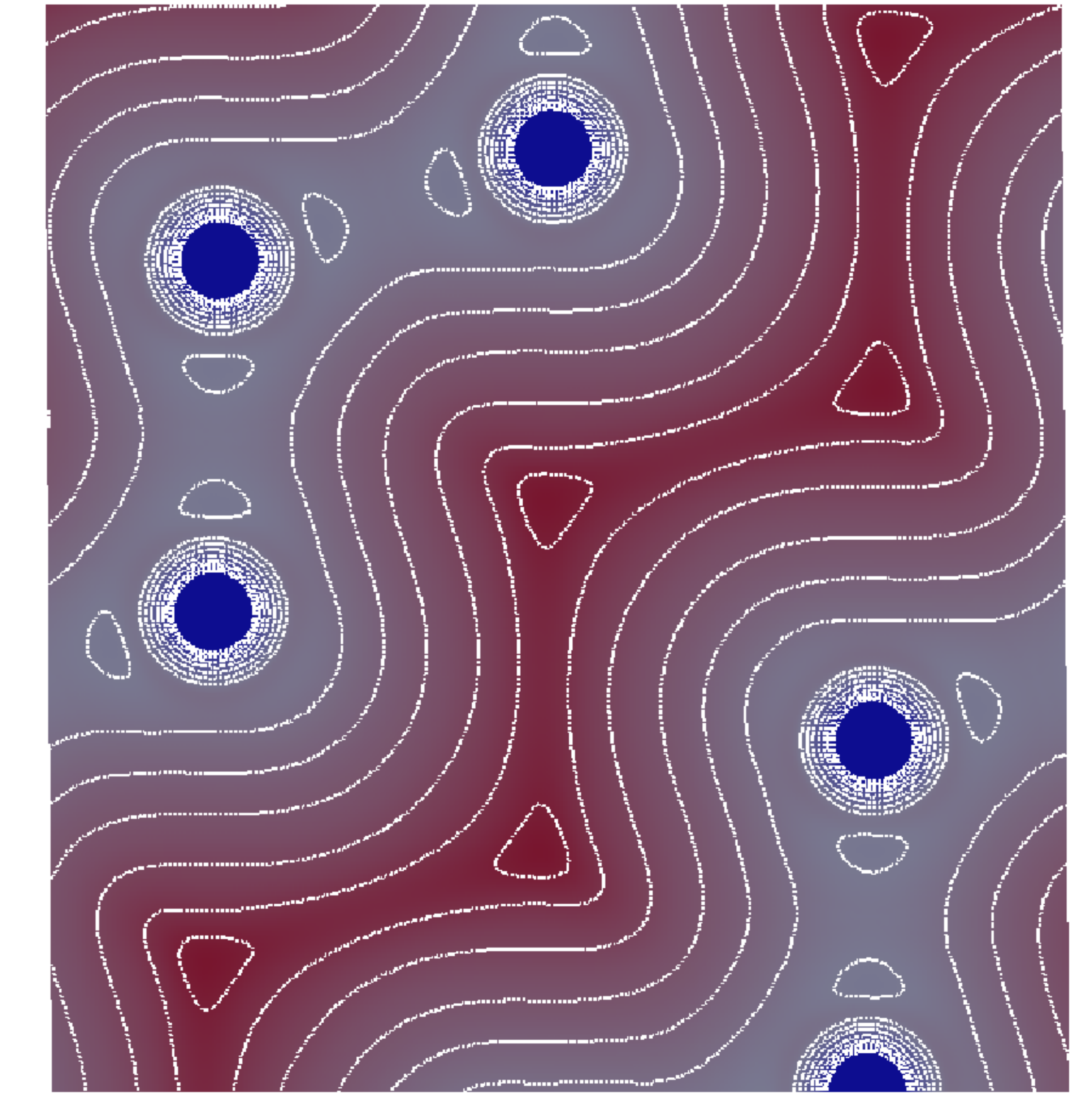}}
\put(4.3,4.35){\epsfxsize=4.0cm \epsfbox{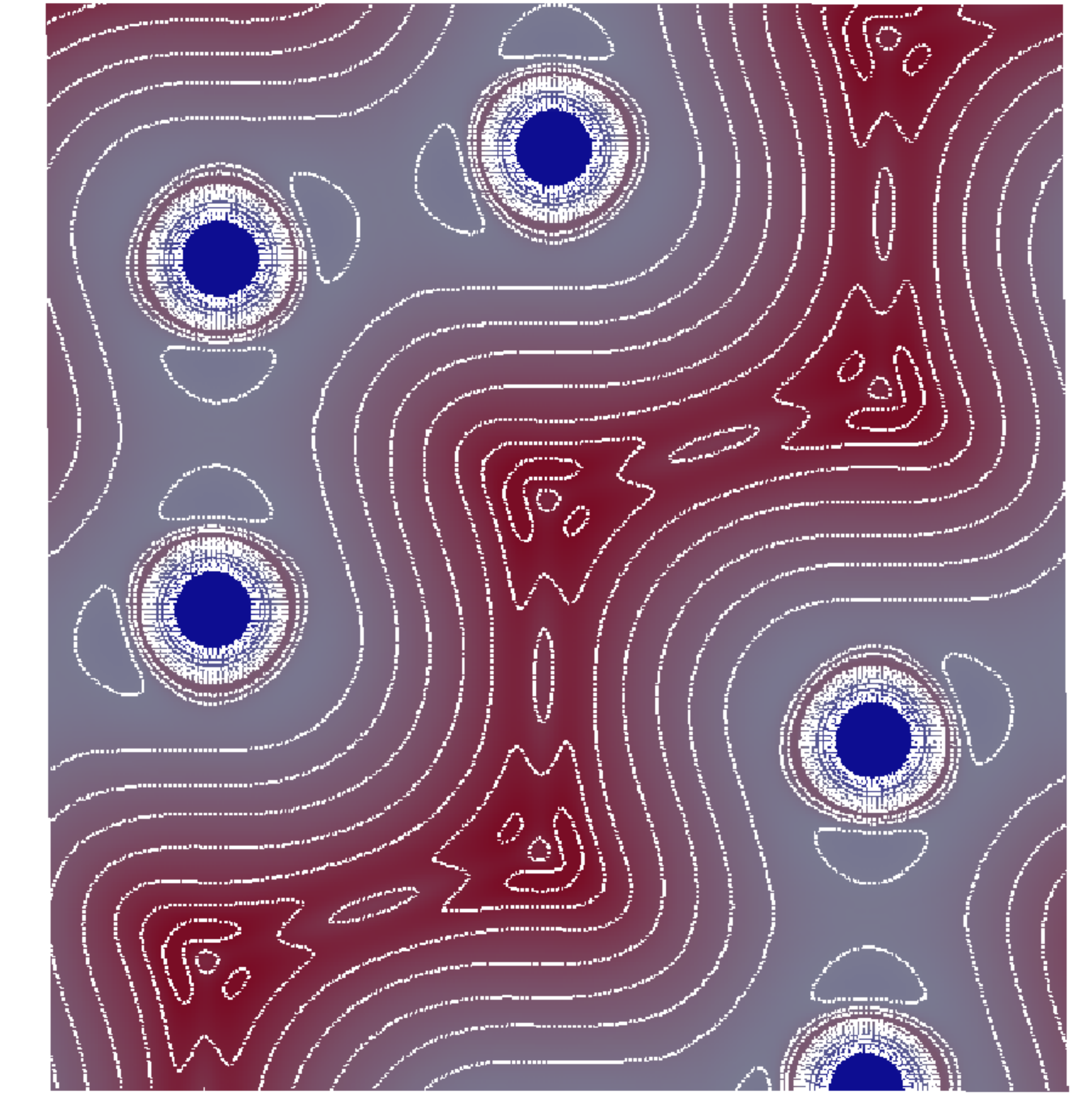}}
\put(0,0){\epsfxsize=4.0cm \epsfbox{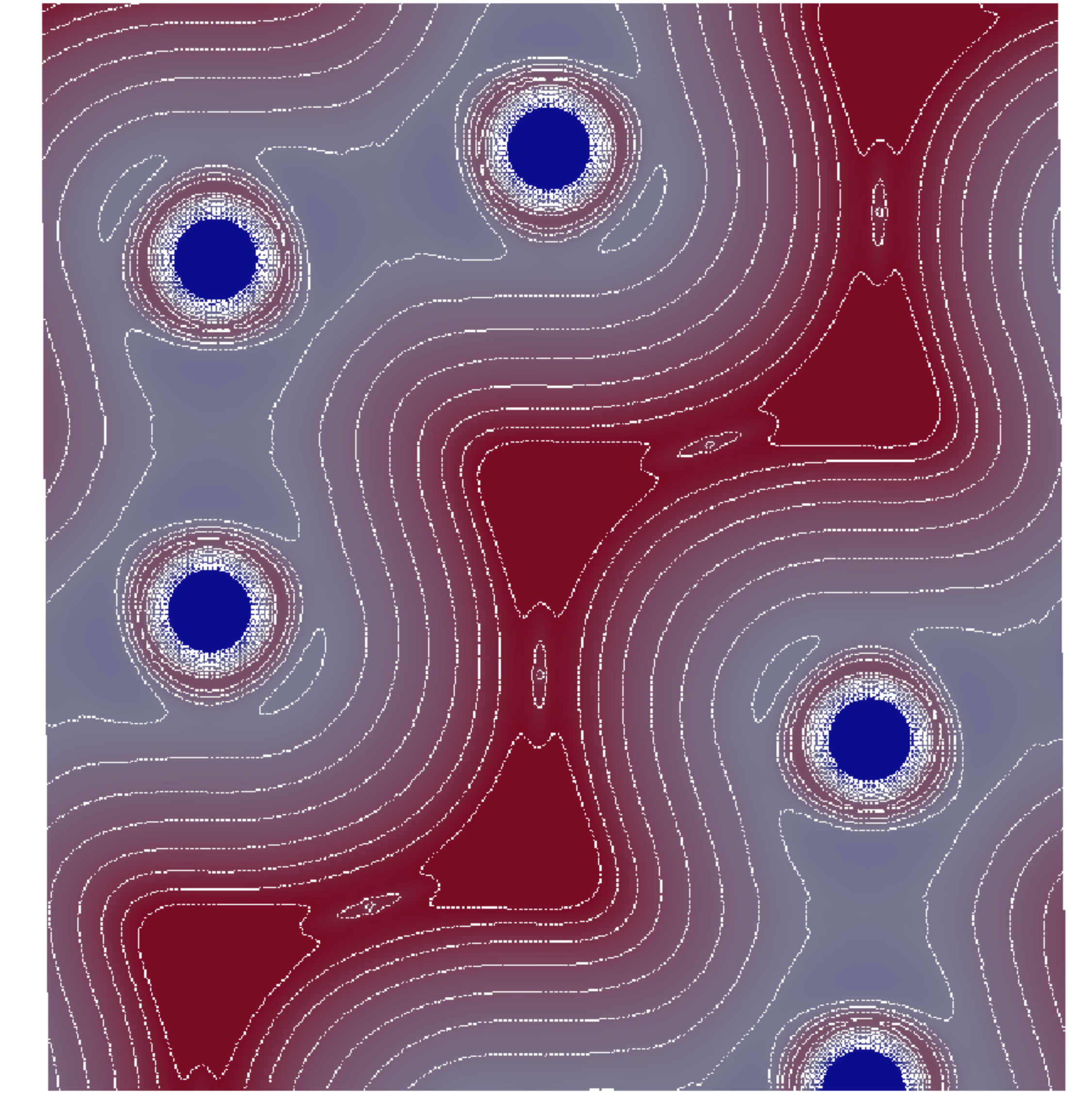}}
\put(4.3,0){\epsfxsize=4.0cm \epsfbox{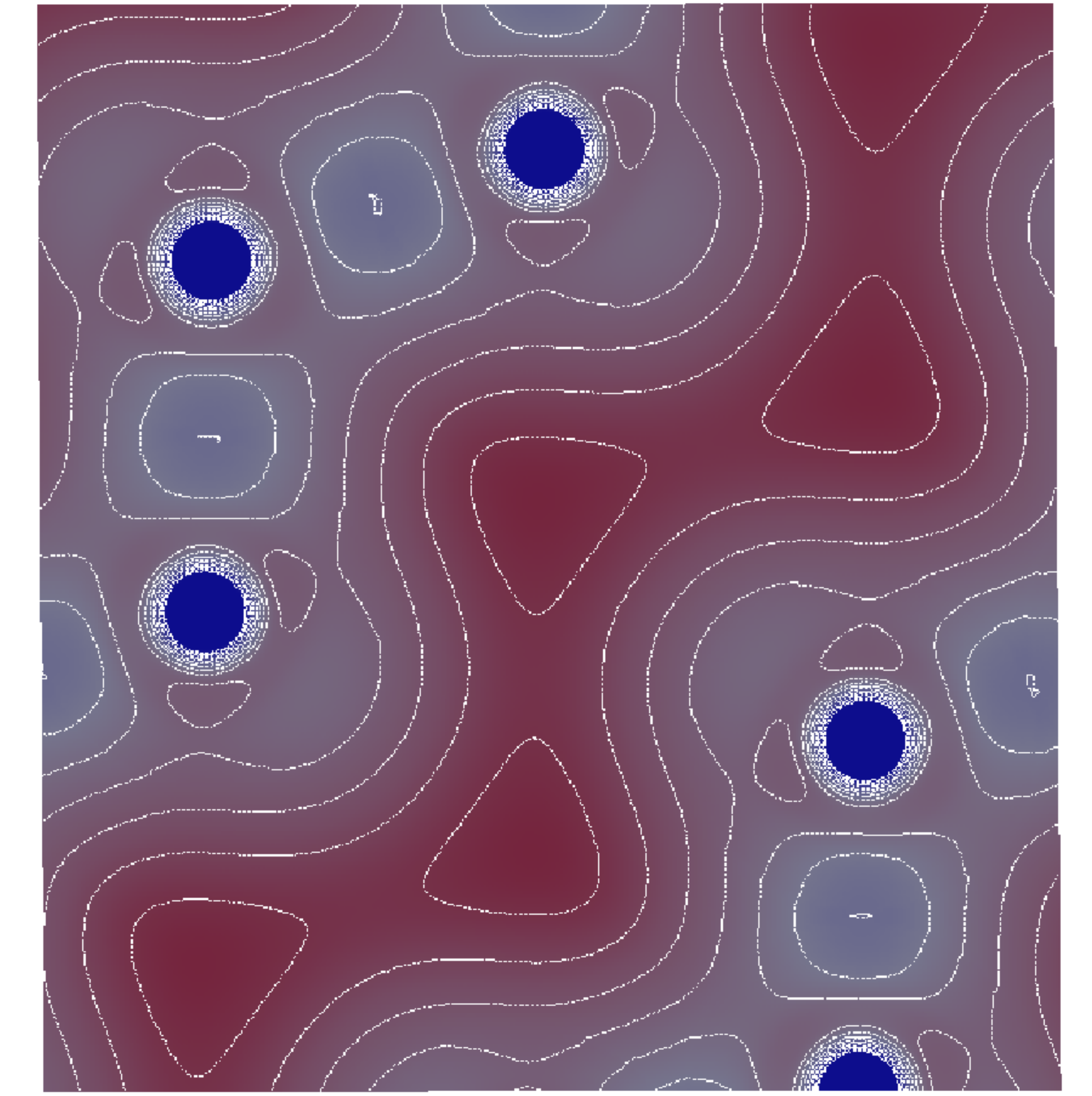}}
\put(0.2,12.65){EXX-OEP}
\put(4.5,12.65){LDA}
\put(0.2,8.36){PBE}
\put(4.5,8.36){EV93}
\put(0.2,4.07){AK13}
\put(4.5,4.07){gBJ$(0.6,1.0,0.60)$}
\end{picture}
\caption{\label{fig6}(Color online) Two-dimensional plots of exchange
potentials $v_{\text{x}}$ in a $(110)$ plane of C. The potentials were shifted
such that $\int_{\text{cell}} v_{\text{x}}(\mathbf{r})d^{3}r=0$. The contour
lines start at $-2$ Ry (blue color) and end at 1 Ry (red color)
with an interval of 0.2 Ry.}
\end{figure}

\begin{figure}
\begin{picture}(8,13.1)(0,0)
\put(0,6.5){\epsfxsize=8cm \epsfbox{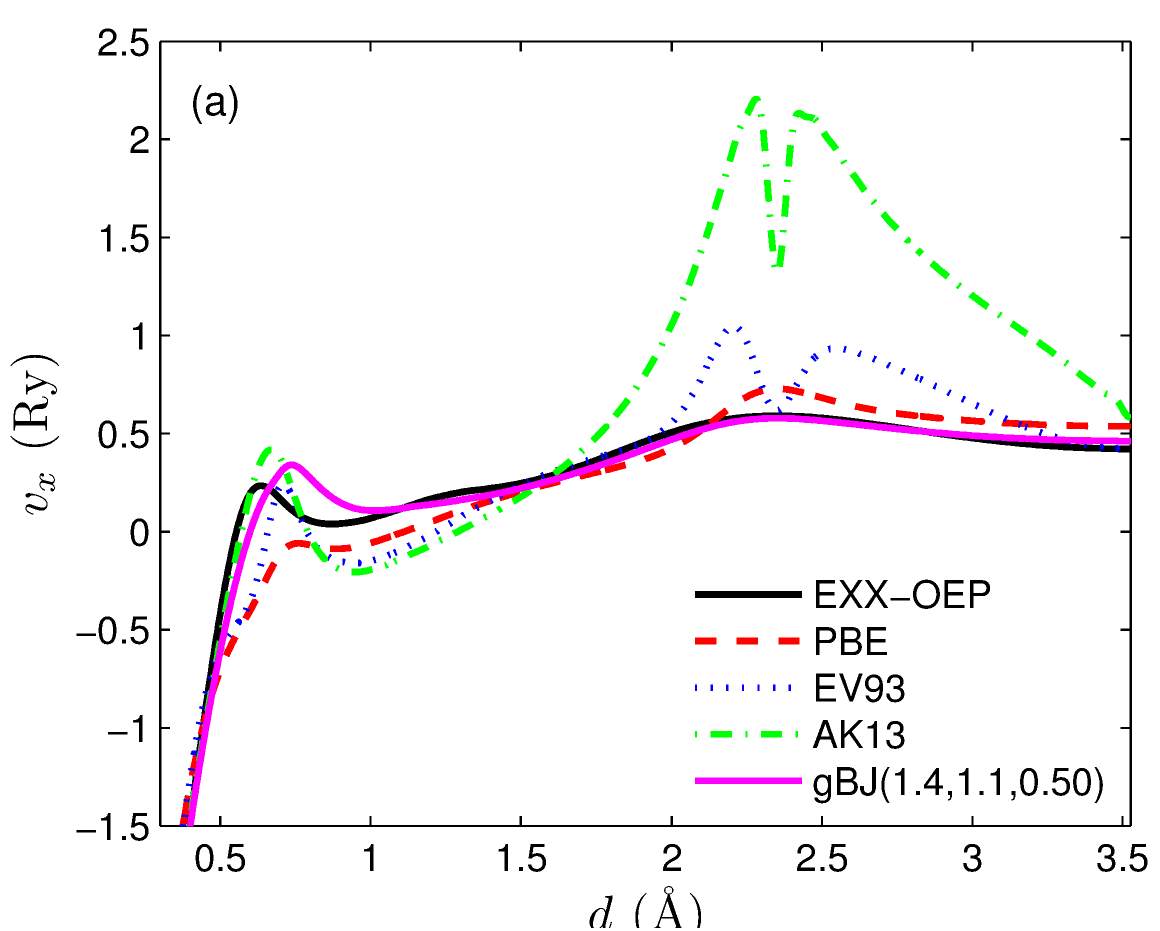}}
\put(0,0){\epsfxsize=8cm \epsfbox{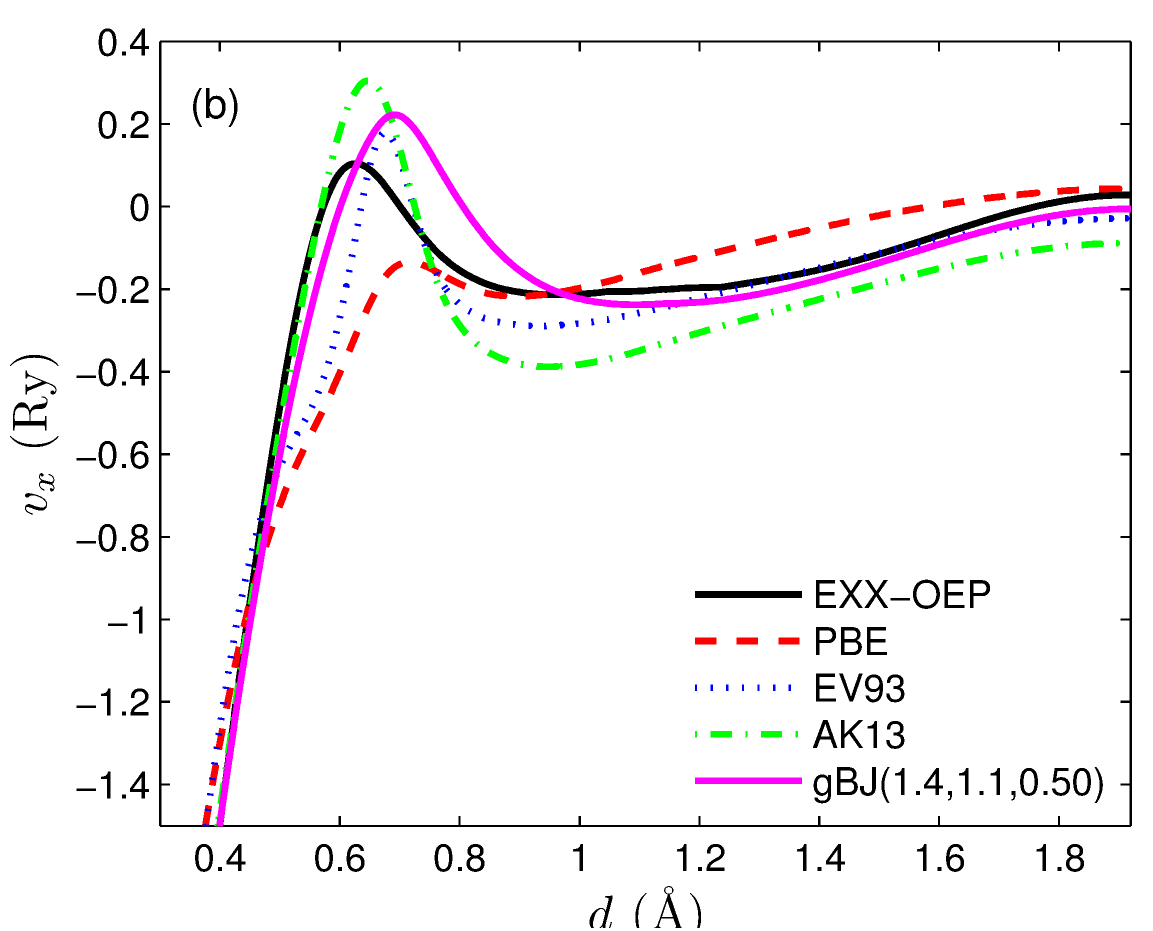}}
\end{picture}
\caption{\label{fig7}(Color online)
Exchange potentials $v_{\text{x}}$ in Si plotted from the vicinity of the atom
at site $(1/8,1/8,1/8)$ ($d=0$) to (a) the center of the unit cell ($d=3.53$ \AA)
or (b) the mid-distance to the atom at site $(5/8,5/8,1/8)$ ($d=2.38$ \AA).
The potentials were shifted such that
$\int_{\text{cell}}v_{\text{x}}(\mathbf{r})d^{3}r=0$.}
\end{figure}

\begin{figure*}
\includegraphics[scale=0.50]{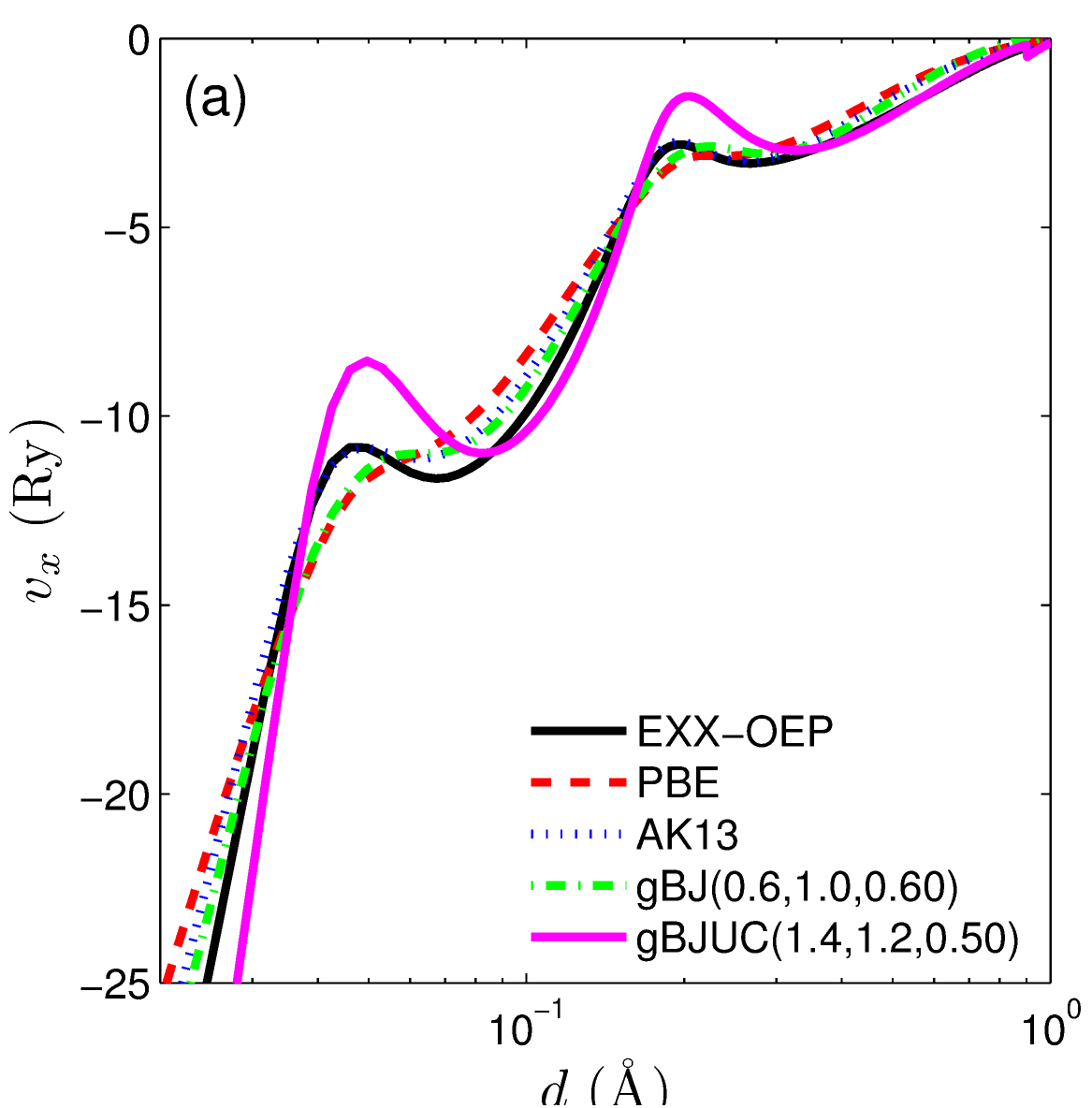}
\includegraphics[scale=0.50]{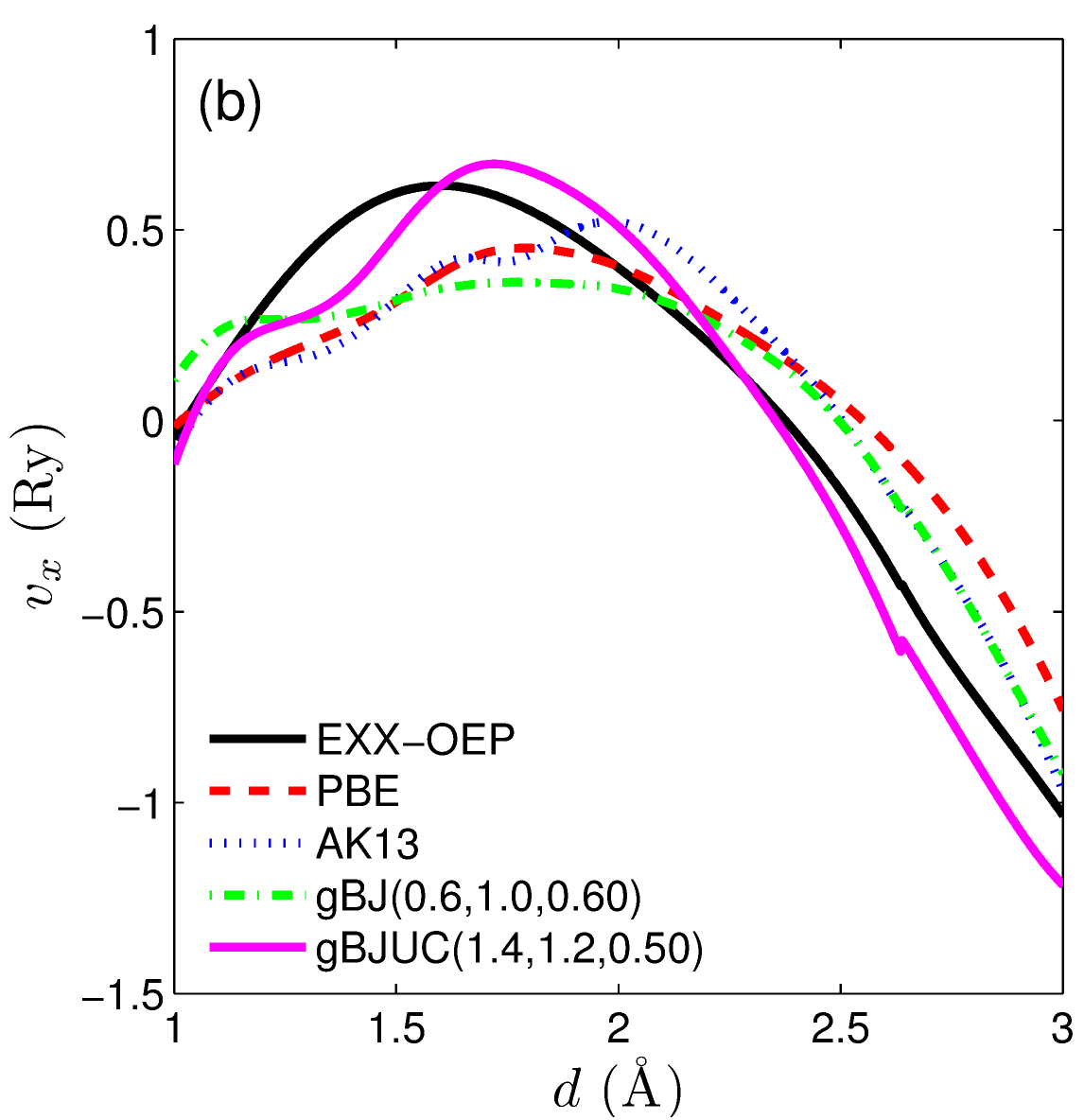}
\includegraphics[scale=0.50]{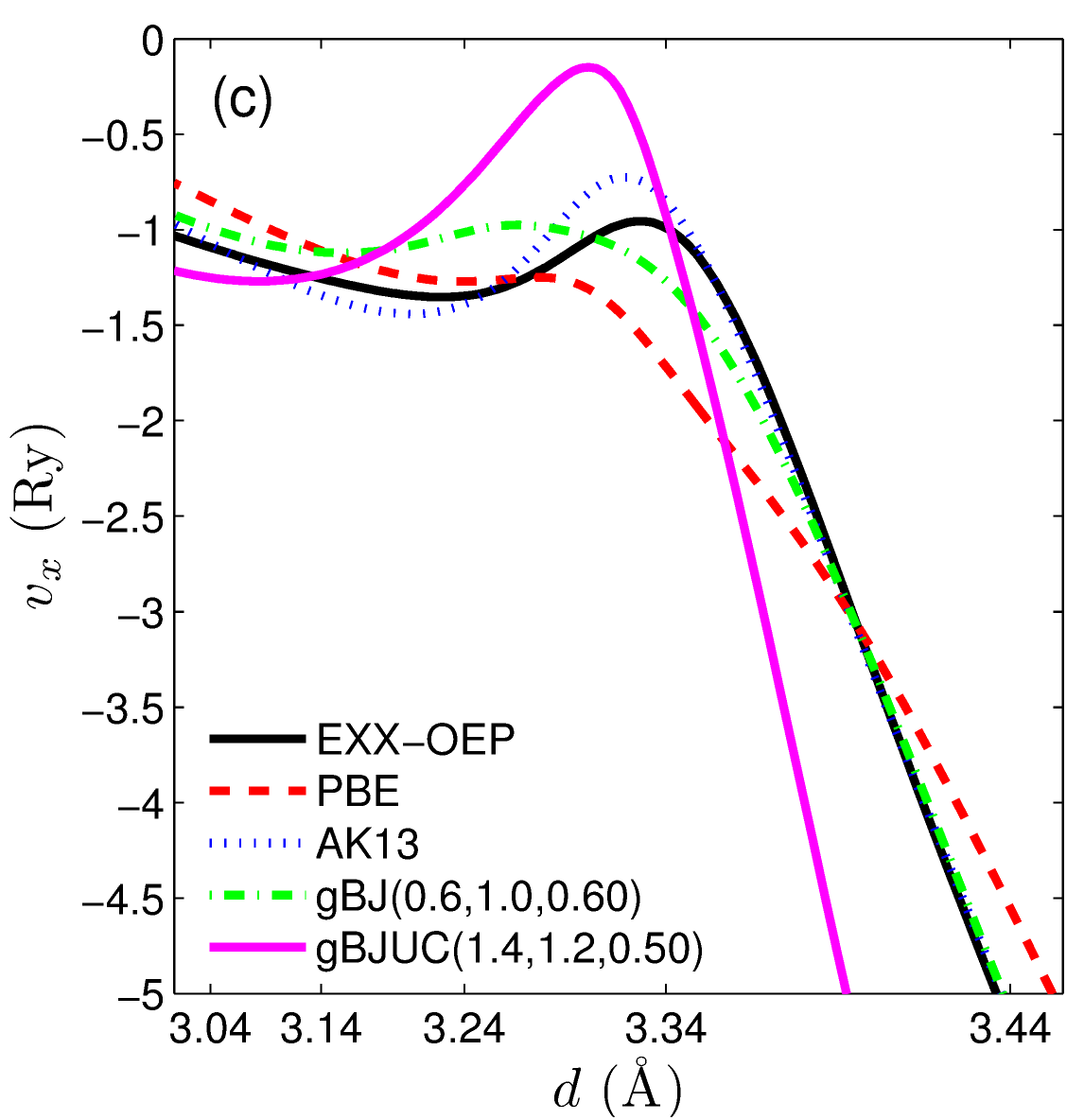}
\caption{\label{fig8}(Color online) Exchange potentials $v_{\text{x}}$ in
Cu$_{2}$O plotted from the Cu atom at site $(1/2,1/2,0)$ ($d=0$) in the
direction of the O atom at site $(3/4,3/4,3/4)$ ($d=3.54$ \AA).
Logarithmic scales on the $x$-axis were
used for panels (a) and (c) which correspond to the vicinity of the Cu and
O atoms, respectively. The potentials were shifted such that
$\int_{\text{cell}}v_{\text{x}}(\mathbf{r})d^{3}r=0$.}
\end{figure*}

\begin{figure}
\begin{picture}(8,14)(0,0)
\put(0,6.9){\epsfxsize=7.5cm \epsfbox{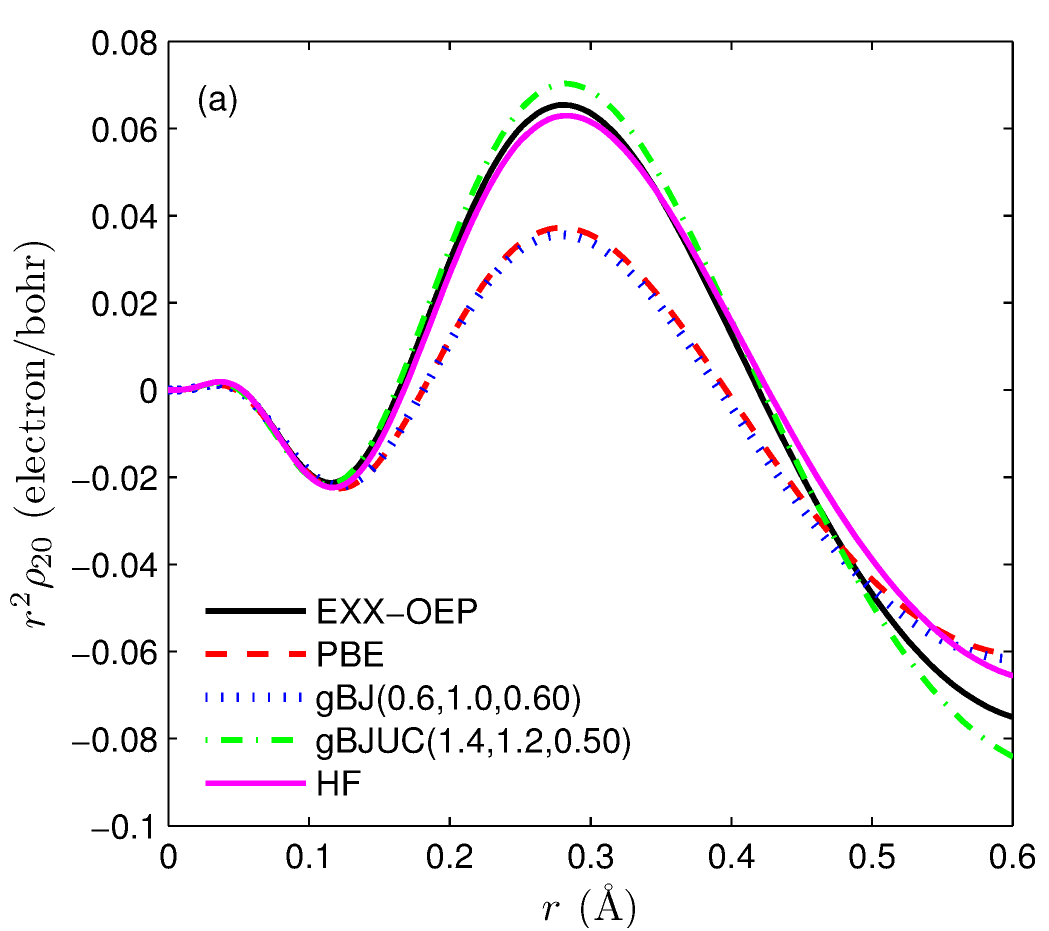}}
\put(0,0){\epsfxsize=7.5cm \epsfbox{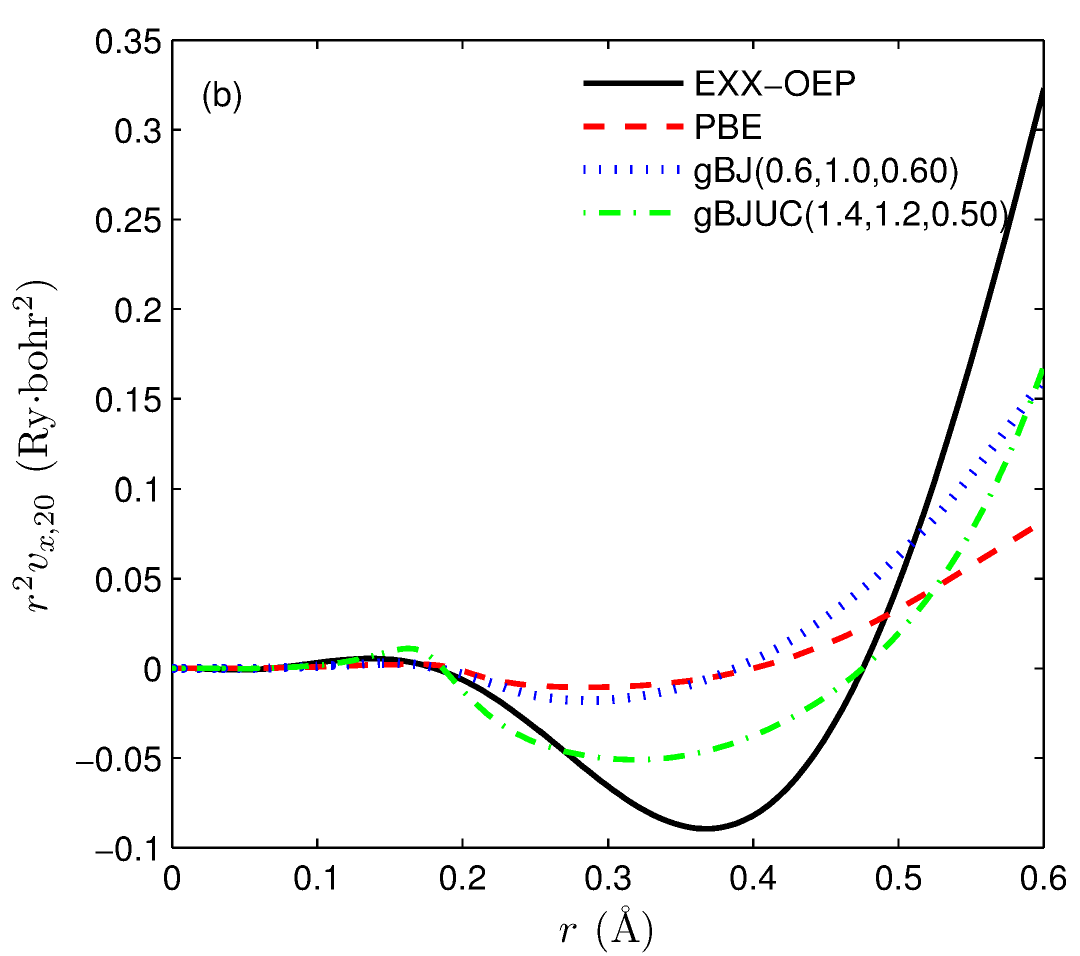}}
\end{picture}
\caption{\label{fig9}(Color online)
Plots of the radial functions $\rho_{20}$ (a) and $v_{\text{x},20}$ (b)
versus the distance from the Cu atom in Cu$_{2}$O. The functions are
multiplied by $r^{2}$.}
\end{figure}

\begin{figure}
\begin{picture}(8.6,14.1)(0,0)
\put(0,9.3){\epsfxsize=4.2cm \epsfbox{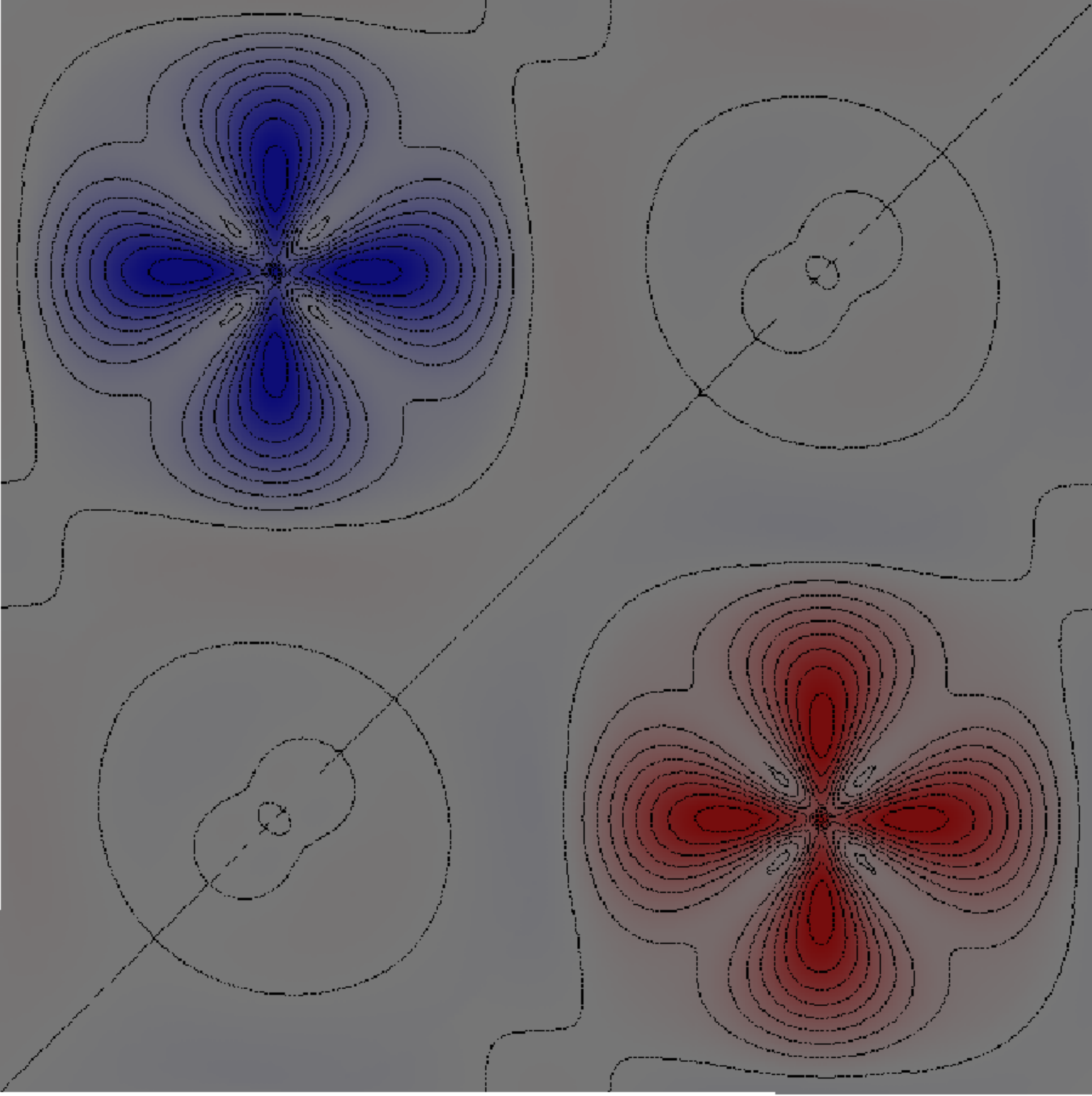}}
\put(4.3,9.3){\epsfxsize=4.2cm \epsfbox{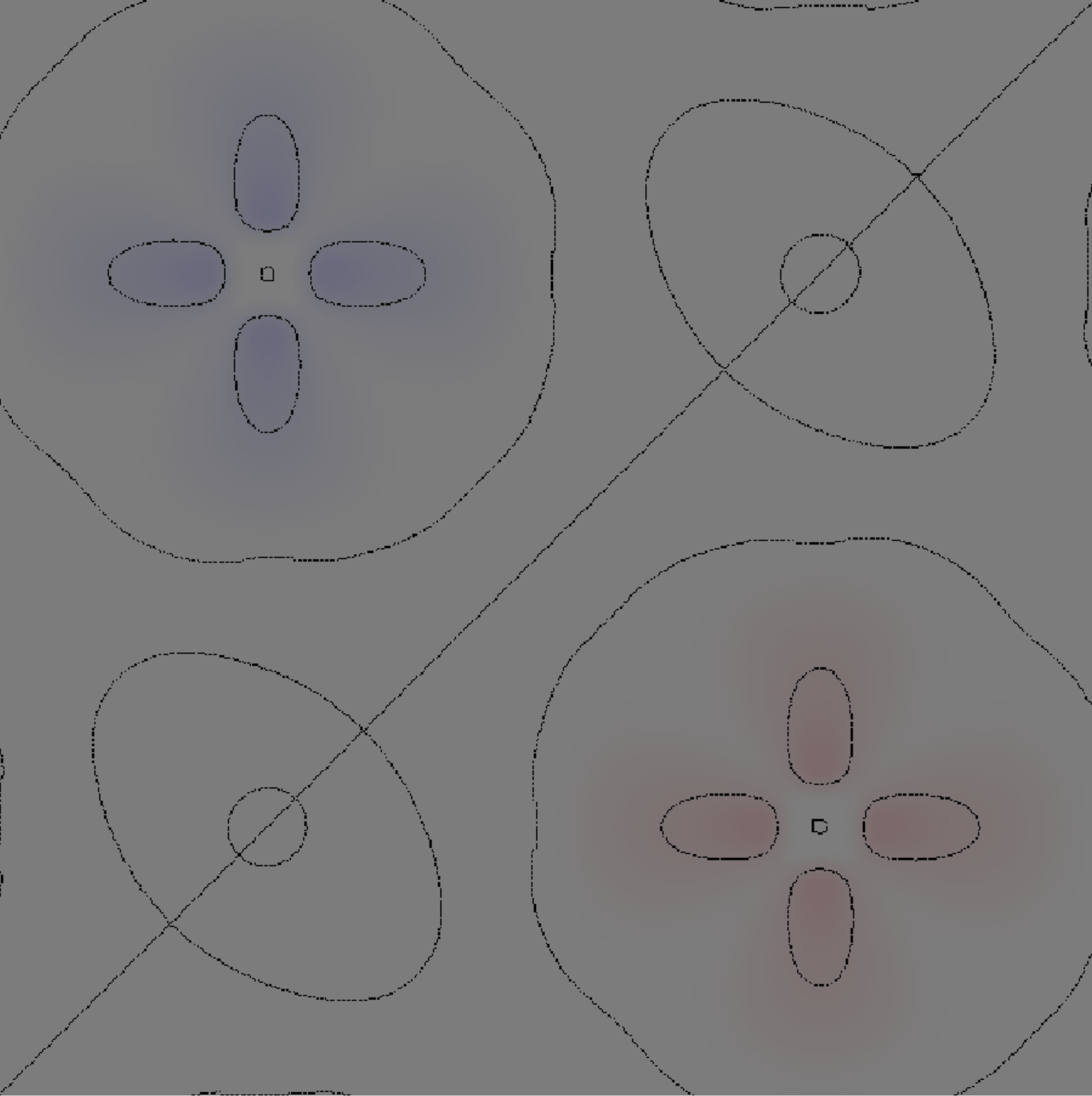}}
\put(0,4.7){\epsfxsize=4.2cm \epsfbox{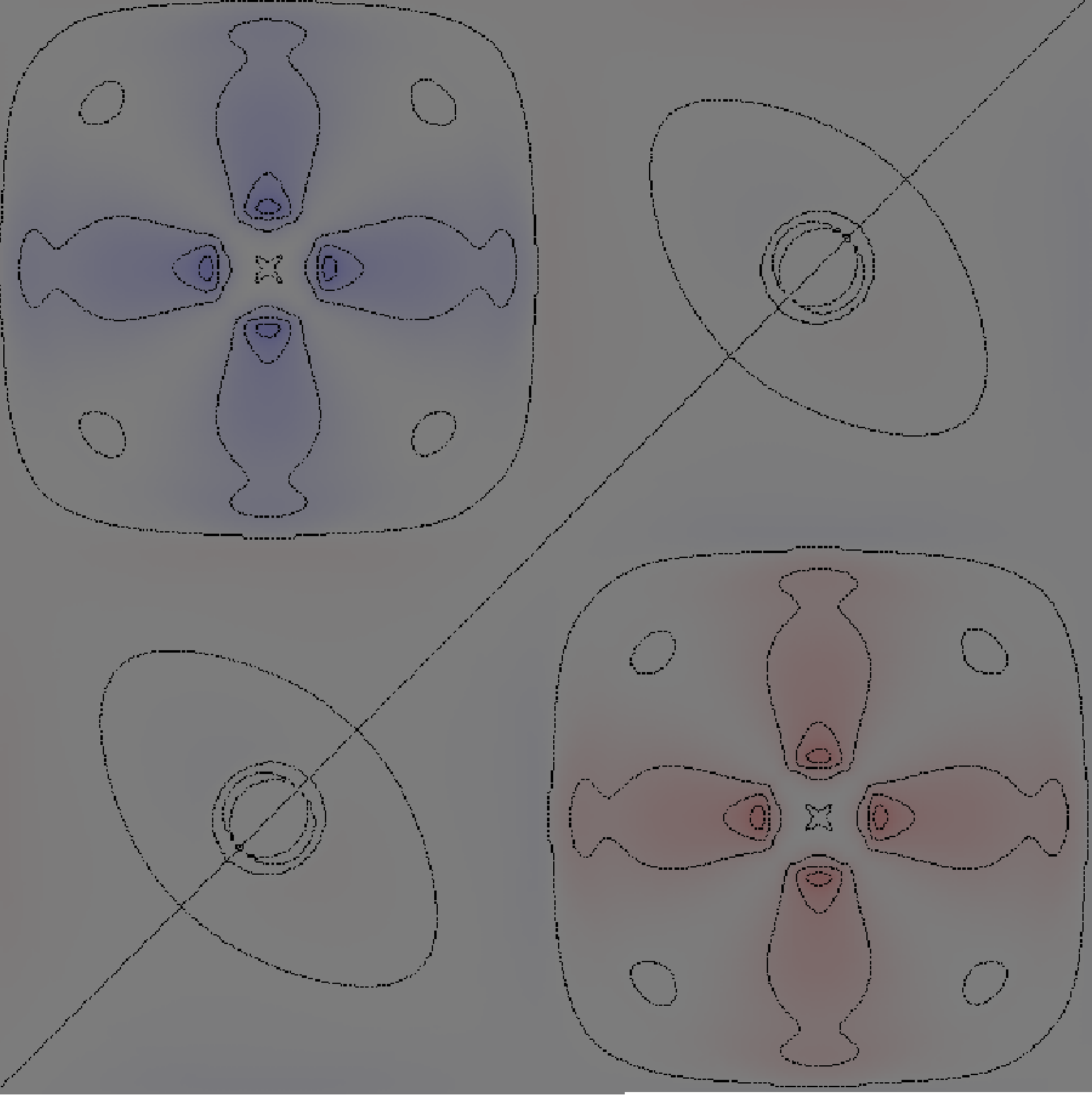}}
\put(4.3,4.7){\epsfxsize=4.2cm \epsfbox{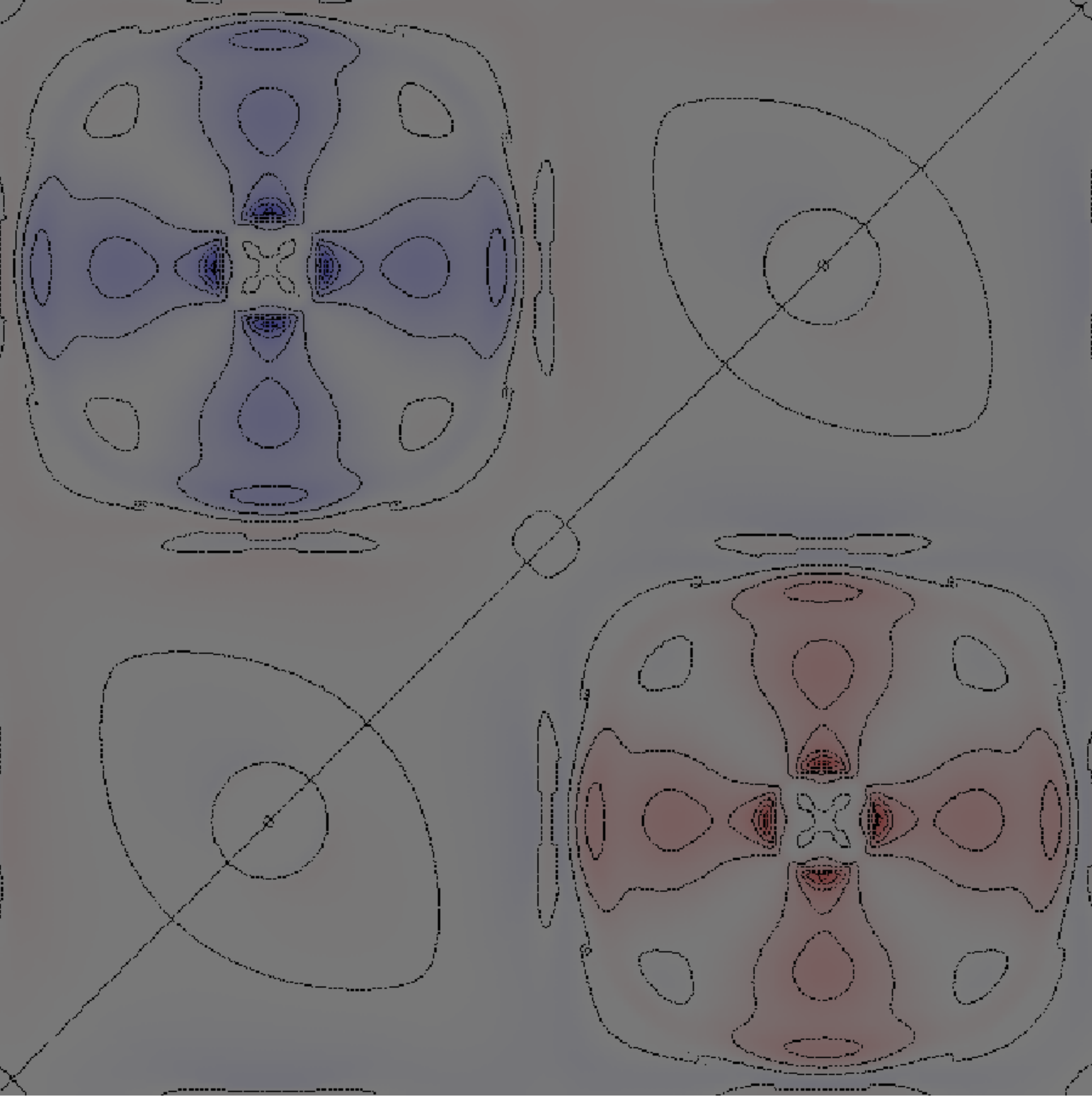}}
\put(0,0){\epsfxsize=4.2cm \epsfbox{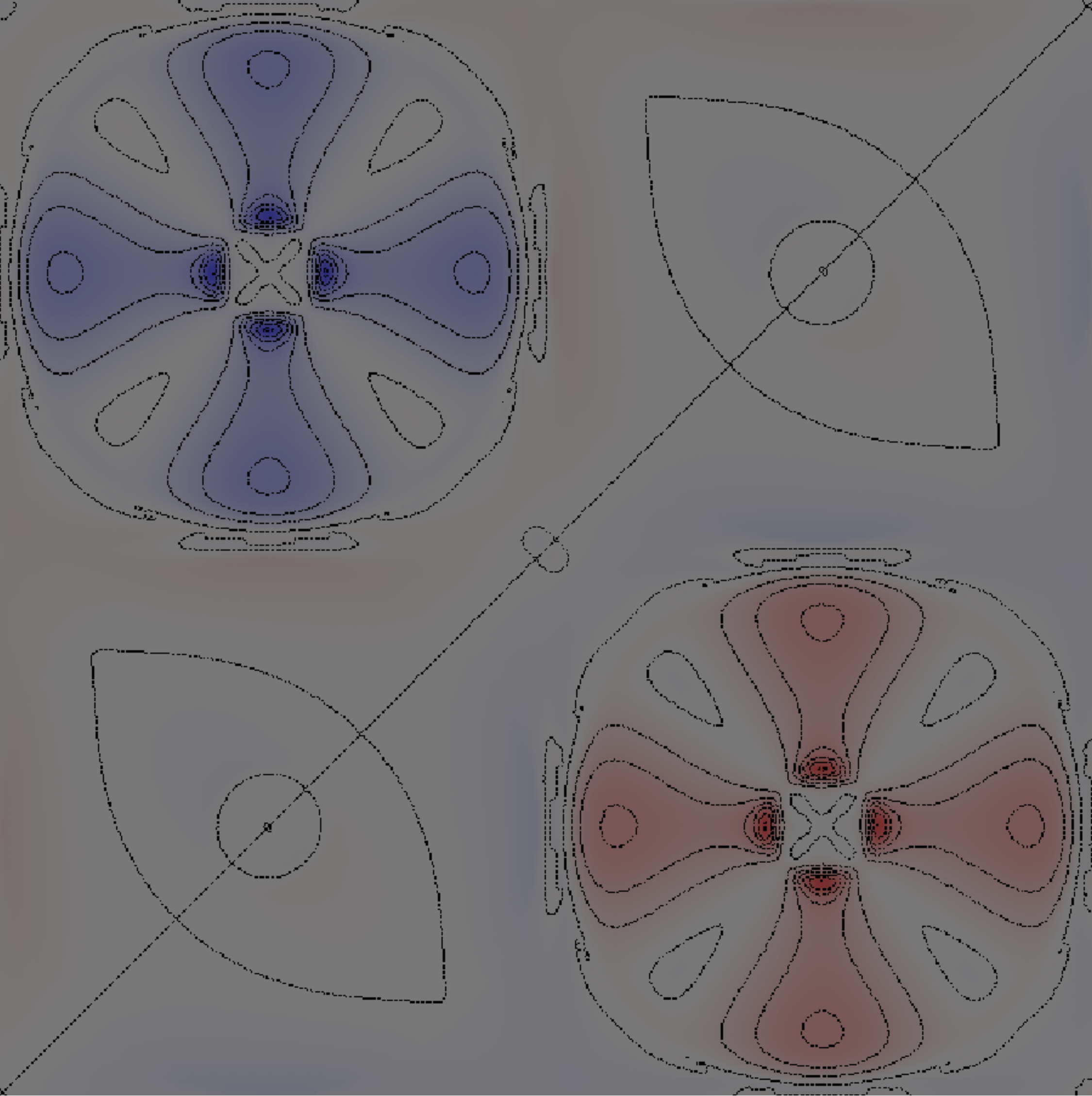}}
\put(4.3,0){\epsfxsize=4.2cm \epsfbox{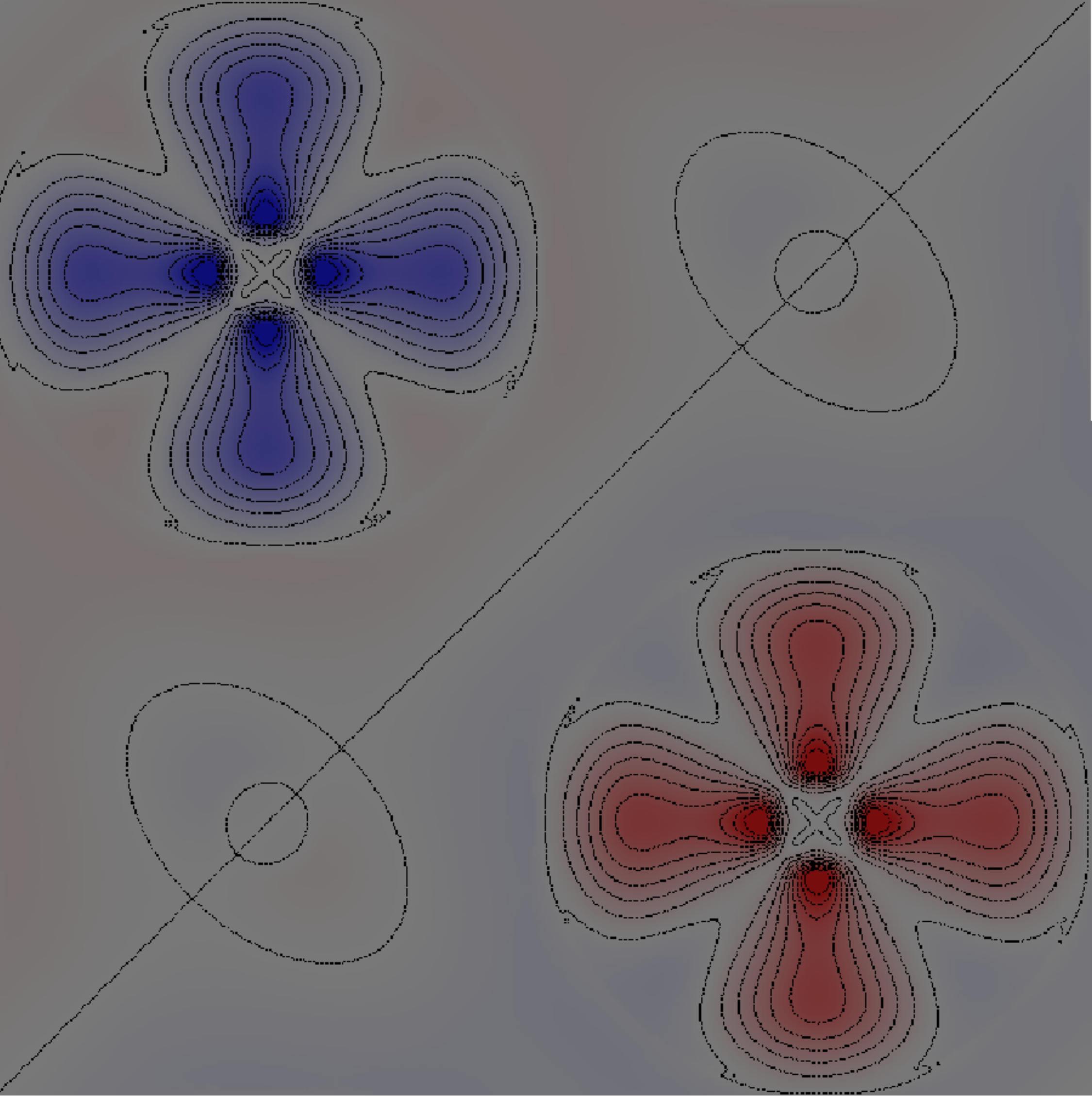}}
\put(0.1,13.7){EXX-OEP}
\put(4.4,13.7){LDA}
\put(0.1,9.0){PBE}
\put(4.4,9.0){EV93}
\put(0.1,4.35){AK13}
\put(4.4,4.35){gBJ$(0.4,1.3,0.65)$}
\end{picture}
\caption{\label{fig10}(Color online) Two-dimensional plots of the difference
between spin-up and spin-down exchange potentials
($v_{\text{x}}^{\uparrow}-v_{\text{x}}^{\downarrow}$) in a $(001)$ plane of
antiferromagnetic NiO. The contour lines start at $-2$ Ry (blue color) and end
at 2 Ry (red color) with an interval of 0.235 Ry. The Ni atom with a full spin-up
$3d$-shell is at the left upper corner.}
\end{figure}

In the following, the results of the previous subsections are set in relation
to the spatial form of the different exchange potentials. We start with
discussing the exchange potential of C in the (110) plane (see Fig.~\ref{fig6}).
Since the LDA potential depends only on the electron density $\rho$,
the corresponding contour plot is the most structureless. In comparison to
the other potentials it exhibits a more spherical shape around the C atoms
and is less corrugated in the interstitial region. It is less attractive
(i.e., negative) than EXX-OEP in the bonding region, but more attractive
in the interstitial. Therefore, compared to LDA there is a transfer of
electrons from the interstitial to the bonding region with EXX-OEP
(see Sec.~\ref{analysisrho}). The GGA potentials (PBE,
B88, EV93, and AK13), that depend additionally on the first and second
derivatives of $\rho$, show stronger spatial variations. For example, the PBE
potential is more undulated than LDA, but features seen in the EXX-OEP are
still reproduced too weakly or completely missing. (B88 is not shown since
its contour plot is very similar to the PBE plot.) The GGA potentials EV93
and AK13 as well as the gBJ potential are more anisotropic. The gBJ
potentials leads to an improved agreement with EXX-OEP both in the bonding
and interstitial regions, whereas the EV93 and AK13 potentials show too much
variation in the interstitial region. We note that similar conclusions can
be drawn for Si, BN, and MgO.

However, it is also clear from Fig.~\ref{fig6} that the agreement between
EXX-OEP and the best semilocal potentials (gBJ) is not perfect and that
differences are still present.
For a more detailed analysis we show in Fig.~\ref{fig7} one-dimensional potentials
plots for Si. It becomes evident from Fig.~\ref{fig7}(a) that AK13 (and to a lesser
extent also EV93) leads to completely unphysical oscillations in the
interstitial region of Si, while the EXX-OEP and gBJ potentials are rather
flat and very similar to each other in this region. Actually, we have observed
that in general the AK13 and EV93 potentials show such large oscillations in
the wide interstitial regions present in such open structures,
which is due to their enhancement factors
$F_{\text{x}}\left(s\right)$ in Eq.~(\ref{ExGGA}) whose magnitudes are much
larger than for PBE and B88 (see Fig.~\ref{fig1}).
The direct effect of these much more positive values of
the AK13 potential in the interstitial region is to shift up the unoccupied
orbitals (located mainly in the interstitial) relative to the occupied ones,
thus explaining the positive (or less negative than for most other potentials)
AK13 values for the ME on the transition energies (Table~\ref{table2}).

In Fig.~\ref{fig7}(b) we can see that the height of the intershell peak at
$d\sim0.6$ \AA~is strongly underestimated and washed-out by PBE, whereas EV93
and AK13 tend to overestimate the peak height for Si. However, with
increasing atomic number the ability of the AK13 and EV93 potentials to
reproduce the height and position of the intershell peaks seems to improve.
As shown in Fig.~\ref{fig8}, in the vicinity of the Cu atom both AK13 and EV93 mimic
the EXX-OEP quite accurately, while substantial differences are present at
the O atom [see Fig.~\ref{fig8}(c)]. Similar observations hold for the intershells
peaks in NiO. This is in agreement with Refs.~\onlinecite{EngelPRB93,ArmientoPRL13}
which show that AK13 and EV93 reproduce very accurately the position and height of the
intershells peaks in transition-metal atoms. As already discussed in
Sec.~\ref{semilocal} and shown in Fig.~\ref{fig2}, the height and position of the
peaks with gBJ depend strongly on the three parameters $\gamma$, $c$, and $p$.
As shown in Fig.~\ref{fig7}(b), the intensity of the peak is too large with
$(\gamma,c,p)=(1.4,1.1,0.50)$, but too weak with
$(\gamma,c,p)=(0.6,1.0,0.60)$ (not shown).

Concerning the BJ-based potential with the UC,
gBJUC, the results are very bad for the EXX total energy and energy position of
core states as discussed above (see Tables~\ref{table1} and \ref{table3}).
This is a consequence of the very inaccurate gBJUC potential in the region
close to the nuclei as shown in Figs.~\ref{fig8}(a) and \ref{fig8}(c).
As discussed in Sec.~\ref{semilocal}, the effect of replacing the kinetic-energy
density $t$ by $D$ in the second term of Eq.~(\ref{vxgBJ}) is very large in the
core region of atoms with the consequence that the potential becomes too attractive.
On the contrary,
without the UC the gBJ potential resembles the EXX-OEP very closely
in the core region [except at the position of the intershell peaks, see
Figs.~\ref{fig8}(a) and \ref{fig8}(c)].

Though, for Cu$_{2}$O it was mandatory to use the UC in order to
obtain qualitative agreement with EXX-OEP for the transition energies and EFG.
For the EFG in particular, this could seem puzzling that the gBJUC potential
gives good results despite it looks quite inaccurate close to the Cu
nucleus. However, as already mentioned in Sec.~\ref{efg}, the EFG is determined
by the non-sphericity of the electron density $\rho$ near the nucleus.
More specifically, the EFG is determined essentially by the
radial function $\rho_{LM}$ with $\left(L,M\right)=\left(2,0\right)$ of the
spherical harmonics expansion of the electron density inside the atomic
sphere.\cite{BlahaPRB88} Figure~\ref{fig9}(a) shows the (expected) very good
agreement between the gBJUC, EXX-OEP, and HF methods
for $\rho_{20}$ (and also for $\rho_{40}$ but not $\rho_{00}$).
By looking at the corresponding radial function $v_{\text{x},20}$ of the
exchange potential [see Fig.~\ref{fig9}(b)], we can observe a rather good
agreement between gBJUC and EXX-OEP in the region beyond 0.2 \AA, which
mainly concerns the $d$-$d$ component of the EFG (see Table~\ref{table4}).
We mention that the radial functions $\rho_{20}$ and $v_{\text{x},20}$ obtained
with B88, EV93, and AK13 are qualitatively similar to PBE and gBJ without UC.
For the $p$-$p$ component, the agreement between gBJUC and EXX-OEP
also comes from the valence region of the Cu atom and the interstitial,
and, as shown in Fig.~\ref{fig8}(b), these two potentials are
relatively close to each other in this region.
Actually, the correct description of the Cu-$p$
states far away from the Cu nucleus affects the anisotropy of the
Cu-$p$-states close to the Cu nucleus. These similarities observed in the
EXX-OEP and gBJUC potentials can also explain the agreement
for the transition energies.

Turning to antiferromagnetic NiO, the difference
$v_{\text{x}}^{\uparrow}-v_{\text{x}}^{\downarrow}$ between the spin-up and
spin-down exchange potentials is shown in Fig.~\ref{fig10}.
The angular $e_{g}$-shape around the Ni atoms is the most pronounced
with the EXX-OEP and gBJ [with $(\gamma,c,p)=(0.4,1.3,0.65)$] potentials,
thus leading to the large band gaps between the $t_{2g}$ and $e_{g}$
states of the minority spin (see DOS in Fig.~\ref{fig5}) in comparison to the other
potentials. However, it can also be observed that the magnitude of
$v_{\text{x}}^{\uparrow}-v_{\text{x}}^{\downarrow}$ is the largest with EXX-OEP
(i.e., compared to gBJ there are a couple of additional isolines between the Ni
and O atoms), which could explain the large exchange splitting between the
spin-up and spin-down states observed in Fig.~\ref{fig5} for EXX-OEP.

\subsection{\label{analysisrho}Analysis of the electron density}

\begin{figure}
\begin{picture}(8.6,13.9)(0,0)
\put(0,9.2){\epsfxsize=4.2cm \epsfbox{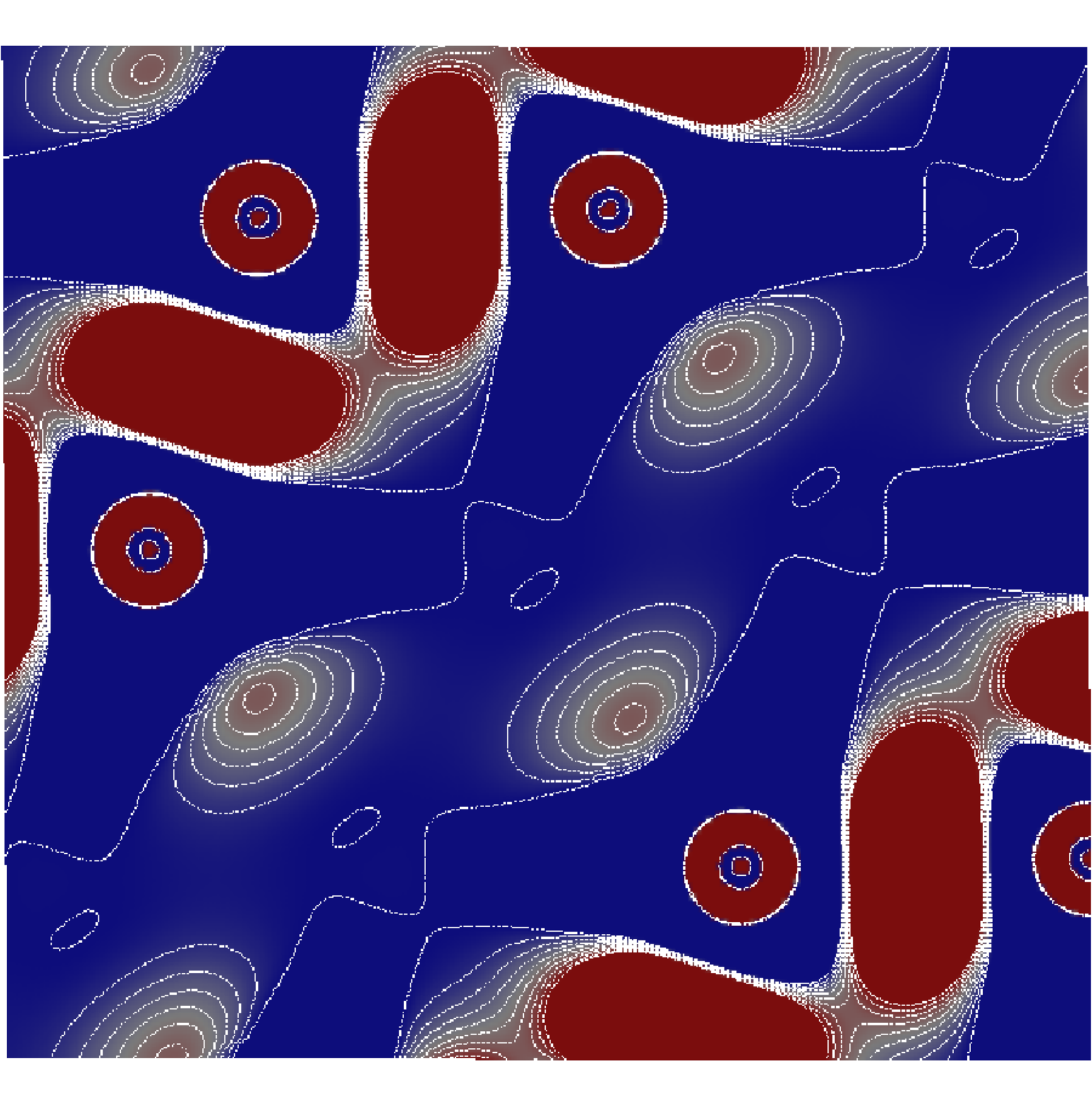}}
\put(4.3,9.2){\epsfxsize=4.2cm \epsfbox{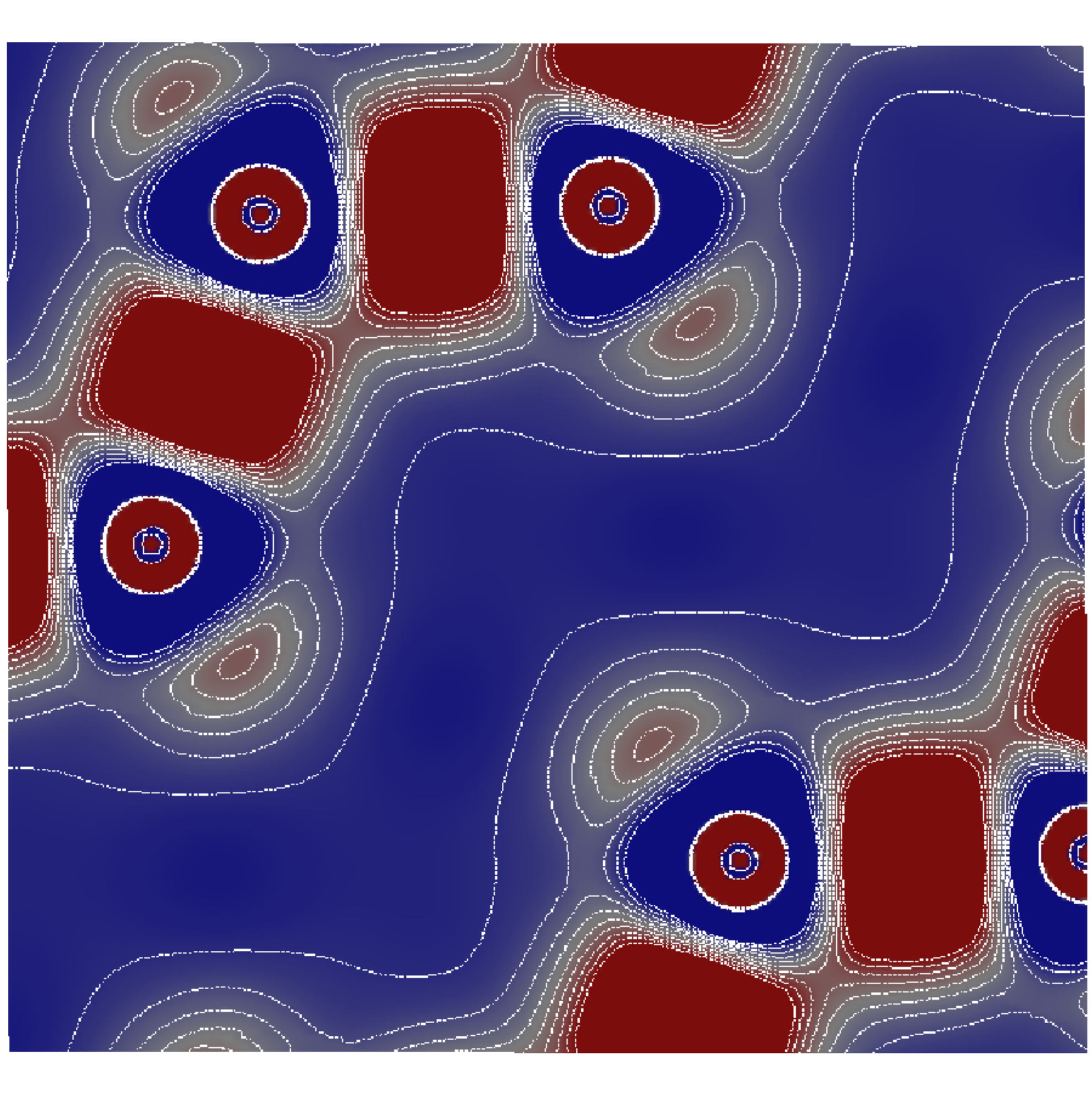}}
\put(0,4.6){\epsfxsize=4.2cm \epsfbox{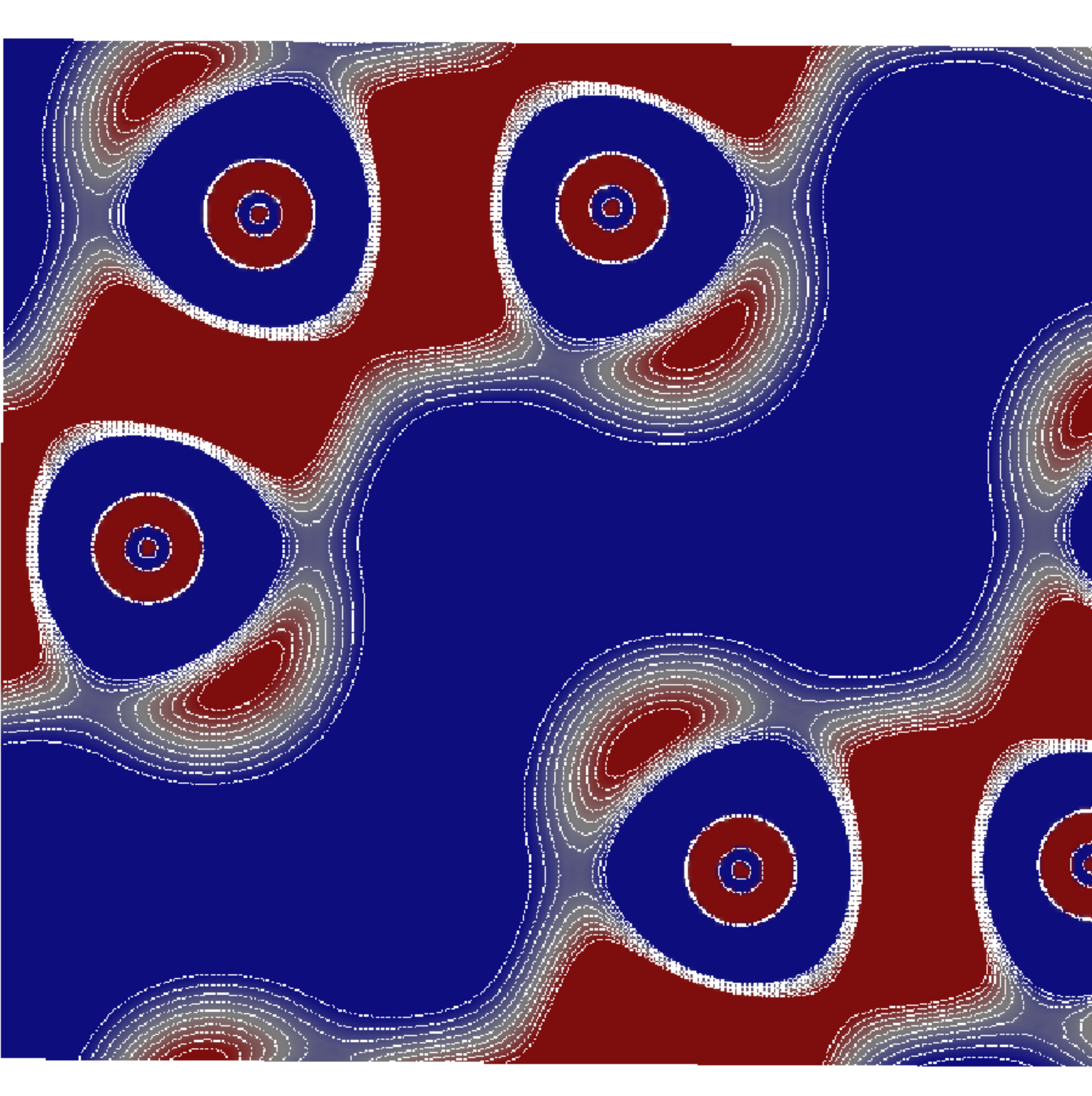}}
\put(4.3,4.6){\epsfxsize=4.2cm \epsfbox{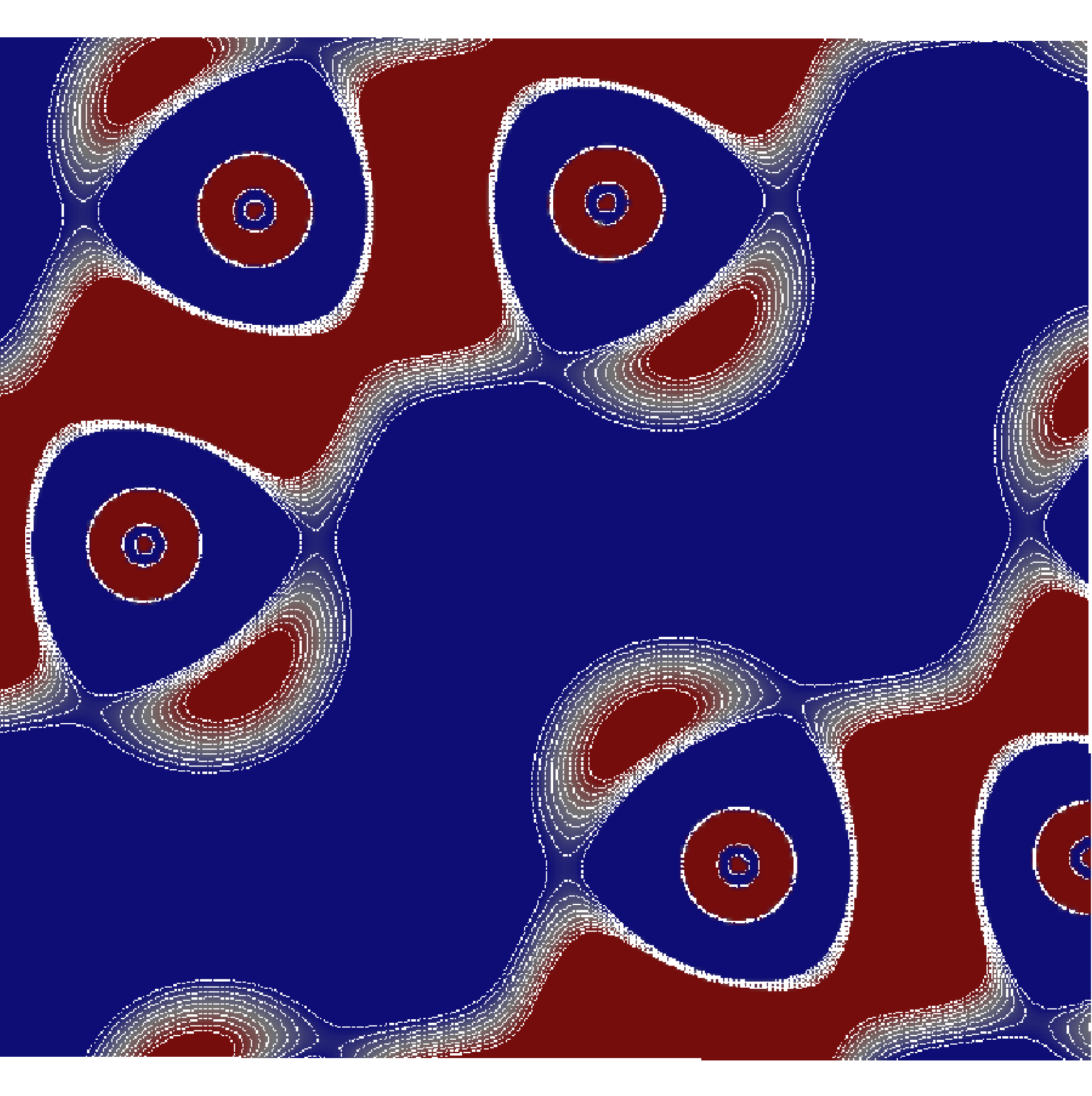}}
\put(0,0){\epsfxsize=4.2cm \epsfbox{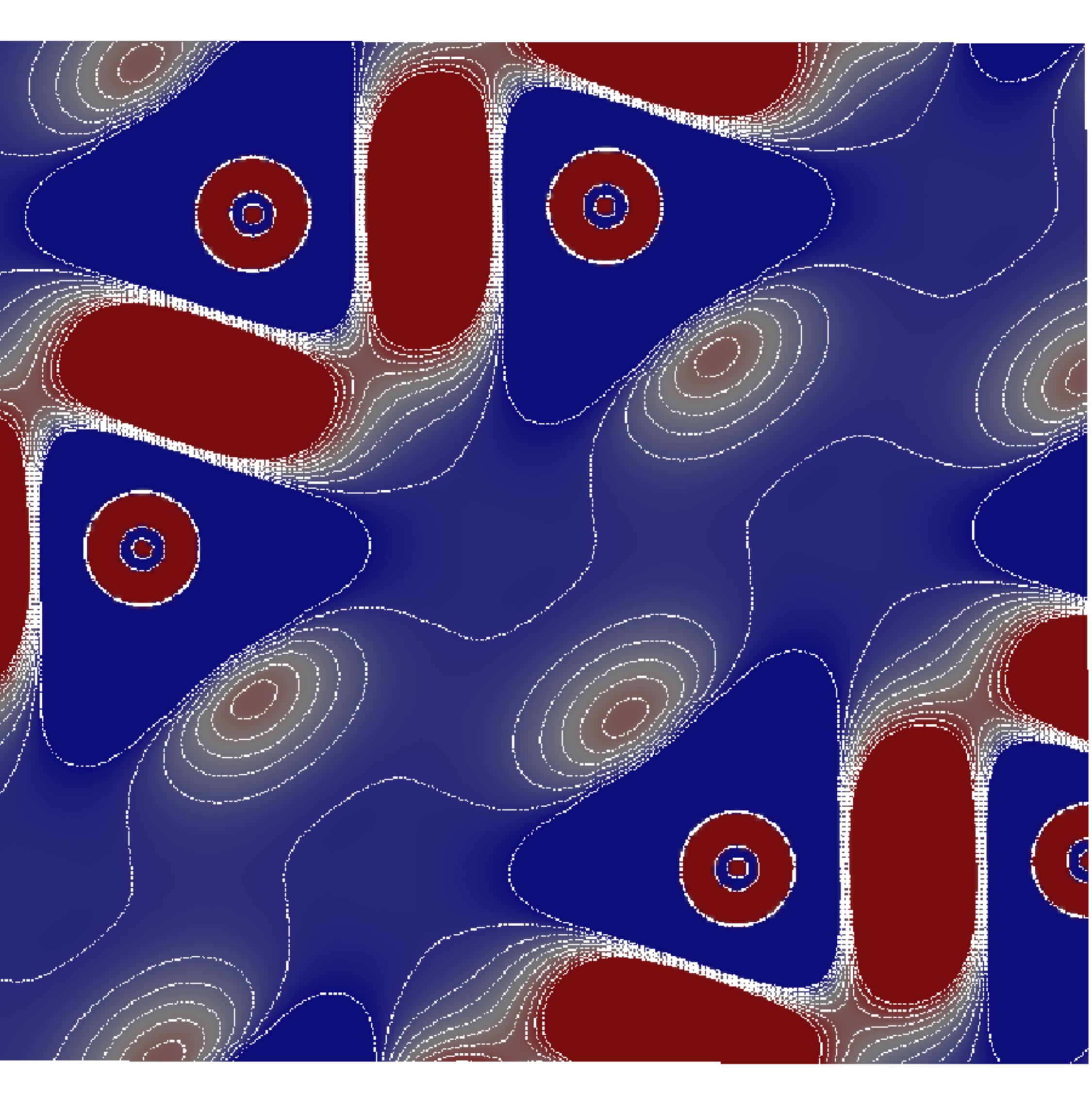}}
\put(4.3,0){\epsfxsize=4.2cm \epsfbox{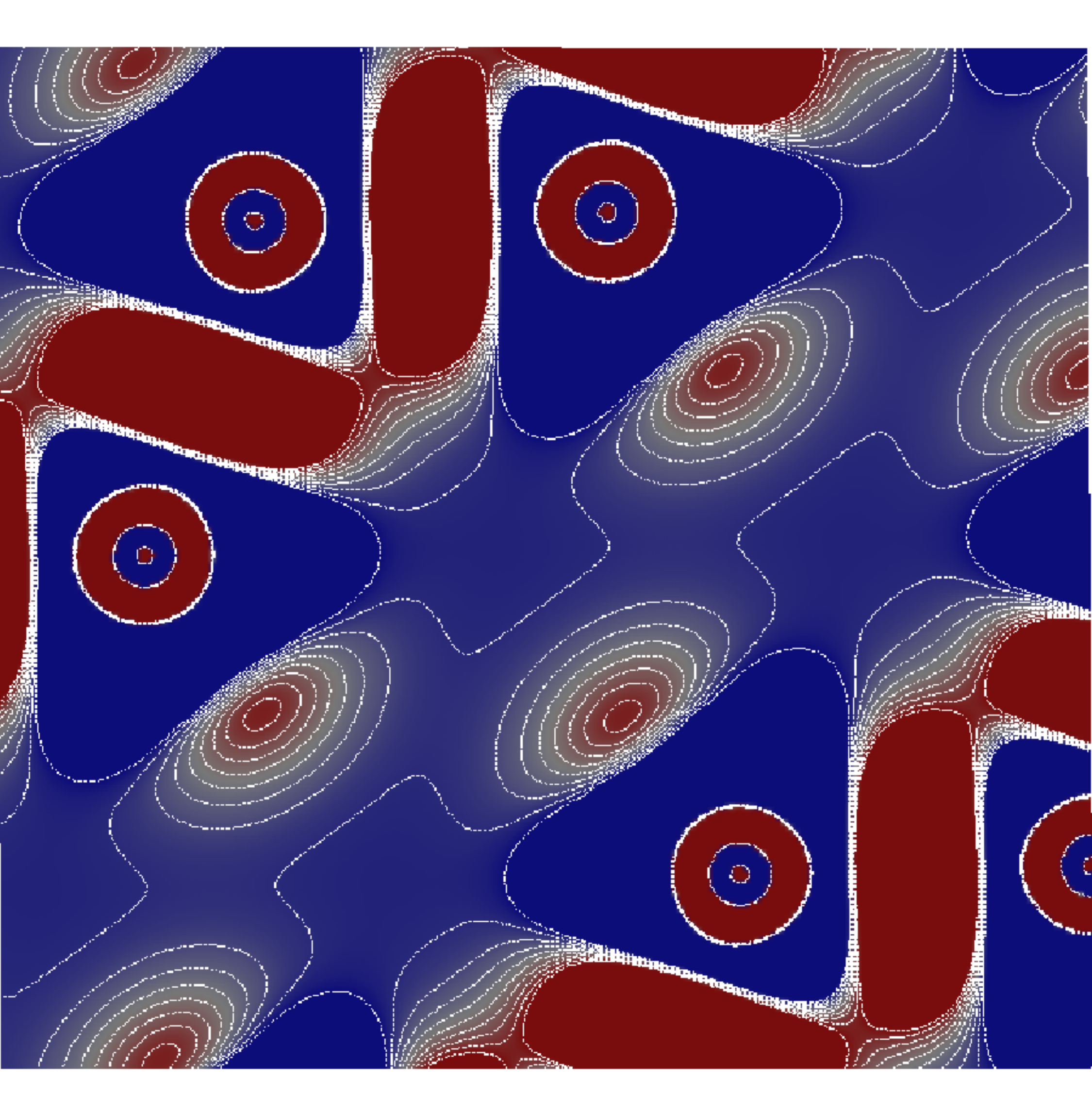}}
\put(0.1,13.5){EXX-OEP$-$LDA}
\put(4.4,13.5){PBE$-$LDA}
\put(0.1,8.9){EV93$-$LDA}
\put(4.4,8.9){AK13$-$LDA}
\put(0.1,4.3){gBJ$(0.6,1.0,0.60)$$-$LDA}
\put(4.4,4.3){gBJ$(1.4,1.1,0.50)$$-$LDA}
\end{picture}
\caption{\label{fig11}(Color online)
Electron density $\rho$ obtained with different exchange potentials minus
$\rho^{\text{LDA}}$ plotted in a $(110)$ plane of Si. The contour lines start
at $-0.001$ electron/bohr$^{3}$ (blue color) and end at 0.001
electron/bohr$^{3}$ (red color) with an interval of 0.0002 electron/bohr$^{3}$.}
\end{figure}

\begin{figure}
\begin{picture}(8.6,14.4)(0,0)
\put(0,9.6){\epsfxsize=4.2cm \epsfbox{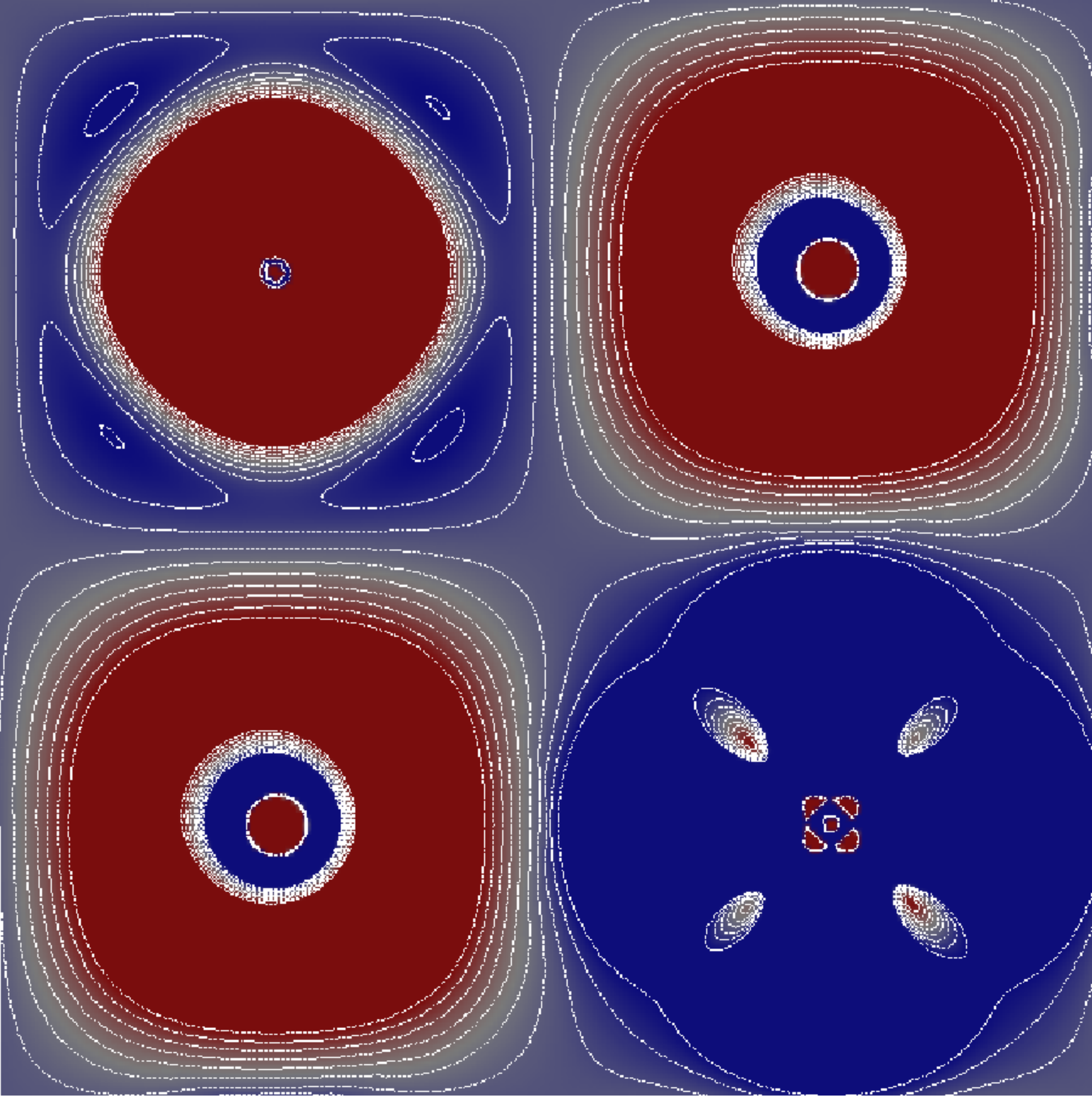}}
\put(4.3,9.6){\epsfxsize=4.2cm \epsfbox{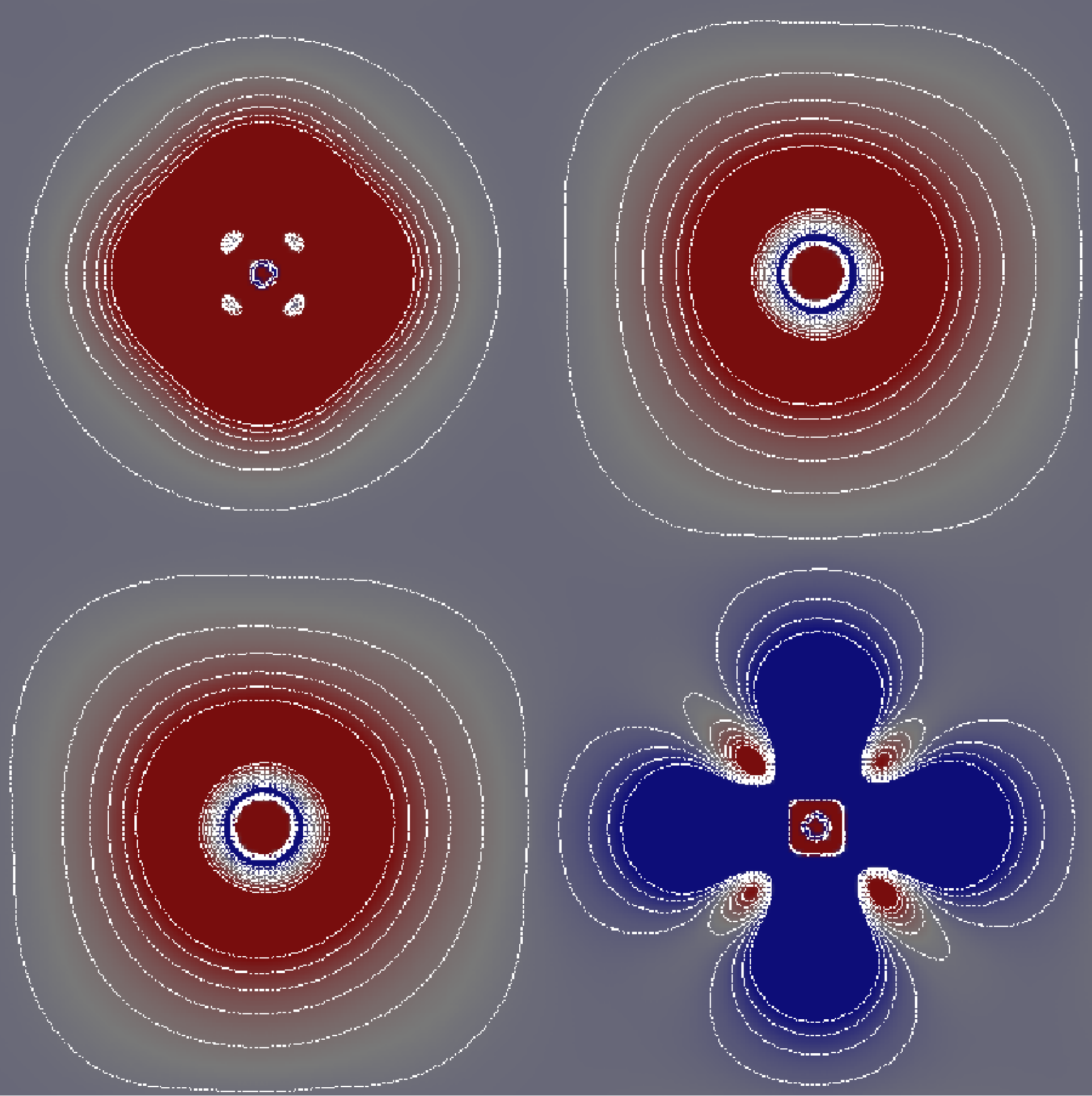}}
\put(0,4.8){\epsfxsize=4.2cm \epsfbox{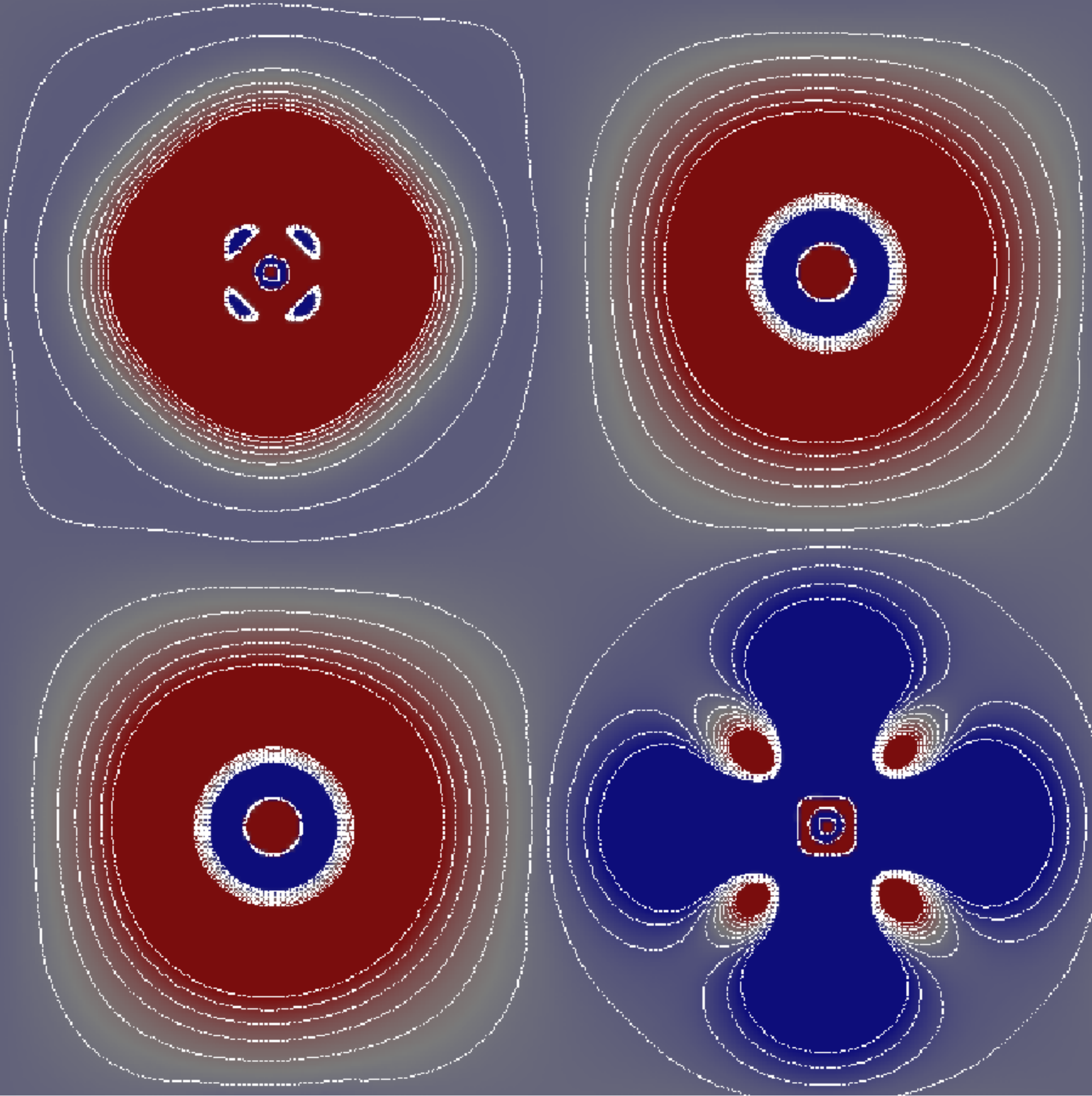}}
\put(4.3,4.8){\epsfxsize=4.2cm \epsfbox{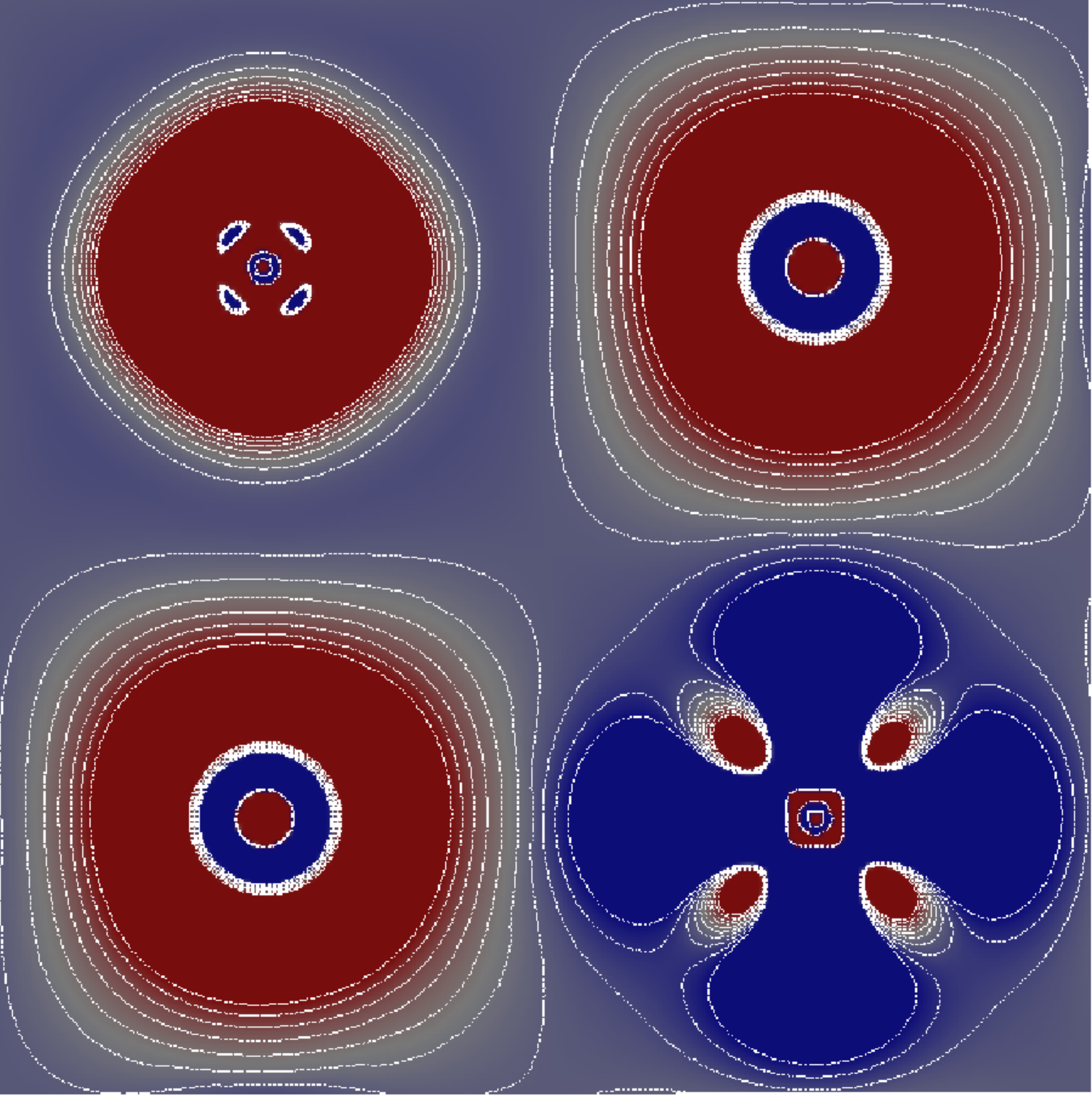}}
\put(0,0){\epsfxsize=4.2cm \epsfbox{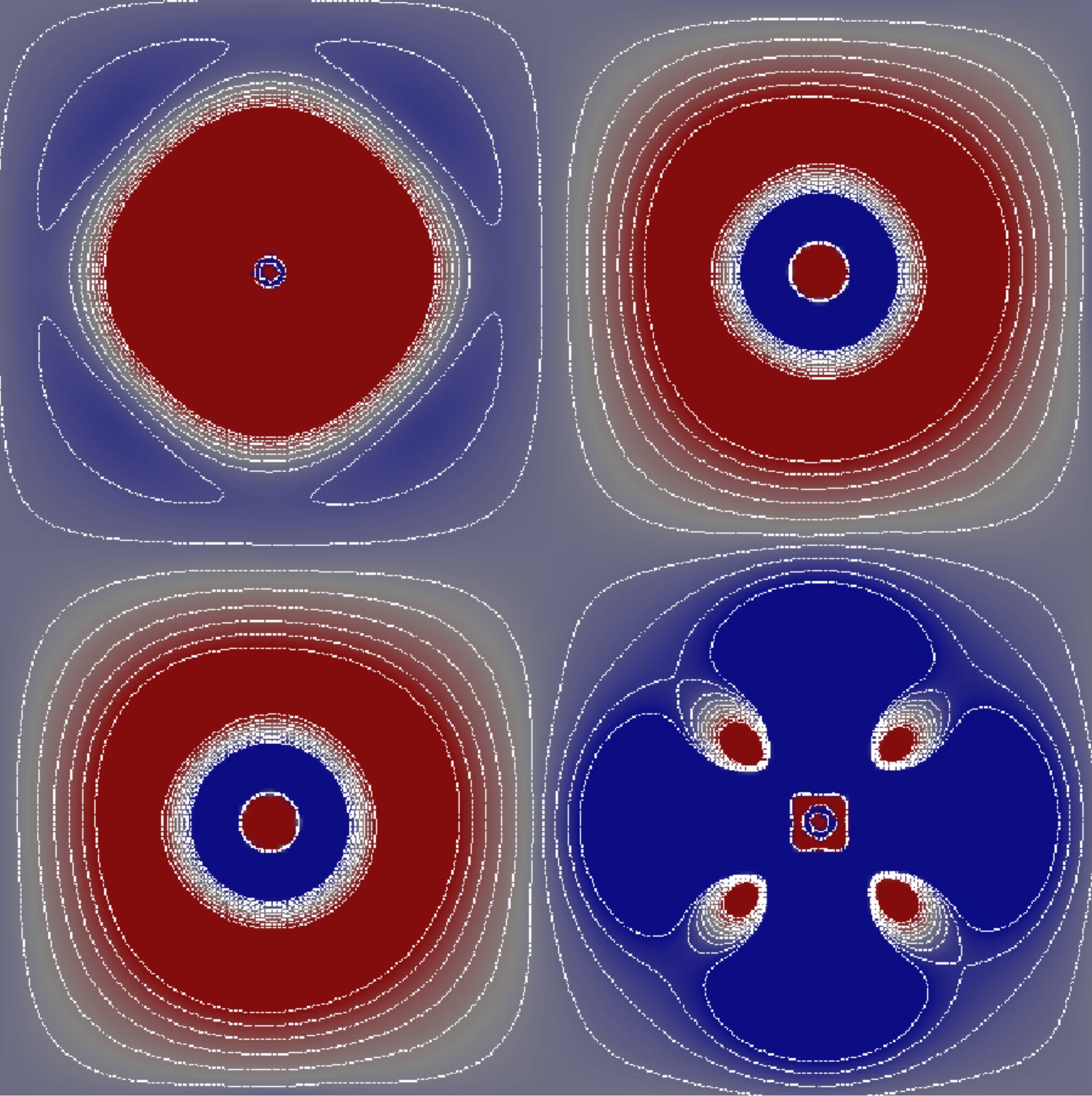}}
\put(4.3,0){\epsfxsize=4.2cm \epsfbox{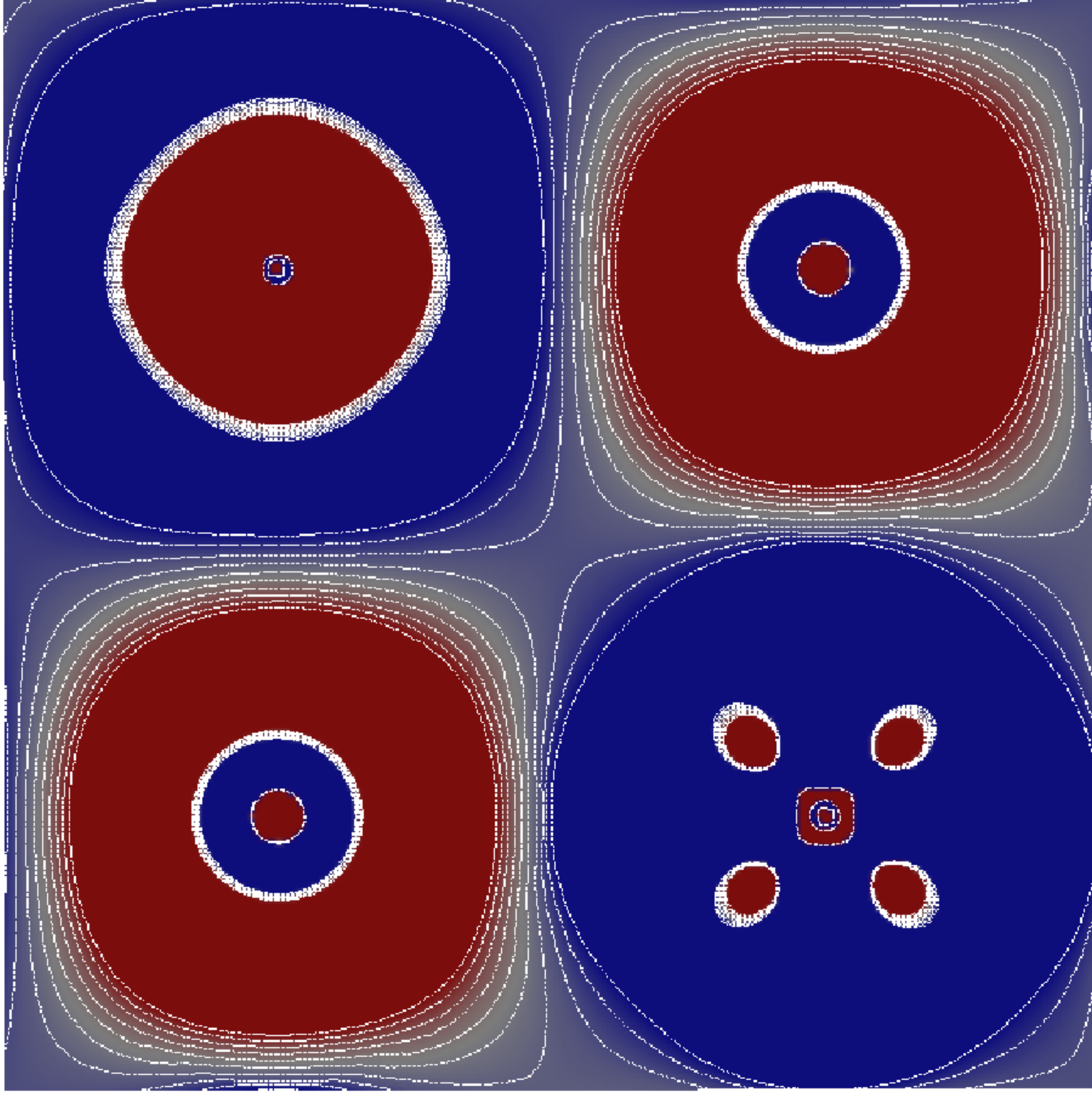}}
\put(0.1,14){EXX-OEP$-$LDA}
\put(4.4,14){PBE$-$LDA}
\put(0.1,9.2){EV93$-$LDA}
\put(4.4,9.2){AK13$-$LDA}
\put(0.1,4.4){gBJ$(0.6,1.0,0.60)$$-$LDA}
\put(4.4,4.4){gBJ$(0.4,1.3,0.65)$$-$LDA}
\end{picture}
\caption{\label{fig12}(Color online)
Spin-up electron density $\rho_{\uparrow}$ obtained with different exchange
potentials minus $\rho_{\uparrow}^{\text{LDA}}$ plotted in a $(001)$ plane of
antiferromagnetic NiO. The contour lines start at $-0.005$ electron/bohr$^{3}$
(blue color) and end at 0.005 electron/bohr$^{3}$ (red color) with an interval
of 0.001 electron/bohr$^{3}$. The Ni atom with a full spin-up $3d$-shell is at
the left upper corner.}
\end{figure}
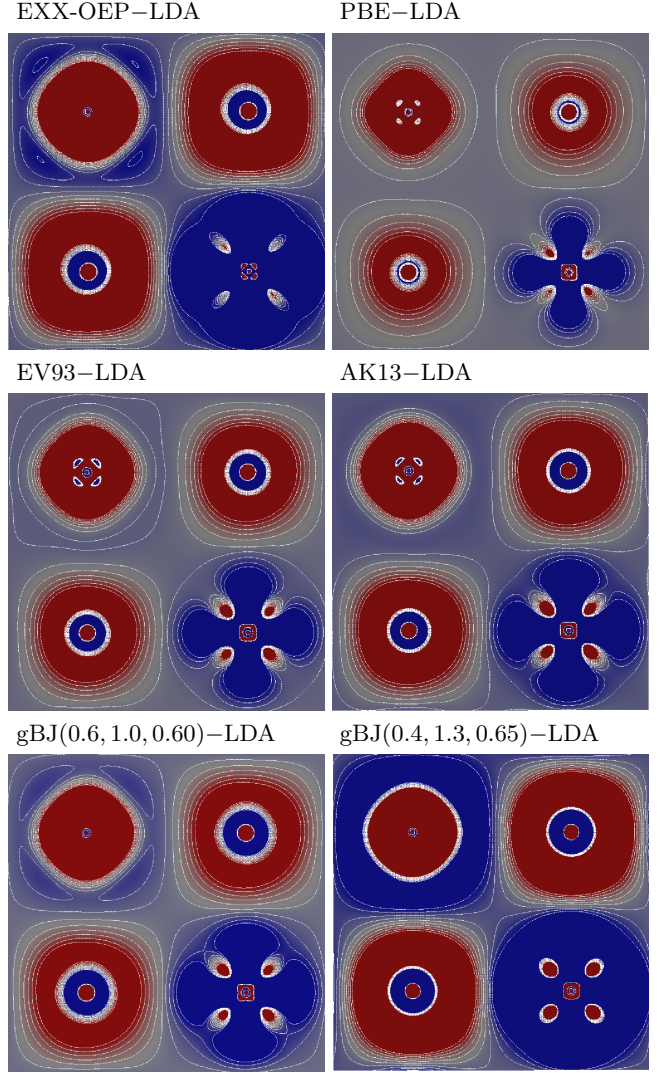

\begin{figure}
\includegraphics[width=\columnwidth]{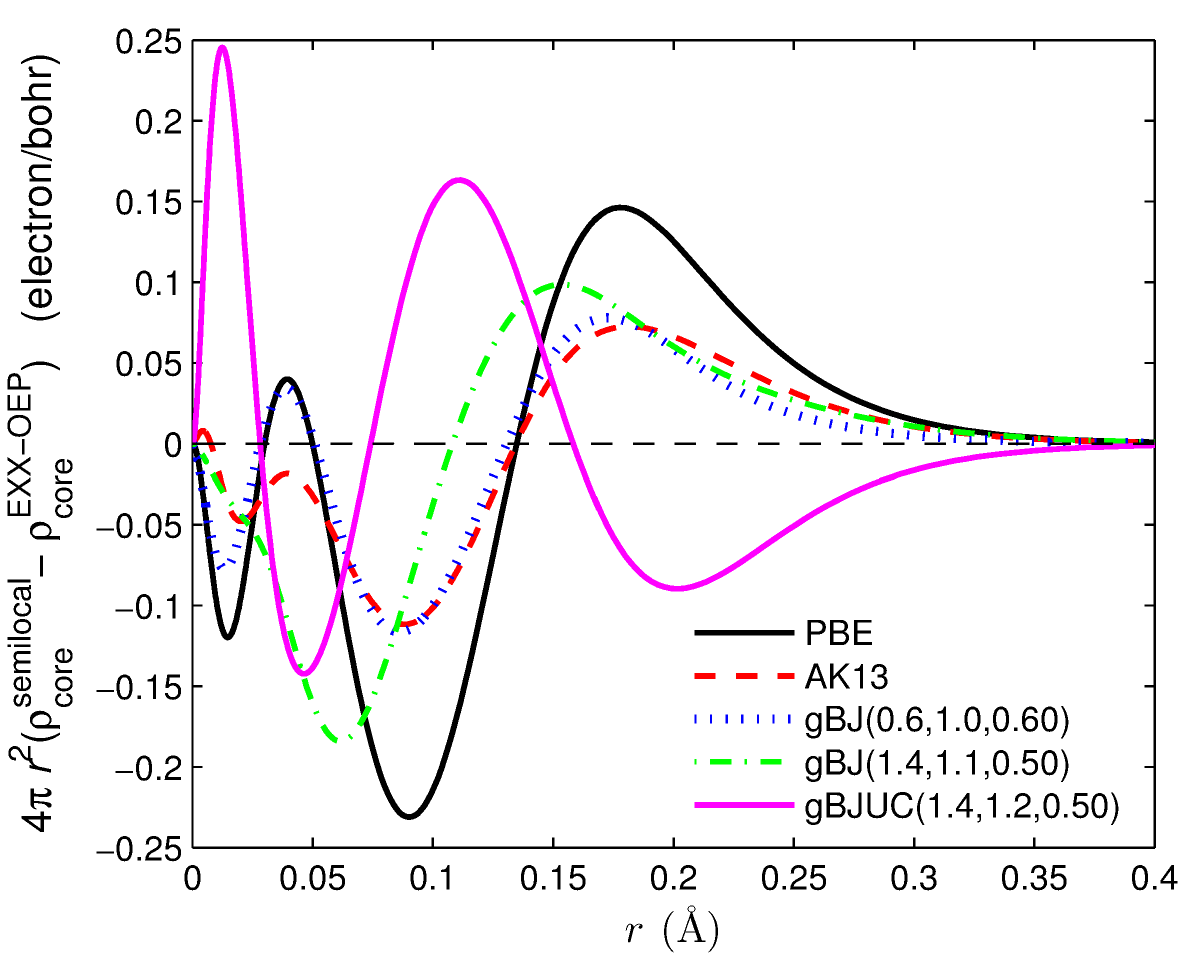}
\caption{\label{fig13}(Color online) The difference between the density
$\rho_{\text{core}}$ of the core electrons of Cu in Cu$_{2}$O calculated with
semilocal functionals and $\rho_{\text{core}}$ from EXX-OEP (multiplied by
$4\pi r^{2}$).}
\end{figure}

Figure~\ref{fig11} shows the electron density in Si
obtained from various potentials minus the LDA density, which
serves as reference. As discussed above for the case of C (Fig.~\ref{fig6}),
which is very similar to Si and BN, the GGA, gBJ, and EXX-OEP potentials are
more attractive (repulsive) than LDA in the bonding (interstitial) region of Si.
Consequently, the electron density is increased (decreased) in the bonding
(interstitial) region. From Fig.~\ref{fig11} it is rather clear that the
gBJ and EXX-OEP potentials lead to very similar electron densities, while the
EV93 and AK13 densities are, compared to EXX-OEP, too large (small) in the
bonding (interstitial) regions.

As shown in Fig.~\ref{fig12} for NiO, the effect of using a beyond-LDA potential
is to increase the spin-up electron density $\rho_{\uparrow}$ on the Ni atom
with a full spin-up $3d$-shell (red regions around the Ni atom
at the left upper corner) and to
decrease the spin-down density $\rho_{\downarrow}$ on the same Ni atom
(which corresponds to $\rho_{\uparrow}$ around the other Ni atom).
This results in an increase of the magnetic moment of the Ni atom as
discussed above (see Table~\ref{table4}). The other effect of using a
beyond-LDA potential is to increase the ionicity (red regions around the
O atoms). More quantitatively, compared to LDA the number of electrons
inside the sphere of the Ni atom changes by $+0.01$ (PBE), $-0.01$ (EV93, AK13),
$-0.14$ [gBJ$(0.4,1.3,0.65)$], and $-0.11$ (EXX-OEP), while for the O atom
the changes are $+0.10$ (PBE), $+0.16$ (EV93), $+0.23$ (AK13),
$+0.47$ [gBJ$(0.4,1.3,0.65)$], and $+0.35$ (EXX-OEP), which shows that
gBJ reproduces quite accurately the trends of EXX-OEP.
From Fig.~\ref{fig12}, it is also rather clear that the gBJ$(0.4,1.3,0.65)$
and EXX-OEP electron densities are overall very similar (note in particular the
asymmetry of the $3d$ density around the Ni atom with partially filled
spin-up electrons).

In Fig.~\ref{fig13} we show the density of the core electrons of the Cu atom in
Cu$_{2}$O. Compared to the density $\rho_{\text{core}}$ obtained with EXX-OEP,
the PBE core density is less contracted since it has smaller values for
$r<0.14$ \AA~but is larger for $r>0.14$ \AA. The reverse is true for the gBJUC
potential since the positive values of
$\rho_{\text{core}}^{\text{gBJUC}}-\rho_{\text{core}}^{\text{EXX-OEP}}$ are on
average closer to the nucleus than the negative values, which is a consequence
of the fact that close to the nuclei gBJUC is much more attractive than the
EXX-OEP and all other potentials (as discussed in Secs.~\ref{semilocal} and
\ref{analysispot}). The AK13 and gBJ potentials lead to trends similar to PBE,
however the discrepancies with respect to EXX-OEP are reduced. Note also that
the gBJ potential with the parameters $(\gamma,c,p)=(0.6,1.0,0.60)$, which
are more appropriate for the EXX total energy, leads to a slightly more
accurate core density than with the values $(1.4,1.1,0.50)$ that were
determined for the transition energies.

\section{\label{summaryoutlook}Summary and outlook}

In this work, we have compared several approximate semilocal
exchange potentials to the exact EXX-OEP. The closeness between
the semilocal and EXX-OEP potentials was quantified by considering
the EXX total energy and electronic band structure for
various solids, as well as the EFG in Cu$_{2}$O and the magnetic moment in NiO.
An attempt to parameterize a semilocal BJ-based potential has also been made
and we have shown that by the introduction of a few
parameters, it was possible to improve substantially the
agreement with EXX-OEP compared to the GGA and original BJ potentials.
However, it became also obvious that there is
no universal set of parameters that leads to satisfying results for all
properties and solids at the same time. For instance, if a set of parameters
is appropriate for the EXX total energy (or electronic structure) of C, Si, BN,
and MgO, then it will not work so well for antiferromagnetic NiO, and vice-versa.
Another example was Cu$_{2}$O for which it is mandatory
to use the UC to obtain qualitative agreement with EXX-OEP for the
band gap and EFG, while the UC is very detrimental for the EXX total energy
and energy position of the core states in all solids.

From the results, it is clear but not surprising that although the BJ-based
potentials lead to interesting results, the semilocal approximations show
limitations and, furthermore, there is no systematic way to improve their
accuracy. Beyond the semilocal level of theory, there is the group of exchange
potentials which consist of the nonlocal Slater potential $v_{\text{x}}^{\text{S}}$
[Eq.~(\ref{vxS})] plus a term which is either nonlocal [like in the
Krieger-Li-Iafrate (KLI)\cite{KriegerPRA92a} or localized
HF (LHF)\cite{DellaSalaJCP01,GritsenkoPRA01} potentials] or local
with an eventual dependency on the energies of the occupied
orbitals.\cite{HarbolaIJQC02,GritsenkoPRA95}
The computational cost of these potentials is rather high since the
Slater potential and the nonlocal terms require the calculation of HF-type
integrals (see Ref.~\onlinecite{KohutJCP14} for a summary
of the expression of these potentials). These nonlocal potentials avoid some of
the technical difficulties of EXX-OEP as their construction involves only
the occupied orbitals. There are numerous studies on the Slater-based potentials
and we just mention Ref.~\onlinecite{RyabinkinPRL13} where it was shown that the
KLI- and LHF-generated orbitals lead to EXX total energy of atoms which are much
lower than with BJ orbitals. However, Engel (Ref.~\onlinecite{EngelPRL09}) noted
that the KLI approximation is not able to open the band gap in antiferromagnetic
FeO, while a band gap of 1.66 eV is obtained with EXX-OEP (augmented by LDA for
correlation).

On the other hand, as already mentioned in Sec.~\ref{semilocal}, the semilocal
BR potential\cite{BeckePRA89} [Eq.~(\ref{vxBR})] seems to reproduce (at least
visually) quite accurately the features of the Slater potential in atoms,
\cite{BeckePRA89,BeckeJCP06} however, the agreement is not perfect (see
Ref.~\onlinecite{KarolewskiPRA13}) and the comparison of the EXX total energies
evaluated with BJ(Slater) and BJ(BR) orbitals shows non-negligible differences
in some cases.\cite{BeckeJCP06} Avoiding the calculation of the Slater
potential by using the BR potential instead would certainly be advantageous,
but more comparison studies between the BR and Slater potentials are needed.

A possible way of improving the reliability of a semilocal
potential (e.g., gBJ) to reproduce EXX-OEP, could be to use a similar
parameterization as the one used for the constant $c$ in the mBJ
potential\cite{TranPRL09} [Eq.~(\ref{vxmBJ})]:
\begin{equation}
c = \alpha + \beta\left(\frac{1}{V_{\text{cell}}}\int\limits_{\text{cell}}
\frac{\left\vert\nabla\rho(\mathbf{r})\right\vert}{\rho(\mathbf{r})}
d^{3}r\right)^{1/2},
\label{c}
\end{equation}
where $\alpha$ and $\beta$ are parameters and $V_{\text{cell}}$ is the unit
cell volume. It has been shown that with the optimized values $\alpha=-0.012$
and $\beta=1.023$ bohr$^{1/2}$, mBJ reproduces with rather great accuracy the
experimental band gap of many solids.
\cite{TranPRL09,SinghPRB10,KollerPRB11,KollerPRB12}
Actually, the use of an integral expression like
Eq.~(\ref{c}) is a way to introduce nonlocality (similar as with the Slater
potential), but in a cheap way since there is no summation over orbitals like
in the Slater potential. However, the drawback of using the average of
$\left\vert\nabla\rho\right\vert/\rho$ in the unit cell is that this
quantity is infinite for systems with an infinite vacuum (e.g., isolated
molecule or surface). An alternative to Eq.~(\ref{c}) which can be applied to any
kind of systems might be helpful in improving the universal character of a
potential like gBJ. For instance, as suggested by Marques \textit{et al}. in
Ref.~\onlinecite{MarquesPRB11}, a possibility would be to make
$c$ $\mathbf{r}$-dependent and the integrand in Eq.~(\ref{c})
localized around $\mathbf{r}$
by multiplying $\left\vert\nabla\rho\right\vert/\rho$ by
a function of $\left\vert\mathbf{r}-\mathbf{r}'\right\vert$ which goes
to zero at $\left\vert\mathbf{r}-\mathbf{r}'\right\vert\rightarrow\infty$.

In order to adjust the parameters of an approximate functional for each system,
an approach as suggested in Ref.~\onlinecite{YangPRL02} might be helpful.
The free parameters of the potential are adapted at each iteration such that
the EXX total energy becomes minimal. However, such a procedure is rather
expensive since the equations to determine the parameters involve HF-like
matrix elements between occupied and unoccupied orbitals. Nevertheless, this
would be a way to adjust the parameters for each solid and therefore improve
the universality of the potential.

Moreover, we mention the work of Staroverov and co-workers
\cite{RyabinkinPRL13,KohutJCP14} who proposed an expression for a multiplicative
exchange potential (making no use of unoccupied orbitals) which leads to
results that are quasi-identical to the EXX-OEP results. However,
this approach requires the HF orbitals which reduces
its use for solids and large scale applications.

Finally, we note the very few studies reporting OEP calculations including
correlation like the random-phase approximation (RPA) in addition to EXX (see
Refs.~\onlinecite{KotaniJMMM98,KotaniJPCM98,GruningJCP06,GruningPRB06,KlimesJCP14}
for results on solids). The RPA-OEP potentials could certainly also serve as
reference for the modelling of realistic multiplicative exchange-correlation
potentials $v_{\text{xc}}$ including correlation. 

\begin{acknowledgments}

This work was supported by the project SFB-F41 (ViCoM) of the Austrian Science
Fund. M. B. gratefully acknowledges financial support from the Helmholtz
Association through the Hemholtz Postdoc Programme (VH-PD-022).

\end{acknowledgments}

\bibliography{/area52/tran/divers/references}

\end{document}